\documentclass[%
  preprint, 
  amsmath,amssymb,
  aps, physrev,
]{revtex4-2}

\usepackage{graphicx}% Include figure files
\usepackage{dcolumn}% Align table columns on decimal point
\usepackage{bm}% bold math
\usepackage{hyperref}% add hypertext capabilities
\usepackage[ruled,vlined,linesnumbered]{algorithm2e}
\hypersetup{
  colorlinks,
  linkcolor={blue!100!black},
  citecolor={blue!100!black},
  urlcolor={blue!80!black}
}
\usepackage{subcaption}
\usepackage{caption}
\usepackage{amsthm,amssymb,amsfonts}
\usepackage{algpseudocode}
\usepackage{enumitem}
\usepackage{algorithmicx}
\usepackage{graphicx}
\usepackage{textcomp}
\usepackage{xcolor}
\usepackage[normalem]{ulem}
\usepackage{cancel}
\usepackage{changes}
\setdeletedmarkup{\textcolor{red}{\sout{#1}}}

\newcommand{\cA}{\mathcal{A}}

\newcommand{\cS}{\mathcal{S}}
\newcommand{\cT}{\mathcal{T}}

\newcommand{\bolds}{\mathbf{s}}

\newcommand{\boldA}{\mathbf{A}}
\newcommand{\boldB}{\mathbf{B}}
\newcommand{\boldC}{\mathbf{C}}

\newcommand{\boldS}{\mathbf{S}}

\newcommand{\boldX}{\mathbf{X}}

\newtheorem{axiom}{Axiom}

\newtheorem{remark}{Remark}
\newtheorem{definition}{Definition}
\newtheorem{lemma}{Lemma}
\newtheorem{corollary}{Corollary}
\newtheorem{property}{Property}
\newtheorem{theorem}{Theorem}
\newtheorem{observation}{Observation}

\begin{document}

\preprint{APS/123-QED}

\title{Multivariate Partial Information Decomposition: Constructions, Inconsistencies, and Alternative Measures}

\author{Aobo Lyu}
\affiliation{Department of Electrical and Systems Engineering, Washington University in St. Louis, St. Louis, MO, USA}%Lines break automatically or can be forced with \\
\author{Andrew Clark}
\affiliation{Department of Electrical and Systems Engineering, Washington University in St. Louis, St. Louis, MO, USA}%Lines break automatically
\author{Netanel Raviv}%
\email{Contact author: aobo.lyu@wustl.edu}
\affiliation{%
Department of Computer Science and Engineering, Washington University in St. Louis, St. Louis, MO, USA}%

\date{\today}% It is always \today, today,

\begin{abstract}
  While mutual information effectively quantifies dependence between two variables,
  it does not by itself reveal the complex, fine-grained interactions among variables, i.e., how multiple sources contribute redundantly, uniquely, or synergistically to a target in multivariate settings.
  The Partial Information Decomposition (PID) framework was introduced to address this by decomposing the mutual information between a set of source variables and a target variable into fine-grained information atoms such as redundant, unique, and synergistic components.
  In this work, we review the axiomatic system and desired properties of the PID framework and make three main contributions.
  First, we resolve the two-source PID case by providing explicit closed-form formulas for all information atoms that satisfy the full set of axioms and desirable properties.
  Second, we prove that for three or more sources, PID suffers from fundamental inconsistencies:
  we review the known three-variable counterexample where the sum of atoms exceeds the total information, and extend it to a comprehensive impossibility theorem showing that no lattice-based decomposition can be consistent for all subsets when the number of sources exceeds three.
  Finally, we deviate from the PID lattice approach to avoid its inconsistencies, and present explicit measures of multivariate unique and synergistic information.
  Our proposed measures, which rely on new systems of random variables that eliminate higher-order dependencies, satisfy key axioms such as additivity and continuity,
  provide a robust theoretical explanation of high-order relations, and show strong numerical performance in comprehensive experiments on the Ising model.
  Our findings highlight the need for a new framework for studying multivariate information decomposition.
  % \footnote{Parts of this work appeared in the 2024 IEEE International Symposium on Information Theory~\cite{lyu2024explicit}.}

\end{abstract}

\maketitle

%{\let\thefootnote\relax\footnote{}}

%\tableofcontents
\section{Introduction}\label{sec:intro}
Since its inception by Claude Shannon~\cite{shannon2001mathematical}, mutual information has remained a pivotal measure in information theory, which finds extensive applications across multiple other domains.
Extending mutual information to three-or-more source variable systems has attracted significant academic interest, but no widely agreed upon generalization exists to date. For instance, the so-called \textit{interaction information}~\cite{watanabe1960information} emerged in 1960 as an equivalent notion for mutual information in three-or-more source variable systems, and yet, it provides negative values in many common systems, contradicting Shannon's viewpoint of information measures as nonnegative quantities.

Partial Information Decomposition (PID), introduced by Williams and Beer~\cite{williams2010nonnegative}, addresses this by decomposing the mutual information $I(S_1,\dots, S_N;T)$ between a set of source variables $S_1,\dots, S_N$ and a target $T$ into distinct “information atoms” that describe how the sources contribute to the target both individually and in combination.
The implementation of this framework is based on a structure called  \textit{the redundancy lattice}~\cite{crampton2001completion}, which is inspired by and closely aligned with the principles of classical set theory.
In essence, PID partitions the mutual information into components such as unique information (contributed by one source alone), redundant or shared information (common to multiple sources), and synergistic information (emerging only from specific source combinations), etc.
{For broader background and tutorial expositions on PID and multivariate information, we refer readers to \cite{lizier2018information,finn2020generalised,gutknecht2021bits}.}

Since its proposal, PID has been widely used in various fields.
In brain network analysis, PID (or similar ideas) has been instrumental in measuring correlations between neurons~\cite{schneidman2003synergy} and understanding complex neuronal interactions in cognitive processes \cite{varley2023partial}.
For privacy and fairness studies, the synergistic concept provides insights about data disclosure mechanisms~\cite{rassouli2019data,hamman2023demystifying}.
In the field of causality, information decomposition can be used to distinguish and quantify the occurrence of causal emergence \cite{rosas2020reconciling}, and more.

Despite these successes, the theoretical underpinnings of PID remain incomplete.
Ref.~\cite{williams2010nonnegative} proposed a set of axioms that the above information atoms should satisfy in order to provide said insights, and follow-up works in the field identified several additional properties \cite{ince2017partial,mediano2019beyond,lyu2023system,varley2023generalized}. Yet, in spite of extensive efforts~\cite{griffith2014intersection, ince2017measuring, bertschinger2013shared, harder2013bivariate, bertschinger2014quantifying}, a comprehensive quantitative definition of information atoms that satisfies all these axioms and properties is yet to be found (e.g., non-negativity and consistency with marginal and conditional mutual information).

This difficulty points to deeper conceptual issues in the framework itself.
A key assumption inherent (albeit implicitly) in the classical PID framework is that \emph{the Whole Equals the Sum of its Parts} (abbrv. WESP, often referred to as the set-theoretic principle) when decomposing information.
In other words, one expects that if the mutual information is fully broken down into independent atoms, then summing those atoms should recover the total information of the whole system, {and similarly for any its subsystems}.

However, we argue that this persistent difficulty arises primarily from conceptual flaws inherent to the PID framework, since information does not always behave like a set measure: higher-order \emph{synergistic} effects can cause violations of the WESP principle. Such violations have previously been noted~\cite{kolchinsky2022novel} through the perspective of the inclusion-exclusion principle~\cite{ting1962amount,yeung1991new}.
In this paper, we will show that the PID framework, as originally formulated on a lattice of set partitions, inherently permits situations where ``the whole is \emph{less} than the sum of its parts,'' meaning the sum of PID components differs from the total information of the system.
This reveals an inconsistency in the traditional lattice-based PID approach for systems with three or more variables.

Our contributions are threefold. \textit{(i)}
We derive an explicit (closed-form, optimization-free) formula for two-source PID, which, to the best of our knowledge, is the first explicit instance satisfying all axioms and properties listed in Section~\ref{sec:pid-framework}
\textit{(ii)} Turning to the case with three or more sources, we show some negative results. We first review the known result that there is no set of functions that satisfies all the desired axioms and properties for three or more source variable PID by explicitly demonstrating the breakdown of the WESP assumption using a counterexample in a three-source system \cite{rauh2014reconsidering}, highlighting a fundamental \emph{subsystem inconsistency} in the PID framework. We then show that for general multivariate systems, \emph{no} decomposition based on the PID lattice can avoid such inconsistencies, thereby underscoring the inadequacy of current lattice-based formulations for large systems.
\textit{(iii)} Motivated by these findings, we propose an alternative approach to the multivariate scenario that focuses on defining measures of \emph{unique}, and \emph{synergistic} information for multivariate systems without relying on the lattice structure. By doing so, our method circumvents the contradictions inherent in PID frameworks.
Finally, we validate the proposed measures through comparative experiments and theoretical analysis, showing that they satisfy the desired axiomatic properties and are able to reveal high-order informational relationships among multiple variables.
Beyond their information-theoretic interest, our definitions connect naturally to standard observables in physical systems.
In the Ising model, the redundant and unique terms track the development of shared local structure, while the synergistic term captures genuinely collective dependencies whose role changes qualitatively near criticality.
These information-theoretic signatures exhibit systematic relations with established physical quantities, such as magnetic susceptibility and specific heat.
Section~\ref{sec:experiments} confirms these links empirically across complementary experimental settings: in local predictive tasks, redundancy and unique information peak near criticality while synergy is suppressed, whereas when collective spatial order is treated as the target, synergy becomes dominant after the phase transition.
Together, these results position our framework as a quantitative tool for characterizing both predictive and emergent aspects of collective behavior in statistical-physics systems.

The remainder of this paper is organized as follows. In Section~\ref{sec:pid-framework}, we review the PID framework, including its axioms and the redundancy lattice structure.
In Section~\ref{sec:two-source-formula}, we present our proposed operational definitions for two-source PID, providing closed-form formulas for unique, redundant, and synergistic information that satisfy the PID axioms.
Section~\ref{sec:limitations} then details the limitations of the PID framework in multivariable (three or more sources) settings: we review a three-variable counterexample violating WESP and prove a general impossibility result for four or more sources under lattice-based partial decompositions.
In Section~\ref{sec:multi-measures}, we introduce the extended definitions of multivariate unique and synergistic information that apply without relying on the PID lattice, defining higher-order synergy measures for general source sets.
Section~\ref{sec:experiments} provides case studies: we compare our measures with existing PID approaches on synthetic examples and verify the effectiveness of our measure in simulation experiments of the Ising model.
Finally, Section~\ref{sec:discussion} offers a discussion and concludes the paper.

Parts of this work appeared in the 2024 IEEE International Symposium on Information Theory~\cite{lyu2024explicit}.

\section{Partial Information Decomposition Framework}\label{sec:pid-framework}
Williams and Beer \cite{williams2010nonnegative} introduced the concept of Partial Information Decomposition (PID) to address the challenge of decomposing multivariate information.
Rather than being grounded in a specific algorithm, the PID framework adopts an axiomatic approach.
At its core, they proposed three fundamental axioms to guide the decomposition of information contributed by multiple source variables toward a target variable.

\begin{figure}[htbp]
  \centering
  \fbox{\includegraphics[width=.5\linewidth]{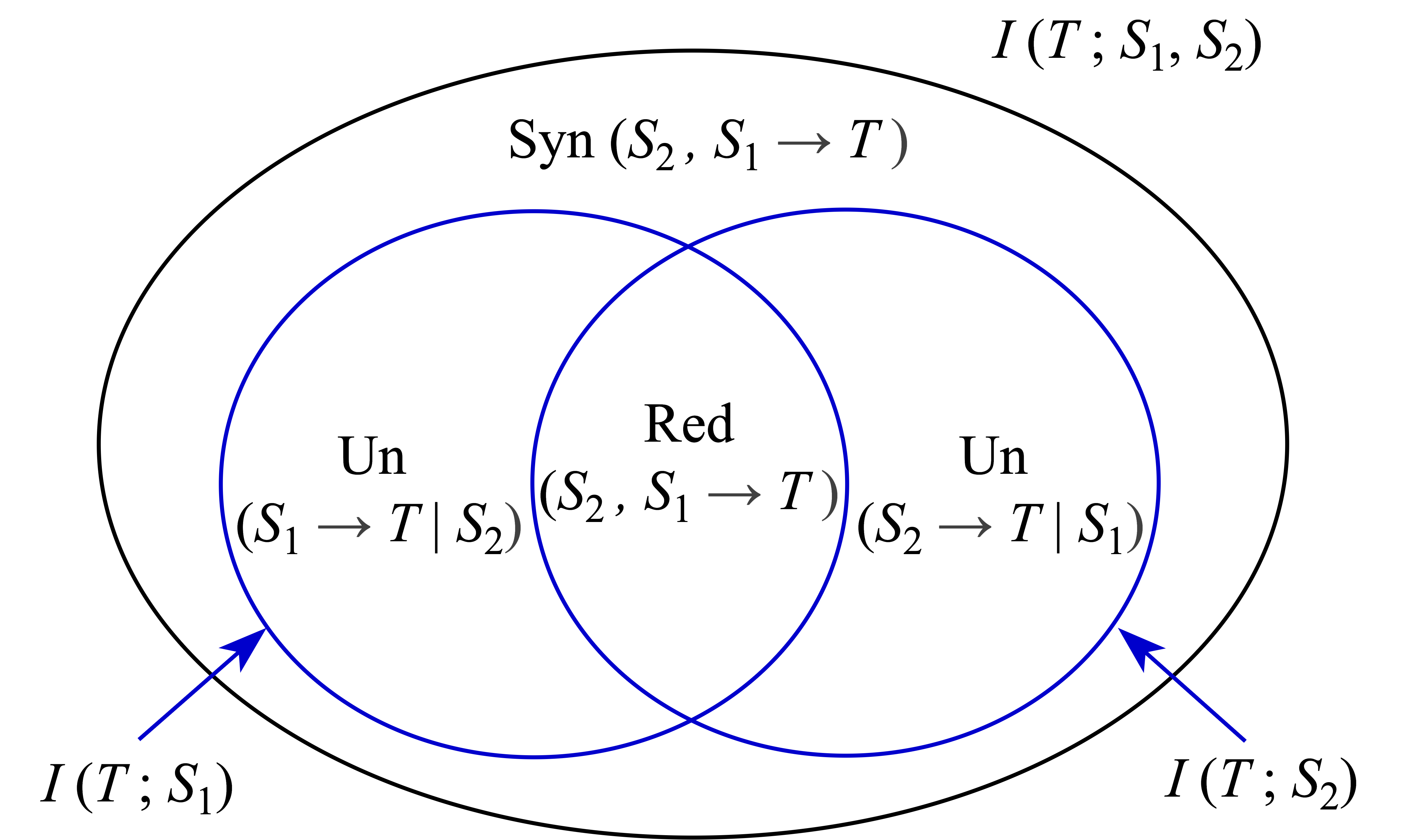 }}
  \caption{A pictorial representation of Partial Information Decomposition~\eqref{equ:Information Atoms' relationship_1}, where~$I((S_1,S_2);T)$ is decomposed to its finer information atoms, the synergistic $\operatorname{Syn}(S_1,S_2\to T)$ (also called ``complementary''), the redundant $\operatorname{Red}(S_1,S_2\to T)$ (also called ``shared''), and the two directional unique components~$\operatorname{Un}(S_1\to T|S_2)$ and~$\operatorname{Un}(S_2\to T|S_1)$. The summation of the redundant atom and one of the unique atoms must be equal to the corresponding mutual information, as described in Eq.~\eqref{equ:Information Atoms' relationship_2}.}
  \label{fig:PID}
\end{figure}

For a system with random variables~$S_1,S_2,T$, the quantity~$I(S_1,S_2;T)$ (hereafter all logarithms are in base-2 and all information quantities are reported in bits) captures the amount of information that one \textit{target variable}~$T$ shares with the \textit{source variables}~$(S_1,S_2)$, but provides no further information regarding finer interactions between the three variables.
To gain more subtle insights into the interactions between~$T$ and~$(S_1,S_2)$, \cite{williams2010nonnegative} proposed to further decompose~$I(S_1,S_2;T)$ into \textit{information atoms}.
Specifically, the shared information between~$T$ and~$(S_1,S_2)$ should contain a \textit{redundant} information atom, two \textit{unique} information atoms, and one~\textit{synergistic} information atom (see Fig.~\ref{fig:PID}).

The redundant information atom $\operatorname{Red}(S_1,S_2\to T)$ (also called ``shared'') represents the information which either~$S_1$ or~$S_2$ imply about~$T$.
The unique information atom $\operatorname{Un}(S_1 \to T|S_2)$ represents the information individually contributed to~$T$ by~$S_1$, but not by~$S_2$ (similarly $\operatorname{Un}(S_2 \to T|S_1)$).
The synergistic information atom $\operatorname{Syn}(S_1,S_2\to T)$ (also called ``complementary''), represents the information that can only be known about~$T$ through the \textit{joint} observation of~$S_1$ and~$S_2$, but cannot be provided by either one of them separately.
Together, we must have that
\begin{align}
  I(S_1,S_2;T) = \operatorname{Red}(S_1,S_2 \to T) +\operatorname{Syn}(S_1,S_2 \to T) +\operatorname{Un}(S_1\to T\vert S_2)+ \operatorname{Un}(S_2\to T\vert S_1).\label{equ:Information Atoms' relationship_1}
\end{align}
We refer to~\eqref{equ:Information Atoms' relationship_1} as \textit{Partial Information Decomposition} (PID).

Moreover, since the redundant atom together with one of the unique atoms constitute all information that one source variable implies about the target variable, it must be the case that their summation equals the mutual information between the two, i.e.,
\begin{align}
  I(S_1;T) &= \operatorname{Red}(S_1,S_2 \to T) + \operatorname{Un}(S_1\to T\vert S_2), \mbox{ and}\nonumber\\
  I(S_2;T) &= \operatorname{Red}(S_1,S_2 \to T) + \operatorname{Un}(S_2\to T\vert S_1).\label{equ:Information Atoms' relationship_2}
\end{align}

In a similar spirit, the synergistic information atom and one of the unique information atoms measure shared information between the target variable and one of the source variables, while excluding the other source variable. Therefore, the summation of these quantities should coincide with the well-known definition of conditional mutual information, i.e.,
\begin{align}
  I(T;S_1|S_2) &= \operatorname{Syn}(S_1,S_2 \to T) + \operatorname{Un}(S_1\to T\vert S_2),\mbox{ and}\nonumber\\
  I(T;S_2|S_1) &= \operatorname{Syn}(S_1,S_2 \to T) + \operatorname{Un}(S_2\to T\vert S_1).\label{equ:Information Atoms' relationship_3}
\end{align}

For general systems with source variables $\boldS = \{S_1,\dots,S_n\}$ and target $T$, PID uses \textit{the redundancy lattice}~$\mathcal{A}(\boldS)$~\cite{williams2010nonnegative,crampton2001completion}, which is the set of antichains formed from the power set of $\boldS$ under set inclusion with a natural order~$\preceq_\boldS$.

\begin{definition}[PID Redundancy Lattice]
  \label{definition:PID lattice}
  For the set of source variables $\boldS$, the set of antichains is:
  \begin{align}
    \mathcal{A}(\boldS) = \{\alpha \subseteq \mathcal{P}(\boldS)\setminus \{\emptyset\}:\alpha \ne \emptyset, \forall \boldA_i,\boldA_j \in \alpha, \boldA_i \not \subset \boldA_j\},
  \end{align}
  where $\mathcal{P}(\boldS)$ is the powerset of $\boldS$, and
  for every~$\alpha,\beta\in\mathcal{A}(\boldS)$, we say that~$\beta \preceq_\boldS \alpha$ if for every~$\boldA\in\alpha$ there exists~$\boldB\in \beta$ such that~$\boldB\subseteq \boldA$.
\end{definition}
For ease of exposition, we denote elements of~$\mathcal{A}(\boldS)$ using bracketed expressions that contain only the indices of the corresponding sources. For example, write $\{\{S_1,S_2\}\}$ as $\{12\}$, $\{\{S_1\}\{S_2\}\}$ as $\{1\}\{2\}$ and $\{\{S_1\}\}$ as $\{1\}$.
Based on Def.~\ref{definition:PID lattice}, the PID of a system can be expressed as a set of functions used to assign values to all nodes of the lattice for every subset of source variables in a system, which is defined as follows:

\begin{definition} [Partial Information Decomposition Framework]
  \label{pid def:PIDF}
  Let~$\boldS$ be a collection of sources and let~$T$ be the target.
  The set of PI-atoms is defined as a family of partial information functions (PI-function) $\Pi^{T}_{\boldA}:\mathcal{A}(\boldA) \rightarrow \mathbb{R}$ for all $\boldA \subseteq \boldS$.

  % from the perspective of system $(\boldA,T)$.
\end{definition}
Intuitively, for every~$\alpha\in\cA(\boldA)$, the atom $\Pi^{T}_{\boldA}(\alpha)$ measures the amount of information provided by each set in the anti-chain~$\alpha$ to~$T$ and is not provided by any subset~$\beta \preceq_\boldA \alpha$.
%$\beta\in\cA(\boldA)$ such that~$\beta \preceq_\boldA \alpha$.
For simplicity we denote $\Pi^{T}_{i\dots}(\cdot)$ for $\Pi^{T}_{\{S_i\dots\}}(\cdot)$, e.g., $\Pi^{T}_{12}(\{\{1\}\})=\Pi^{T}_{\{S_1,S_2\}}(\{\{S_1\}\})$.
%And for the set of sources variables $\boldS=\{S_1,S_2\}$ and target variable $T$, the PI-atoms~$\operatorname{Red}, \operatorname{Un}$, and~$\operatorname{Syn}$ are associated with following notations.
Note that in the case~$\boldS=\{S_1,S_2\}$, Def.~\ref{pid def:PIDF} reduces to
\begin{align*}
  \operatorname{Red}(S_1,S_2\to T) & =  \Pi^{T}_{12}( \bigl\{ \{ 1 \} \{ 2 \} \bigl\} ), \operatorname{Un}(S_1 \to  T\vert S_2 )  =  \Pi^{T}_{12}( \bigl\{ \{1\} \bigl\} ),\\
  \operatorname{Syn}(S_1,S_2  \to  T)  &=  \Pi^{T}_{12}( \bigl\{ \{12\} \bigl\} ),
  \operatorname{Un}(S_2 \to  T\vert S_1)  =  \Pi^{T}_{12}( \bigl\{ \{2\} \bigl\} ),
\end{align*}
recovering the terms in~\eqref{equ:Information Atoms' relationship_1} and~\eqref{equ:Information Atoms' relationship_2}.

{For three-or-more source variable systems, set of source variables $\boldS=\{S_1,\dots,S_N\}$ and target variable $T$, there will be more types of information atoms. For the those three representations, $\operatorname{Red}$, $\operatorname{Un}$, and~$\operatorname{Syn}$, we specifically refer to the following three types of information atoms:
  \begin{align*}
    \operatorname{Red}(S_1,\dots,S_N \!\to\! T) &= \Pi^{T}_{\boldS}(\bigl\{\{1\}\dots,\{N\}\bigl\}),\\
    \operatorname{Syn}(S_1,\dots,S_N \!\to\! T) &= \Pi^{T}_{\boldS}(\bigl\{\{1,\dots,N\}\bigl\}),\text{ and} \\
    \operatorname{Un}(S_i\!\to\! T\vert \boldS \setminus S_i) &= \Pi^{T}_{\boldS}(\bigl\{\{i\}\bigl\}), \forall i\in [N].
\end{align*}}
\begin{figure}[htbp]
  \centering
  \fbox{\includegraphics[width=.6\linewidth]{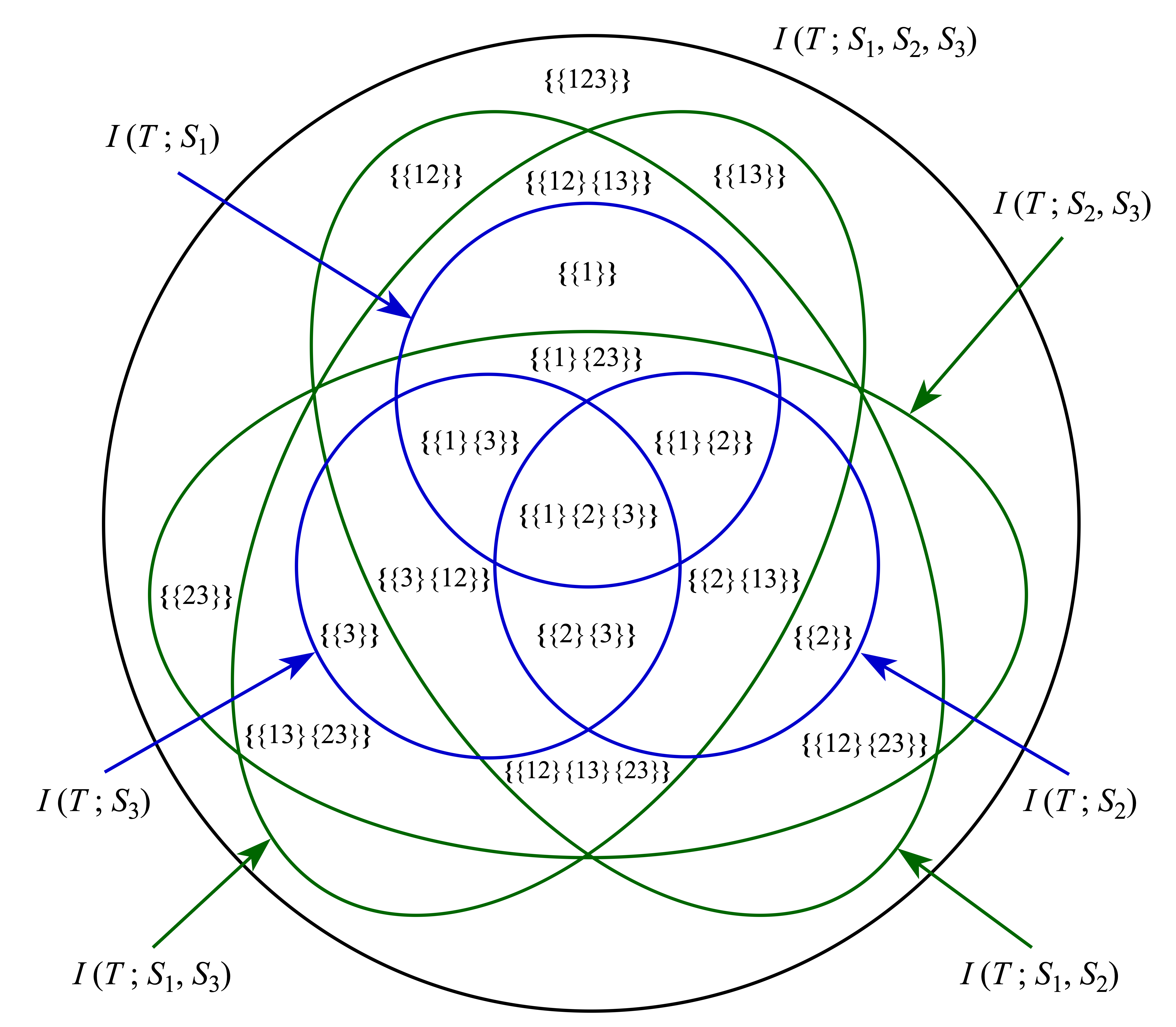 }}
  \caption{The structure of PID with 3 source variables. }
  \label{fig:ThreeCase}
\end{figure}

For general systems, PID requires the following mutual information constraints~\cite{williams2010nonnegative} (i.e., the equivalent of~\eqref{equ:Information Atoms' relationship_1} and~\eqref{equ:Information Atoms' relationship_2}).

\begin{axiom}
  \label{axiom:mutual constrains}
  For any subsets $\boldA$, $\boldB$ of sources~$\boldS$ with $\boldA\subseteq \boldB$, the sum of PI-atoms decomposed from system $\boldB$ satisfies
  \begin{align}
    \label{equ:PID Information Atoms}
    I(\boldA;T) = \sum_{\beta \preceq_{\boldB} \{\boldA\} } \Pi^{T}_{\boldB}(\beta),
  \end{align}
  % \begin{align}
  % \label{equ:PID Information Atoms}
  %     I(\boldA;T) = \sum_{\beta \in\mathcal{O}_{\boldA} } \Pi^{T}_{\boldB}(\beta),
  % \end{align}
  where $\{\boldA\}$ is the antichain with a single element~$\boldA$.
\end{axiom}
It is worth noting that \eqref{equ:PID Information Atoms} constrains the consistency of the sum of PI-atoms decomposed from different subsystems \cite{ince2017partial,chicharro2017synergy,rosas2020operational,lizier2013towards} by taking different sets $\boldB$ on the right-hand side.

\begin{lemma}[Subsystem Consistency]
  \label{lemma: subsystem consistency} For the system with target $T$ and sources~$\boldA,\boldB,\boldC\subseteq \boldS$ such that~$\boldC\subseteq\boldA\cap \boldB$, let $\Pi_{\boldS^*}^T$, $\boldS^* \in \{\boldA,\boldB,\boldC\}$, be defined as in Def.~\ref{pid def:PIDF} and satisfy Axiom~\ref{axiom:mutual constrains}. Then, we have that
  \begin{align}
    \label{equ:subsystem}
    \sum_{\beta \preceq_\boldA \{\boldC\}} \Pi^{T}_{\boldA}(\beta)=\sum_{\beta \preceq_\boldB \{\boldC\}} \Pi^{T}_{\boldB}(\beta).
  \end{align}
\end{lemma}
% The decomposition in \eqref{equ:subsystem} constrains the consistency of the sum of PI-atoms decomposed from different subsystems.
% This consistency of subsystem decomposition is considered part of the axioms that should be satisfied in the field of information decomposition \cite{ince2017partial,chicharro2017synergy,rosas2020operational,lizier2013towards}.
Consider the system in Fig.~\ref{fig:PID}.
For the atoms decomposed from the system $(S_1,T)$, the quantity $\Pi^{T}_{1}(\bigl\{\{1\}\bigl\})$ reflects the (redundant) information that $S_1$ provides about~$T$.
If we add a source~$S_2$ to this system, this information will be further decomposed into the redundant information from $S_1,S_2$ and the unique information only from $S_1$ but not $S_2$. This observation is formally written by the following corollary, which is proved in Appendix~\ref{app: proof of corollary}.
\begin{corollary}
  \label{corollary: two result}
  For the system $(S_1, S_2,S_3,T)$ and its subsystems $(S_1,\!S_2,\!T)$ and $(S_1,\!T)$, the decomposed PI-atoms from the different subsystems have the following relation:
  \begin{align}
    \label{equ:cross scale_0}
    \Pi^{T}_{1}(\!\bigl\{\!\{1\}\!\bigl\}\!) = \Pi^{T}_{12}(\!\bigl\{\!\{1\}\{2\}\!\bigl\}\!) +\Pi^{T}_{12}(\!\bigl\{\!\{1\}\!\bigl\}\!),
  \end{align}
  similarly, for the systems $(S_1,S_2,S_3,T)$ and $(S_1,S_2,T)$,
  \begin{align}
    \label{equ:cross scale}
    \Pi^{T}_{12}(\!\bigl\{\!\{1\}\{2\}\!\bigl\}\!) &= \Pi^{T}_{123}(\!\bigl\{\!\{1\}\!\{2\}\!\{3\}\!\bigl\}\!) +\Pi^{T}_{123}(\!\bigl\{\!\{1\}\!\{2\}\!\bigl\}\!),\\
    \Pi^{T}_{12}(\!\bigl\{\!\{1\}\!\bigl\}\!) &= \Pi^{T}_{123}(\!\bigl\{\!\{1\}\{3\}\!\bigl\}\!)+\Pi^{T}_{123}(\!\bigl\{\!\{1\}\{23\}\!\bigl\}\!) +\Pi^{T}_{123}(\!\bigl\{\!\{1\}\!\bigl\}\!).\label{equ:cross scale_2}
  \end{align}
\end{corollary}

\begin{observation}
  Following Axiom~\ref{axiom:mutual constrains}, we may rewrite~\eqref{equ:Information Atoms' relationship_1} and~\eqref{equ:Information Atoms' relationship_2} as:
  \begin{align*}
    I(S_1,S_2;T) &= \Pi^T_{\boldS}(\{1,2\}) +\Pi^T_{\boldS}(\{1\}\{2\}) + \Pi^T_{\boldS}(\{1\}) + \Pi^T_{\boldS}(\{2\}), \\
    I(S_1;T) &= \Pi^T_{\boldS}(\{1\}\{2\}) + \Pi^T_{\boldS}(\{1\}), \\
    I(S_2;T) &= \Pi^T_{\boldS}(\{1\}\{2\}) + \Pi^T_{\boldS}(\{2\}),
  \end{align*}
  where $\boldS = \{1,2\}$.
\end{observation}

The above definitions and axioms are the basis of an axiomatic approach to the definition of the information atoms. These equations lead to the first of a series of additional axioms regarding the redundant information $\operatorname{Red}(S_1,\dots,S_N\to T)$ (or $\Pi^{T}_{\boldS}(\bigl\{\{1\}\dots\{N\}\bigl\})$) for any multivariate system~$\boldS$, which were raised in previous works on the topic \cite{williams2010nonnegative,williams2011information,griffith2014quantifying}.

The first additional axiom is \textit{commutativity} of the source variables, which implies that the order of the source variables must not affect the value of the redundant information.

\begin{axiom} [Commutativity]
  \label{axiom: commutativity}
  Redundant information is invariant under any permutation $\sigma$ of sources, i.e., $\operatorname{Red}(S_1,\dots,S_N\to T) = \operatorname{Red}(S_{\sigma(1)}, \dots, S_{\sigma(N)}\to T)$.
\end{axiom}

The second is \textit{monotonicity}, which implies that the redundant information is non-increasing when adding a source variable, since the newly added variable cannot increase the redundancy between the original variables.

\begin{axiom} [Monotonicity]
  \label{axiom: Monotonicity}
  Redundant information decreases monotonically as more source variables are included, i.e.,
  $\operatorname{Red}(S_1,\dots,S_{N},S_{N+1} \to T) \le \operatorname{Red}(S_1,\dots,S_{N} \to T)$.
\end{axiom}
Axiom~\ref{axiom: Monotonicity} also implies another lemma, as follows.

\begin{lemma} [Nonnegativity]
  \label{Lemma: Nonnegativity}
  PID satisfies $\operatorname{Red}(S_1,\dots,S_N \to T) \ge 0$.
  \begin{proof}
    Add a constant variable $S^*$ and obtain $\operatorname{Red}(\boldA \to T) \ge \operatorname{Red}(\boldA,S^* \to T) =0$.
  \end{proof}
\end{lemma}

The third axiom is \textit{self-redundancy}, which defines the redundant information from \textit{one} source variable to the target variable as the mutual information between the two.
\begin{axiom} [Self-redundancy]
  \label{axiom: Self-redundancy}
  Redundant information for a single source variable $S_i$ equals the mutual information $\operatorname{Red}(S_i \to T) = I(S_i;T)$.
\end{axiom}

Besides, subsequent to~\cite{williams2010nonnegative,lizier2013towards}, studies suggested two additional properties, \textit{additivity} and \textit{continuity}~\cite{bertschinger2014quantifying,rauh2023continuity}. Additivity implies that whenever independent variable systems are considered, the joint information measures should be the sum of the information measures of each individual system. This is the case, for instance, in joint entropy of two independent variables.

\begin{property} [Additivity]
  \label{property: Additivity}
  Partial Information Decomposition of two independent systems
  $(S_1,\dots,S_N , T)$ and $(X_1,\dots,X_N,\bar{T})$ satisfy
  \begin{align*}
    &\operatorname{F}((S_1,X_1),\dots,(S_N,X_N)\to (T,\bar{T})) =\operatorname{F}(S_1,\dots,S_N \to T) + \operatorname{F}(X_1,\dots,X_N\to \bar{T}), and \\
    &\operatorname{Un}((S_i,X_i)\to (T,\bar{T})|(S_1,X_1),\dots,(\hat{S}_i,\hat{X}_i),\dots,(S_N,X_N)) \\&=\operatorname{Un}(S_i\to T|S_1,\dots,\hat{S}_i,\dots,S_N) + \operatorname{Un}(X_i\to \bar{T}|X_1,\dots,\hat{X}_i,\dots,X_N)
  \end{align*}
  for every~$\operatorname{F}\in\{\operatorname{Red},\operatorname{Syn}\}, i\in\{1,\dots,N\}$, and~$\hat{\cdot}$ denotes omission.
\end{property}

\begin{property} [Continuity]
  \label{property: Continuity}
  $\operatorname{Red}$, $\operatorname{Un}$, and~$\operatorname{Syn}$ are continuous functions from the underlying joint distributions of $(S_1,\dots,S_N , T)$ to~$\mathbb{R}$.
\end{property}

Additionally, a well-known property called \textit{independent identity}~\cite{ince2017measuring} asserts that in a system where two source variables are independent and the target variable is equal to their joint distribution, the redundant information should be zero.

\begin{property}[Independent Identity]
  \label{property: Independent Identity}
  If $I(S_1;S_2)=0$ and $T=(S_1,S_2)$, then $\operatorname{Red}(S_1,S_2\to T) = 0$.
\end{property}

\begin{remark}
  \label{remark:def equal}
  From an information theoretic perspective, there is no difference between ``$T=(S_1,S_2)$'' in Property~\ref{property: Independent Identity} and ``$H(T|S_1,S_2)=H(S_1,S_2|T)=0$.''
  For brevity we denote $H(T|S_1,S_2)=H(S_1,S_2|T)=0$ by $T \overset{\text{det}}{=} (S_1, S_2)$.
\end{remark}

In summary, the PID framework provides a conceptual decomposition of information into redundant, unique, and synergistic components that obey natural axioms. It is built on a lattice structure that mirrors set-theoretic inclusion relationships between source sets.

\section{Two-Source PID: An Explicit Formula}\label{sec:two-source-formula}
In this section, we present our \emph{explicit formula} for the PID components $(\text{Red}, \text{Un}, \text{Syn})$ that satisfies all the axioms in the previous section for two source variable systems.
Prior approaches offers two paths: dependency-constraint approaches, i.e.,~\cite{bertschinger2014quantifying}, that are axiom-compliant but optimization-based, or explicit formulae that do not satisfy all required axioms/properties. By contrast, our method delivers closed-form expressions for all two-source PID while meeting every axiom in Section~\ref{sec:pid-framework}. Section~\ref{sec:experiments} then compares mainstream alternatives side-by-side to document these property differences empirically.

In order to give our definition of information atoms, we construct a new distribution as follows. Given a system with sources $S_1,S_2$ and target $T$, with respective alphabets~$\cS_1,\cS_2$, and~$\cT$, define a new random variable~$S_1'$ over~$\mathcal{S}_1$ via its conditional joint distribution as follows:
\begin{align} \label{equation: S_1'S_2T}
  \Pr(S_1'=s_1,S_2=s_2|T=t)\triangleq\Pr(S_1=s_1|T=t)\Pr(S_2=s_2|T=t).
\end{align}
% \begin{align}\label{equation:S_1'T}
% \Pr(S_1'=s_1,T=t|S_2=s_2) \triangleq\Pr(S_1=s_1|T=t)\Pr(T=t|S_2=s_2),
% \end{align}
meaning, for every~$s_1\in\mathcal{S}_1$,
\begin{align*}
  \Pr(S_1'=s_1)=\sum_{(s_2,t)\in\mathcal{S}_2\times \mathcal{T}}\Pr(S_1=s_1|T=t)\Pr(S_2=s_2|T=t)\Pr(T=t).
\end{align*}
The variable~$S_1'$ is well-defined since all probabilities are non-negative, and since it can be readily shown that
\begin{align*}    \sum_{s_1\in\mathcal{S}_1}\Pr(S_1'=s_1)&=\sum_{(s_1,s_2,t)\in\mathcal{S}_1\times\mathcal{S}_2\times \mathcal{T}}\Pr(S_1=s_1|T=t)\Pr(S_2=s_2|T=t)\Pr(T=t) =1.
\end{align*}

Eq.~\eqref{equation: S_1'S_2T} intuitively illustrates the rationale behind constructing a new joint probability distribution. It preserves the original pairwise joint probability distributions between each source variable and the target variable (i.e., $(S_1,T)$ and $(S_2,T)$), while also ensuring that the source variables are conditionally independent given the target variable. The purpose of this operation is to eliminate higher-order relationships in the original joint probability distribution while preserving the pairwise relationships between each source variable and the target variable.
Based on this logic, our definitions of information atoms are as follows.

\begin{definition}[Unique Information in two source variable systems]\label{definition:un}

  The unique information from~$S_1$ to~$T$ given~$S_2$ is $\operatorname{Un}(S_1 \to T | S_2) = I(S_1';T|S_2)$.
\end{definition}

Def.~\ref{definition:un} instantiates the framework in Def.~\ref{pid def:PIDF} by specifying $\Pi$ via the unique atom.
Following that, the redundant and synergistic atoms are presented as the following lemmas directly from Axiom~\ref{axiom:mutual constrains}.

\begin{lemma}[Redundant Information in two source variable systems]
  \label{lemma:red}
  The Redundant Information from $S_1$ and $S_2$ to $T$ is:
  \begin{align*}
    \operatorname{Red}(S_1,S_2 \to T) = I(S_1;T) - \operatorname{Un}(S_1 \to T | S_2).
  \end{align*}
\end{lemma}

\begin{lemma}[Synergistic Information in two source variable systems]
  \label{lemma:syn}
  The synergistic information from $S_1$ and $S_2$ to $T$ is defined as:
  \begin{align*}
    \operatorname{Syn}(S_1,S_2 \to T) = I(S_1;T|S_2) - \operatorname{Un}(S_1 \to T | S_2).
  \end{align*}
\end{lemma}

In addition to using the unique information and Axiom~\ref{axiom:mutual constrains} to obtain the definition of redundant and synergistic information, we can also give equivalent forms directly from~\eqref{equation: S_1'S_2T} as follow, which is proved in Appendix~\ref{app: proof of def2}.

\begin{lemma} \label{le:redefinition of redundant and synergistic information}
  Redundant information (Lemma~\ref{lemma:red}) can alternatively
  be written as $\operatorname{Red}({S_1},{S_2}\to T) = I({S'_1}; {S_2})$.
  Synergistic information (Lemma~\ref{lemma:syn})  can alternatively
  be written as $\operatorname{Syn}({S_1}, {S_2} \to T) = H(T|S_1', S_2)-H(T|S_1,S_2)$.
\end{lemma}

Also, the closed-form of the above information atoms is given in Appendix~\ref{app:closed-form formulation}.
% \subsection{Satisfaction of the Formulas}
Then, from the definitions of Red, Un, and Syn above, we have the following theorem proved in Appendix~\ref{app: proof of satisfaction of two}.
\begin{theorem}
  \label{theorem:proof of two}
  The definitions of redundant, unique, and synergistic information in two source variable systems (Defs.~\ref{definition:un}, Lemma~\ref{lemma:red}, and~\ref{lemma:syn}) satisfy Axioms~\ref{axiom:mutual constrains}--\ref{axiom: Self-redundancy}, and Properties~\ref{property: Additivity}--\ref{property: Independent Identity}.
\end{theorem}

So far, we have completed the proof that our PID operation definition for two source variable systems satisfies all the required axioms and properties.
Beyond satisfying the axioms and properties as stated in Thm.~\ref{theorem:proof of two}, our two–source atoms have a direct and intuitive meaning.
Eq.~\eqref{equation: S_1'S_2T} constructs an auxiliary variable $S'_1$ that preserves the $S_1\to T$ channel ($\Pr(S'_1=s_1\mid T= t)=\Pr(S_1=s_1\mid T= t)$) while removing higher–order dependencies by enforcing conditional independence of the sources given the target ($S'_1 \perp S_2 \mid T$).
Among all joint distributions that match these two marginals, this  $S'_1$ is the maximum-entropy solution. Equivalently, it retains all pairwise relations present in $\Pr(S_1=s_1,S_2=s_2,T=t)$ and filters out higher-order (non-pairwise) dependencies, which is the basis of synergy.
Quantitatively, the atoms reduce to standard Shannon terms under $\Pr(S'_1=s_1,S_2=s_2,T=t)$, and we use the mutual information, conditional mutual information, and conditional entropy to compute the information atoms (Def.~\ref{definition:un} and Lemma~\ref{le:redefinition of redundant and synergistic information}).
This “pairwise-preserving, max-entropy decoupling” explains why our atoms satisfy all two-source axioms and properties while remaining directly computable from mutual-information and entropy terms, and makes the PID framework defined based on our information atoms more practical in the two-source variable scenario.

However, as we elaborate next, this framework encounters inherent contradictions when extended to more than two source variables.
In particular, it is often the case that higher-order synergy terms violate the principle that the whole equals the sum of parts in every subsystem.
These limitations motivate the development of a new approach, which we introduce in subsequent sections.

\section{Limitations of the PID Framework in Multivariate Systems}
\label{sec:limitations}
While the two-source PID can be handled with ease (Section~\ref{sec:two-source-formula}), fundamental problems surface when extending to systems of three or more sources.
In this section, we first review a concrete three-source counterexample that demonstrates an inconsistency in how the sum of the PID atoms, however defined, can be greater than the total information of the full system.
Then, for the general cases where the target is not equal to the joint distribution of the sources, we generalize this insight by proving that for any decomposition based on the lattice of antichains, such inconsistency cannot be avoided.

\subsection{Three-Variable Counterexample: When the Whole Is Less Than the Sum of Its Parts}

The incompatibility between the
PID axioms (Lemma~\ref{Lemma: Nonnegativity}, non-negativity) and Property~\ref{property: Independent Identity} (Independent Identity) was first established by Rauh et al \cite[Thm.~2]{rauh2014reconsidering}.
To set the stage for our results, we briefly recall this finding and interpret it through the lens of information synergy. We retain the counterexample construction explicitly because it will serve as a recurrent part in what follows; the next subsection builds on it to derive a more general PID inconsistency.

Following \cite{rauh2014reconsidering}, we construct the following system (\(\bar{S}_1,\bar{S}_2,\bar{S}_3,\bar{T}\)), whose inherent contradiction is proved in the subsequent Lemma~\ref{lemma: counter example}:
Let \(x_1\) and \(x_2\) be two independent \(\operatorname{Bernoulli}(1/2)\) variables, and
$x_3 = x_1 \oplus x_2$. Here and throughout, $\oplus$ denotes the exclusive OR (XOR):
for $a,b \in \{0,1\}$,
$a\oplus b=(a+b)\text{ mod }2$. Define $\bar{S}_1 = x_1, \bar{S}_2 = x_2, \bar{S}_3 = x_3$, and $\bar{T} = (x_1, x_2, x_3)$. The proof is similar to \cite{rauh2014reconsidering} and is given in Appendix~\ref{app: proof of counter exp}.

\begin{lemma}[{\cite[Theorem~2]{rauh2014reconsidering}}]
  \label{lemma: counter example}
  For the system $(\bar{S}_1,\bar{S}_2,\bar{S}_3,\bar{T})$, any candidate PID measure $\Pi^{\bar{T}}_{\boldA}:\mathcal{A}(\boldA) \rightarrow \mathbb{R}, \forall \boldA \subseteq \bar{\boldS}$ that satisfies PID Axioms~\ref{axiom: commutativity}, \ref{axiom: Monotonicity}, \ref{axiom: Self-redundancy}, Property~\ref{property: Independent Identity}, and PID Axiom~\ref{axiom:mutual constrains} for any $|\boldA|\le2$,
  violates PID Axiom~\ref{axiom:mutual constrains} for~$|\boldA|=3$, i.e.,
  \begin{align*}
    I(T;\bar{\boldS}) < \sum_{\beta \preceq_{\bar{\boldS}} \{\bar{\boldS}\} } \Pi^{\bar{T}}_{\bar{\boldS}}(\beta).
  \end{align*}
\end{lemma}
The redundancy lattice underlying PID implicitly assumes that the “whole” quantity can be assembled as a sum of disjoint atoms (WESP). Lemma~\ref{lemma: counter example} shows that this may fail in multivariate settings due to higher-order synergy: the sum of synergy-related atoms can exceed the joint entropy because the same $N$-way interaction is recorded under several antichains~\cite{lyu2023system}, which is described rigorously in the following observation.
\begin{observation}[Synergy Overcounting Phenomenon \cite{lyu2023system}]
  \label{observation:HleP}
  For any system $\boldX = \{X_1,\!\dots\!,X_N\}$, denote $\operatorname{Syn} (\boldX \setminus X_{i}\to\! X_i)$ as the information that can only be provided jointly from sources {$\boldX \setminus X_{i}$} to target $X_i$ (i.e., consider a system in which $X_i$ is a target and $\{X_j\}_{j\ne i}$ are sources, and follow the PID framework of Section~\ref{sec:pid-framework}).
  Then, the summation of all that information can be greater than the joint entropy of the system, i.e., in general
  {it is \emph{not} always the case that}
  \begin{align}
    \label{equation:synEffectLessThanEntropy}
    \sum_{i \in \{1,\dots,N\}} \operatorname{Syn} (\boldX \setminus X_i \!\to\! X_i) \le H(\boldX).
  \end{align}
  In particular, in the system given in Lemma~\ref{lemma: counter example} we have that the l.h.s of~\eqref{equation:synEffectLessThanEntropy} equals~$3$, whereas the r.h.s equals~$2$.
\end{observation}

The observation above implies that summing all PID atoms as if they were disjoint can exceed the whole information.
Consequently, in multivariate systems, computing unknown atoms by the overall joint entropy together with a subset of known atoms (as done, e.g., in Lemma~\ref{lemma:red} in two-variable systems) may violate some of the PID axioms.
This is because overlapping higher-order interactions may be counted multiple times when we sum over all the atoms within the lattice.
One might hope to repair this by modifying the summation range in Axiom 1 or by adding corrective weights.
In the next subsection, we show that for any antichain-based decomposition satisfying Axioms 2–4 and Property 3 there exist two systems with identical PID atoms but different joint entropies. Therefore no modification of Axiom 1 within the framework can resolve the problem. We formalize this in the next subsection.

\subsection{General Three or More Sources Cases: Unavoidable Inconsistency}
The summation range in Axiom~\ref{axiom:mutual constrains}, which sums all atoms below $\{A\}$ in the lattice of a larger system $B$, aligns with the mereological intuition that information ``about $A$ within $B$'' should be obtained by aggregating all parts that are not larger than $\{A\}$.
This perspective is articulated in recent mereological accounts of information decomposition \cite{gutknecht2021bits}.
However, such accounts do not fix a unique indexing structure with summation range that guarantee subsystem consistency when $N \ge 3$.

In this section, we will use two systems with identical PID atoms but different overall information entropy to prove that any approach that aims to solve the inconsistency mentioned above by adjusting the summation range in Axiom~\ref{axiom:mutual constrains} is impossible.
That is, for any antichain-indexed, lattice-based partial decomposition in which a global quantity is computed from atoms via Axiom-1-type consistency constraints, there exists a 3-source system that violates subsystem consistency.
This conclusion is presented in the form of the following theorem.
\begin{theorem}[Subsystem inconsistency in lattice-based PID]
  \label{theorem:no sub set}
  For any set of functions $\Pi^{T}_{\boldA}:\mathcal{A}(\boldA) \rightarrow \mathbb{R}, \forall \boldA \subseteq \boldS$, which satisfy PID Axioms~\ref{axiom: commutativity}, \ref{axiom: Monotonicity}, \ref{axiom: Self-redundancy}, and Property~\ref{property: Independent Identity}, there is no way to redefine PID Axiom~\ref{axiom:mutual constrains} so that~\eqref{equ:subsystem} (Lemma~\ref{lemma: subsystem consistency}) is satisfied.
  {Specifically, there is no function~$f:\mathbb{R}^{|\cA(\boldS)|}\to\mathbb{R}$ such that
    \begin{align}
      \label{equ:subseto}
      I(\boldS;T) = f((\Pi^{T}_{\boldS}(\beta))_{\beta\in\cA(\boldS)})
    \end{align}
  for \emph{all} systems with~$|\boldS|=3$ source variables and a target~$T$.}
\end{theorem}
We emphasize that Theorem~\ref{theorem:no sub set}
shows that for any PI-function $\Pi_A$, there is {no function $f$} satisfying~\eqref{equ:subseto} for all systems of three-or-more variables~$\boldS$ and a target~$T$.
To prove this theorem, we construct the following two systems (Fig.~\ref{fig:System12}) and the subsequent Lemma~\ref{lemma:NoUniversalSubset}, which is proved in Appendix~\ref{app: proof of two system}.

\paragraph*{System~1 ($\hat{S}_1,\hat{S}_2,\hat{S}_3,\hat{T}$):}
Define six independent bits $x_1,x_2,x_4,x_5,x_7,x_8 \sim \operatorname{Bernoulli}(1/2)$, three additional bits
\begin{align}\label{equation:S1XOR}
  x_3=x_1\oplus x_2,~x_6=x_4\oplus x_5,\mbox{ and } x_9=x_7\oplus x_8,
\end{align}
and let
$
\hat{S}_1=(x_1,x_4,x_7),
\hat{S}_2=(x_2,x_5,x_8),
\hat{S}_3=(x_3,x_6,x_9), \mbox{ and }
\hat{T}=(x_1,x_5,x_9).
$

\paragraph*{System 2 (\(\bar{S}_1,\bar{S}_2,\bar{S}_3,\bar{T}\)):}
Take the three variables \(x_1\), \(x_2\) and \(x_3\) from System~1, where \(x_1,x_2\sim \operatorname{Bernoulli}(1/2)\), and
$x_3 = x_1 \oplus x_2$. Define $\bar{S}_1 = x_1, \bar{S}_2 = x_2, \bar{S}_3 = x_3$, and set the target as $\bar{T} = (x_1, x_2, x_3)$ (this is the system from Lemma~\ref{lemma: counter example}).
\begin{figure}[htbp]
  \centering
  \fbox{\includegraphics[width=.8\linewidth]{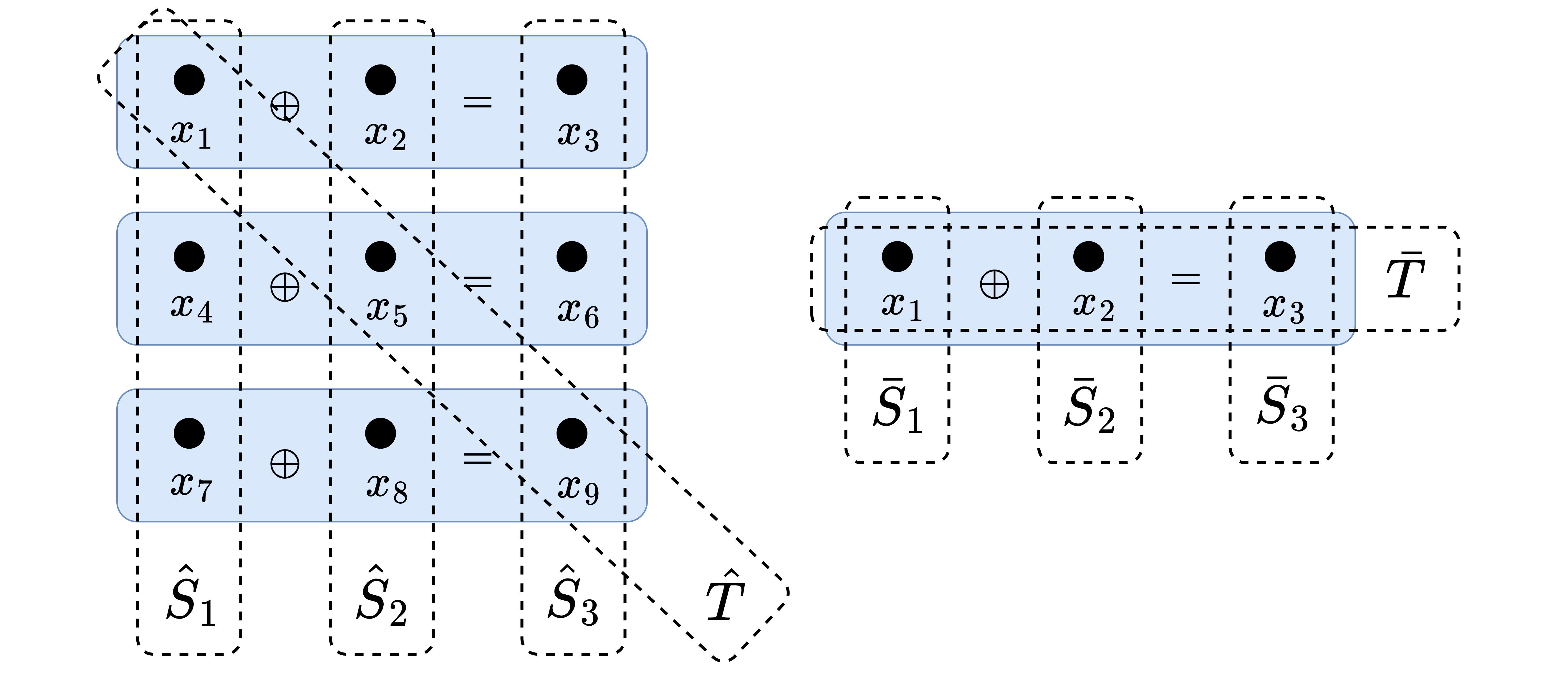}}
  \caption{
    Construction of the two systems
    $(\hat{S}_1,\hat{S}_2,\hat{S}_3,\hat{T})$ and
    $(\bar{S}_1,\bar{S}_2,\bar{S}_3,\bar{T})$
    used in Lemma~\ref{lemma:NoUniversalSubset}.
    Although the two systems induce identical PI-atoms under any lattice-based decomposition satisfying Axioms~\ref{axiom: commutativity}--\ref{axiom: Self-redundancy}, they differ in total mutual information, illustrating the impossibility of a universal reconstruction map from atoms to $I(\boldS;T)$, as formalized in Theorem~\ref{theorem:no sub set}.
  }
  \label{fig:System12}
\end{figure}
\begin{lemma}\label{lemma:NoUniversalSubset}
  For System 1 $(\hat{\boldS}=\{\hat{S}_1,\hat{S}_2,\hat{S}_3\},\hat{T})$ and System 2 $(\bar{\boldS}=\{\bar{S}_1,\bar{S}_2,\bar{S}_3\},\bar{T})$,
  (i) for any family of $\Pi$ functions satisfying Axioms~\ref{axiom: commutativity}, \ref{axiom: Monotonicity}, \ref{axiom: Self-redundancy}, Property~\ref{property: Independent Identity}, and~\eqref{equ:subsystem},
  there exists a bijection $\psi$ form $\mathcal{A}(\hat{\boldS}) \to \mathcal{A}(\bar{\boldS})$ such that $\Pi^{\hat{T}}_{\hat{\boldA}}(\hat{\beta})=\Pi^{\bar{T}}_{\psi(\hat{\boldA})}(\psi(\hat{\beta}))$ for all atoms $\hat{\beta} \in \mathcal{A}(\hat{\boldA})$, and all~$\hat{\boldA}\subseteq\hat{\boldS}$; and
  (ii) $I(\hat{\boldS};\hat{T})\ne I(\bar{\boldS};\bar{T})$.
\end{lemma}

Given Lemma~\ref{lemma:NoUniversalSubset}, Theorem~\ref{theorem:no sub set} is immediate: since all atoms are identical, {it follows that} $f(\Pi^{\hat{T}}_{\hat{\boldS}}(\hat{\beta})_{\hat{\beta} \in \mathcal{A}(\hat{\boldS}) })=f(\Pi^{\bar{T}}_{\psi(\hat{\boldS})}(\psi(\hat{\beta}))_{\psi(\hat{\beta}) \in \mathcal{A}(\psi(\hat{\boldS})) })$ for every function $f$ in the respective lattices, yet~$I(\hat{\boldS};\hat{T})\ne I(\bar{\boldS};\bar{T})$.
This result shows that PID based purely on a fixed collection of antichain-defined atoms will face insurmountable problems when extended beyond three variables.

The above reasoning (formalized in Theorem~\ref{theorem:no sub set}) establishes that no PID measure based on an antichain lattice can universally enforce the principle that “the whole equals the sum of its parts” for all subsets in systems of three or more variables.
This is a strongly negative result, which may explain the difficulty of extending the PID framework to multiple sources.

Several practical PID constructions \cite{williams2010nonnegative,barrett2015exploration,ince2017measuring,finn2018pointwise} have already observed an apparent contradiction between nonnegativity (Lemma~\ref{Lemma: Nonnegativity}) and independent identity (Property~\ref{property: Independent Identity}), and proposed to keep one and discard the other.
Theorem~\ref{theorem:no sub set}, together with Observation~\ref{observation:HleP}, suggests a different diagnosis.
Observation~\ref{observation:HleP} shows a ``whole is less than the sum of its parts'' effect: lattice additivity can reuse the same higher–order synergistic information across different sources, thereby over-counting information when summed.
Lemma~\ref{lemma: counter example}~\cite{rauh2014reconsidering} makes this explicit, and yet our Theorem~\ref{theorem:no sub set} takes it further:
one can build two systems that agree on all partial-information atoms yet differ in their total information. %, which is the situation captured by Theorem~2.
The root cause is that synergy is a global property of the joint system, and hence detecting whether synergy is present requires looking at the system as a whole.
Viewed this way, our Theorem~\ref{theorem:no sub set} points to a route for addressing the inconsistency while keeping both nonnegativity (Lemma~\ref{Lemma: Nonnegativity}) and independent identity (Property~\ref{property: Independent Identity})
This will be further discussed in Section~\ref{sec:discussion}.

In summary, the classic PID framework hits a fundamental roadblock for multivariate systems: it is impossible to assign non-negative, self-consistent redundancy and synergy values to all combinations of sources once $N\ge3$ without sometimes breaking the additive-consistency axiom.
The intuitive cause is the complex nature of high-order synergy, which cannot be localized neatly among lower-order combinations.
These findings motivate the next section, where we outline an alternative approach that sidesteps the problematic lattice approach by focusing on well-defined multivariate unique and synergistic measures, instead of calculating all atoms.
In the next section, we will extend the concepts of unique information and synergistic information to three-or-more variable scenarios with corresponding axioms and prove their satisfaction.
\begin{remark}
  We comment that in a separate follow-up work~\cite{lyu2025sumpartssubsysteminconsistency}, we focused on the problem of decomposing the \emph{entropy} of three variable systems, rather than the mutual information with respect to a target variable. We have found that in that case, a modified summation rule can resolve the problem of synergy over-counting, but no such modification for four-or-more variable systems has been proposed.
\end{remark}

\section{Derived Measures for Multivariate Unique and Synergistic Information}\label{sec:multi-measures}
The failure of lattice-based PID to handle general multivariate interactions encourages a different strategy.
In this section, we define measures of unique and synergistic information for systems with $N\ge3$ sources that do not rely on the PID lattice.
Instead, these measures are \emph{intrinsically} defined for the full multivariate system.
The goal is to quantify how much information is contributed exclusively by each source to the target, and how much information is only revealed by considering certain sets of sources together.

First, we provide axioms that should be satisfied by the unique and synergistic information of multivariate systems, in addition to Property~\ref{property: Additivity} (additivity) and Property~\ref{property: Continuity} (continuity).
In terms of unique information, since it measures the information that can only be provided from a certain variable but not from others to the target, the following requirements should be straightforward.
\begin{axiom}[Axioms for Unique Information]
  \label{axiom:bound for un}
  For a set of source variables $\boldS=\{S_1,\dots,S_N\}$, $N \ge 2$ and target variable $T$, the unique information should satisfy:
  \begin{itemize}
    \item Commutativity: unique information is invariant under any permutation $\sigma$ of the given variables, i.e.,
      \begin{align}
        \operatorname{Un}(S_i \to T|\boldS_{[N]\setminus i}) = \operatorname{Un}(S_i \to T|\sigma(\boldS_{[N]\setminus i})).
      \end{align}
    \item Monotonicity: unique information decreases monotonically as more source variables are added, i.e.,
      \begin{align}
        \operatorname{Un}(S_i \to T|\boldS_{[N]\setminus i}) &\le \min_{\boldA \subsetneq \boldS_{[N]\setminus i}}\{\operatorname{Un}(S_i\to T|\boldA)\}.
      \end{align}
    \item Bound for Unique: unique information from~$S_i$ to~$T$ can neither be greater than all the information from~$S_i$ to~$T$, nor greater than the total entropy of~$T$ given $\boldS_{[N]\setminus i}$, i.e.,
      \begin{align*}
        \operatorname{Un}(S_i \to T|\boldS_{[N]\setminus i}) &\le \min\{I(S_i;T),H(T|\boldS_{[N]\setminus i})\}.
      \end{align*}
  \end{itemize}
\end{axiom}

Similarly, synergistic information measures the information that can only be provided by all source variables, but not from any strict subset, to the target.
Therefore, we write the requirements for synergistic as follows.
\begin{axiom}[Axioms of Synergistic Information]
  \label{axiom:bound for syn}
  For a set of source variables $\boldS=\{S_1,\dots,S_N\}$, $N\ge 2$ and a target variable~$T$, the synergistic information should satisfy
  \begin{itemize}
    \item
      Commutativity: synergistic information is invariant under any permutation $\sigma$ of given variables, i.e.,
      \begin{align}
        \operatorname{Syn}(\boldS \to T) = \operatorname{Syn}(\sigma(\boldS)\to T).
      \end{align}
    \item Monotonicity: synergistic information increases monotonically as some source variables are combined into one, in the nontrivial case where at least two source variables remain after the combination, i.e., for any permutation $\sigma$ of $[N]$,
      \begin{align}
        \operatorname{Syn}(\boldS\to T) &\le \operatorname{Syn}(S_{\sigma(1)},\dots,(S_{\sigma(i)},\dots,S_{\sigma(j)}),\dots,S_{\sigma(N)}\to T),
      \end{align}

    \item Bound for Synergy: synergistic information cannot be provided by any strict subset of the source variables, i.e.,
      \begin{align}
        \operatorname{Syn}(\boldS \to T) \le \min_{\boldA \subsetneq \boldS} \{H(T|\boldA)\}  -H(T|\boldS).
      \end{align}
  \end{itemize}
\end{axiom}
\begin{remark}

  Axioms~\ref{axiom:bound for un} and~\ref{axiom:bound for syn} are stated for $N\ge 2$.
  We do not treat the case $N=1$ separately, since in that degenerate setting the decomposition collapses to mutual information and the notions of unique/synergistic information become trivial.

  Moreover, the monotonicity requirement in Axiom~6 is meant to apply to wrapping operations that still leave at least two source variables after the combination.
  In this regime, wrapping can aggregate synergistic effects of different orders (number of sources): synergistic contributions that were previously attributable to smaller subsets of sources may be absorbed into the synergistic term of the wrapped system.
  As a result, the total synergistic information may increase under wrapping.
  Considering three source variables $S_1,S_2,S_3$ and a target $T$, the third-order synergistic contribution is given by $\operatorname{Syn}(S_1,S_2,S_3\to T)$.
  After wrapping $S_2$ and $S_3$ into a single variable, the quantity $\operatorname{Syn}(S_1,(S_2,S_3)\to T)$ generally includes not only the original third-order synergy, but also contributions corresponding to second-order synergies such as $\operatorname{Syn}(S_1,S_2\to T)$ and $\operatorname{Syn}(S_1,S_3\to T)$, which are absorbed into the wrapped variable.
\end{remark}

We now formally introduce our multivariate unique and synergistic information measures.
For three-or-more source variable system with sources $\boldS_{[N]}\triangleq (S_1,\ldots,S_N)$ and target $T$, where~$S_i$ is over the alphabet~$\mathcal{S}_i$ and~$T$ is over the alphabet~$\mathcal{T}$, we extend Def.~\ref{definition:un} by considering certain variables as a joint variable to get the general definition of unique information.
In what follows, for~$i\in[N]$ denote $\boldS_{[N]\setminus i}\triangleq (S_1,\ldots,\hat{S}_i,\ldots,S_N)$ over the alphabet $\cS_{[N]\setminus i}\triangleq\mathcal{S}_1\times\ldots \times\hat{\mathcal{S}_i} \times\ldots \times\mathcal{S}_N$, i.e., all sources except for~$S_i$.

\begin{definition}[General Unique Information]\label{definition:general un}
  In a system with a target variable $T$ and source variables $\boldS_{[N]}$, the unique information from~$S_i$ to~$T$ given~$\boldS_{[N]\setminus i}$ is
  \begin{align}
    \label{equ:general un}
    \operatorname{Un}(S_i \to T | \boldS_{[N]\setminus i}) \triangleq I(S'_i;T|\boldS_{[N]\setminus i}).
  \end{align}
  where $S'_i$ is constructed similar to~\eqref{equation: S_1'S_2T}, i.e.,
  \begin{align}
    \label{equ:s_1s_N}
    \Pr(S_i'=s_i,\boldS_{[N]\setminus i}=\bolds_{[N]\setminus i}|T=t)\triangleq\Pr(S_i=s_i|T=t)\Pr(\boldS_{[N]\setminus i}=\bolds_{[N]\setminus i}|T=t),
  \end{align}
  meaning, for every~$s_i\in\mathcal{S}_i$,
  \begin{align*}
    \Pr(S_i'=s_i)=\sum_{\bolds_{[N]\setminus i},t\in\mathcal{S}_{[N]\setminus i}\times \mathcal{T}}\Pr(S_i=s_i|T=t)\Pr(\boldS_{[N]\setminus i}=\bolds_{[N]\setminus i}|T=t)\Pr(T=t).
  \end{align*}
  The variable~$S_i'$ is well-defined since all probabilities are non-negative and sum to~$1$, which can be shown that
  \begin{align*}
    \sum_{s_i\in \mathcal{S}_i}\Pr(S_i'=s_i)&=\sum_{\bolds_{[N]\setminus i},t\in\mathcal{S}_{[N]\setminus i}\times \mathcal{T}} \sum_{s_i\in \mathcal{S}_i}\Pr(S_i=s_i|T=t)\Pr(\boldS_{[N]\setminus i}=\bolds_{[N]\setminus i}|T=t)\Pr(T=t) \\
    &=\sum_{\bolds_{[N]\setminus i},t\in\mathcal{S}_{[N]\setminus i}\times \mathcal{T}}\Pr(\boldS_{[N]\setminus i}=\bolds_{[N]\setminus i}|T=t)\Pr(T=t)\\
    &=\sum_{\bolds_{[N]\setminus i},t\in\mathcal{S}_{[N]\setminus i}\times \mathcal{T}}\Pr(\boldS_{[N]\setminus i}=\bolds_{[N]\setminus i},T=t)=1.
  \end{align*}
\end{definition}
Following a similar approach, we extend Lemma~\ref{lemma:syn} of synergistic information of two source variable systems to general three-or-more source variable systems.
However, unlike general unique information, which uses a simple combination of conditional variables, general synergistic information requires an adjustment to the construction of the transformed source variable (the equivalent of~$S_i'$), which is different from~\eqref{equation: S_1'S_2T}.
The reasoning is as follows: Lemma~\ref{le:redefinition of redundant and synergistic information} demonstrates that the calculation of synergistic information involves decoupling the higher-order contribution from two source variables to the target and capturing it through the variation in joint conditional entropy of the target, i.e., $H(T|S_1',S_2) - H(T|S_1,S_2)$.
For three-or-more source variable systems, on one hand, we aim to decouple the higher-order effects from all source variables to the target variable.
On the other hand, we do not wish to decouple the higher-order effects of any proper subset of the source variables to the target.
%, to avoid making the difference of joint conditional entropy miscellaneous.
To achieve this, we utilize the property that repeating conditional variables does not alter conditional entropy, leading to the following variation of~\eqref{equation: S_1'S_2T} and a generalized definition of synergistic information.

For three-or-more source variable systems with sources $\boldS_{[N]}\triangleq (S_1,\ldots,S_N)$ and target $T$, where $\boldS_{[N]\setminus i}\triangleq (S_1,\ldots,\hat{S}_i,\ldots,S_N)$, define a new family of variables $\boldS'_{[N]\setminus 1},\ldots,\boldS'_{[N]\setminus N}$, each over the respective alphabet $\cS_{[N]\setminus i}\triangleq\mathcal{S}_1\times\ldots \times\hat{\mathcal{S}_i} \times\ldots \times\mathcal{S}_N$, via its
conditional joint distribution as follows.
\begin{align} \label{equation: S[N]'T}
  \Pr(\boldS'_{[N]\setminus 1}=\bolds_{[N]\setminus 1},\ldots,\boldS'_{[N]\setminus N}=\bolds_{[N]\setminus N}|T=t) =\prod_{k \in [N]}\Pr(\boldS_{[N]\setminus k}=\bolds_{[N]\setminus k}|T=t),
\end{align}
meaning, for every $\bolds_{[N]\setminus i}\in\mathcal{S}_{[N]\setminus i},$
\begin{align*}
  \Pr(\boldS_{[N]\setminus i}'=\bolds_{[N]\setminus i})=
  \sum_{\overset{\bolds_{[N]\setminus 1},\ldots,\hat{\bolds}_{[N]\setminus i},\ldots,\bolds_{[N]\setminus N},t }{ \in \cS_{[N]\setminus 1}\times \cdots\times \hat{\mathcal{S}}_{[N]\setminus i}\times\cdots \times \cS_{[N]\setminus N}\times \mathcal{T}}}
  \Pr(T=t) \prod_{k \in [N]}\Pr(\boldS_{[N]\setminus k}=\bolds_{[N]\setminus k}|T=t). \nonumber
\end{align*}
Notice that all variables~$\boldS_{[N]\setminus n}',n\in [N]$ are well-defined since all probabilities are non-negative, and since for all~$i\in[N]$,
\begin{align*}
  \sum_{\bolds_{[N]\setminus i}\in\mathcal{S}_{[N]\setminus i}}\!\Pr(\boldS_{[N]\setminus i}'=\bolds_{[N]\setminus i})=
  \sum_{\overset{\bolds_{[N]\setminus 1},\dots,\bolds_{[N]\setminus N},t }{ \in \cS_{[N]\setminus 1}\times \dots \times \cS_{[N]\setminus N}\times \mathcal{T}}}\!
  \Pr(T=t) \prod_{k \in [N]}\Pr(\boldS_{[N]\setminus k}=\bolds_{[N]\setminus k}|T=t) = 1
\end{align*}
Then, general synergistic information is defined as follows.
\begin{definition}[General Synergistic Information]
  \label{definition:general syn}
  In a system with a target variable $T$ and source variables $\boldS_{[N]}$, the synergistic information from~$\boldS_{[N]}$ to~$T$ is
  \begin{align*}
    \operatorname{Syn}(S_1, \dots, S_N\to T) \triangleq H(T|\boldS'_{[N]\setminus 1},\ldots,\boldS'_{[N]\setminus N})-H(T|S_1, \dots, S_N),
  \end{align*}
  where $(\boldS'_{[N]\setminus 1},\ldots,\boldS'_{[N]\setminus N},T)$ is defined by~\eqref{equation: S[N]'T}.
\end{definition}

In this definition, all $\boldS'_{[N]\setminus i}$ ensure that every proper subset of the source variables $\boldS'_{[N]}$ has the opportunity to provide all their information, while also restricting the ability of all source variables to jointly contribute synergistic information.
As a result, the difference in joint conditional entropy captures the highest-order synergistic information $\operatorname{Syn}(S_1, \dots, S_N\to T)$.

The above Definitions~\ref{definition:general un}, and~\ref{definition:general syn} are similar to their two-source variable version, i.e., Def.~\ref{definition:general un} combines multiple conditional variables $(S_1,\ldots,\hat{S}_i,\ldots,S_N)$ as $\boldS_{[N]\setminus i}$ while applying Def.~\ref{definition:un}, and Def.~\ref{definition:general syn} treats $\boldS_{[N]\setminus n},n\in [N]$ as new sources. Those generalized definitions inherit all the proven properties of Def.~\ref{definition:un},  and Lemma~\ref{lemma:syn} i.e., continuity and additivity; the latter is proven in a manner similar to Theorem~\ref{theorem:proof of two}.
In addition, proof that Definitions~\ref{definition:general un},  and~\ref{definition:general syn} satisfy Axioms~\ref{axiom:bound for un}, and~\ref{axiom:bound for syn} respectively is given in Appendix~\ref{app: proof of satisfaction}.
It should be noted that a similar approach can be taken to define a corresponding multivariate redundant information, but its resulting properties are unsatisfactory.
The problem of finding a proper multivariate redundant information remains open.

Moreover, beyond the synergy from all source variables, this approach also allows us to obtain the combined effect of synergistic information of all source subsets of identical cardinality, as well as the total amount of synergy across all subsets.

First, similar to~\eqref{equation: S[N]'T}, for a three-or-more source variable system with sources $\boldS_{[N]}\triangleq (S_1,\ldots,S_N)$ and target $T$, we denote the set of all the indices $\mathcal{K}$ of the subset $\boldS_{\mathcal{K}}$ of $\boldS_{[N]}$ with cardinality $K\in [N]$ by $\binom{{[N]}}{K}$.
Then, we define a new family of variables $\{\boldS'_{\mathcal{K}} \mid \mathcal{K}\in \binom{{[N]}}{K}\}$, each over the respective alphabet $s_{\mathcal{K}}\in\prod_{k\in\mathcal{K}}\mathcal{S}_k$, via its conditional joint distribution as follows.
\begin{align*}
  \Pr\left(\bigcap_{\mathcal{K} \in \binom{ {[N]}}{K}} \{S_{\mathcal{K}}'=s_{\mathcal{K}}\}|T=t\right)=\prod_{\mathcal{K} \in \binom{ {[N]}}{K}}\Pr(S_{\mathcal{K}}=s_{\mathcal{K}}|T=t).
\end{align*}

Then, the combined effect of synergistic information from all the subsets of sources with cardinality~$K$ is defined as follows.
\begin{definition}[$K$-th order Synergistic Effects]
  \label{def:KSE}
  The combined synergistic information from the subset of cardinality $K$ to~$T$ is
  \begin{align}
    \label{equ:kse}
    &\operatorname{SE}_K(S_1, \dots, S_N\to T) \triangleq H\left(T\bigg\vert\bigcap_{\mathcal{K}' \in \binom{[N]}{K-1}} \{S_{\mathcal{K}'}'\}\right)-H\left(T\bigg\vert\bigcap_{\mathcal{K} \in \binom{[N]}{K}} \{S_{\mathcal{K}}'\}\right),
  \end{align}

\end{definition}
Further, the total synergistic effect is the summation of all~$K$-th order synergistic effects with~$K$ ranging from~$2$ to~$N$, as follows.
%which can be simplified as follow.
\begin{definition}[Total Synergistic Effect]
  \label{def: TSE}
  The combined synergistic information from~$S_1, \dots, S_N$ to~$T$ is
  \begin{align}
    \operatorname{TSE}(S_1, \dots, S_N\to T) &\triangleq \sum_{K=2}^N \operatorname{SE}_K(S_1, \dots, S_N\to T) \nonumber\\
    &= H(T|S'_1, \dots, S'_N)-H(T|S_1, \dots, S_N),
  \end{align}
  where $P(S'_1=s_1, \dots, S'_N=s_N|T=t) = \prod_{k \in [N]}\Pr(S_k=s_k|T=t)$.
\end{definition}

In this section, we proposed definitions of unique information and synergistic information in the multivariate scenario. The multivariate unique and synergy measures defined above provide an \emph{independent} extension of PID concepts beyond the confines of the original framework.

One may interpret the operational meaning of our multi source-variable information measures through the following intuitions.
First, generalized unique information measures the amount of information that a specific source variable (or subset) can provide to the target variable.
For any chosen source (or subset) $A\subseteq \boldS$, we treat the complement $\boldS\!\setminus\!A$ as a single super–source and apply the same two–source construction as in Eq.~\eqref{equation: S_1'S_2T}.
This can be understood as dividing multiple source variables into two groups and calculating the unique information of one group (or individuals) under the two-source variable scenario, which is similar to Def.~\ref{definition:un}.

Generalized synergistic information measures the contribution of all source variables to the target variable that must be achieved through the combined action of all $N$ source variables.
The structure of Eq.~\eqref{equation: S[N]'T} is similar to that of Eq.~\eqref{equation: S_1'S_2T}. By constructing a new set of source variables $\boldS_{\mathcal{K}}$, the high-order relationships of all $N-1$ sources (with respect to the target) are retained, while the high-order interactions that only exist among all $N$ sources are filtered out.
This idea extends naturally in Def.~\ref{def:KSE} and~\ref{def: TSE}.
Def.~\ref{def:KSE} ($K$-th Order Synergistic Effects) constructs distributions that preserve all dependencies up to order $K\!-\!1$ but removes exactly the $K$–wise interaction, yielding a measure of the synergy generated by size–$K$ coalitions.
Then, Def.~\ref{def: TSE} (Total Synergistic Effect) aggregates these contributions across orders to summarize the overall multivariate synergy present in~$S$.
These multivariate information measures do not rely on the lattice-based PID framework, avoiding the inherent inconsistencies in the case~$N \ge 3$.
Moreover, they offer an intuitive interpretation of high-order interactions between variables, making them readily and reliably applicable to a variety of multivariate data analysis scenarios.

In the next section, we will demonstrate the effectiveness of our measure through different case studies and compare our measure with existing ones.

\section{Comparison and Case Studies}\label{sec:experiments}
In this section, we demonstrate the performance of our proposed measure, hereafter denoted by $I_{\cap}^{\text{Exp}}$, through a series of experiments. First, we compare it with several widely-recognized measures, examining both their theoretical properties and performance across multiple computational scenarios to highlight our method’s advantages. Next, we test our measure on the Ising Model to illustrate its effectiveness in practical application settings.

\subsection{Comparison Experiments}
We compare our method $I_{\cap}^{\text{Exp}}$ with other existing measures for PID in terms of properties and numerical calculations to demonstrate the advantages of our method. The widely-regarded measures considered in this part include: $I_{\cap}^{\text{*}}$ proposed by Kolchinsky \cite{kolchinsky2022novel}, $I_{\cap}^{\text{WB}}$ proposed by Williams and Beer \cite{williams2010nonnegative}, the ``minimum mutual information'' $I_{\cap}^{\text{MMI}}$ proposed by Barrett \cite{barrett2015exploration}, $I_{\cap}^{\wedge}$ proposed by Griffith et al. \cite{griffith2014intersection}, $I_{\cap}^{\text{Ince}}$ proposed by Ince \cite{ince2017measuring}, $I_{\cap}^{\text{FL}}$ proposed by Finn and Lizier \cite{finn2018pointwise}, $I_{\cap}^{\text{BROJA}}$ proposed by Bertschinger et al. \cite{bertschinger2014quantifying}, $I_{\cap}^{\text{Harder}}$ proposed by Harder et al. \cite{harder2013bivariate}, and $I_{\cap}^{\text{dep}}$ proposed by James et al. \cite{james2018unique}.

First, we compare the following properties in order to highlight the differences between our approach and previous ones.

\begin{table}[htbp]
  \caption{\label{table:comparison}
    Qualitative comparison of existing PID redundancy measures and
    the proposed measure $I_{\cap}^{\text{Exp}}$.
    Each row corresponds to a structural or axiomatic property, and a check mark
    indicates that the measure satisfies the corresponding requirement.
    The table highlights the trade-offs between generality, axiomatic consistency,
  and computational tractability across different approaches.}
  \begin{ruledtabular}
    \begin{tabular}{ccccccccccc}
      & $I_{\cap}^{*}$ & $I_{\cap}^{\text{WB}}$ & $I_{\cap}^{\text{MMI}}$ & $I_{\cap}^{\wedge}$ &
      $I_{\cap}^{\text{Ince}}$ & $I_{\cap}^{\text{FL}}$ & $I_{\cap}^{\text{BROJA}}$ & $I_{\cap}^{\text{Harder}}$ & $I_{\cap}^{\text{dep}}$ & $I_{\cap}^{\text{Exp}}$   \\ \hline
      Three-or-more sources        & \checkmark & \checkmark & \checkmark & \checkmark  & \checkmark & \checkmark &           &           &          &  $\operatorname{Syn} \& \operatorname{Un}$\\
      Monotonicity (Axiom~\ref{axiom: Monotonicity})               & \checkmark & \checkmark & \checkmark & \checkmark  &           &           &  \checkmark    & \checkmark & \checkmark & \checkmark\\
      Additivity (Prop.~\ref{property: Additivity})&\checkmark    &            &            &   \checkmark  &  ? & ?          &    \checkmark        &  &   &  \checkmark \\
      Continuity (Prop.~\ref{property: Continuity})&   & \checkmark            &   \checkmark  &         &?  &    ?       & \checkmark   &    &   \checkmark &  \checkmark \\
      Independent identity (Prop.~\ref{property: Independent Identity})       & \checkmark &            &            & \checkmark & \checkmark &   &   \checkmark        &\checkmark  &  \checkmark & \checkmark\\

      Strong independent identity \footnote{This axiom  states that for two source variable systems, redundancy equals the mutual information in the case that the target equals the sources~\cite{harder2013bivariate}.}      &  &            &            &  &  &   &   \checkmark        &\checkmark  &  \checkmark & \checkmark\\
    \end{tabular}
  \end{ruledtabular}
\end{table}

In Table~\ref{table:comparison}, a check mark ``$\checkmark$'' indicates that the measure has the property, blank entry indicates that it does not have the property, and a question mark ``$?$'' indicates that the respective property was not claimed in the original publication, and we were not able to verify it ourselves.
%we are unable to make an accurate judgment due to our knowledge scope.

The results presented in the table above indicate that our measure satisfies all the required properties except for three-or-more source variable scenarios, which have been proven impossible in Lemma \ref{lemma: counter example}.
Nonetheless, it remains capable of calculating both synergistic and unique information in three-or-more source variable settings. Notably, among the measures considered, the $I_{\cap}^{\text{BROJA}}$ also fulfills all properties except the three-or-more sources.
To have a more detailed comparison of the two methods, we reviewed the following definition of the PID measure $I_{\cap}^{\text{BROJA}}$~\cite{bertschinger2014quantifying}.
\begin{definition}
  Let~$\Delta$ be the set of all random variables over~$\cS_1\times\cS_2\times \cT$ which preserve the marginal distributions~$(S_1,T)$ and~$(S_2,T)$, that is,
  \begin{align}
    \label{equ:broja table}
    \Delta = &\{ (\hat{S}_1,\hat{S}_2,\hat{T}) : \Pr(\hat{T}=t, \hat{S}_1 = s_1) = \Pr(T=t, S_1 = s_1), \nonumber \\ &\text{ and } \Pr(\hat{T}=t, \hat{S}_2 = s_2) = \Pr(T=t, S_2 = s_2), \text{ for all } t \in \mathcal{T}, s_1 \in \mathcal{S}_1, s_2 \in \mathcal{S}_2 \},
  \end{align}
  Then, the unique information is defined as:
  \begin{align}
    \operatorname{Un}^{\text{Broja}}(S_1\to T|S_2)=\min_{(\hat{S}_1,\hat{S}_2,\hat{T}) \in \Delta} I(\hat{S}_1;\hat{T} \mid \hat{S}_2),
  \end{align}
  where $I(\hat{S}_1;\hat{T} \mid \hat{S}_2)$ is the conditional mutual information.
\end{definition}

Our proposed method~$I_{\cap}^{\text{EXP}}$, shares several conceptual similarities with the $I_{\cap}^{\text{BROJA}}$ measure.
Both approaches construct the new joint distribution preserving the marginal distributions of each source–target pair, i.e., $(S_1, T)$ and $(S_2, T)$.
Furthermore, both methods rely on optimizing entropy-related quantities to derive these distributions.
However, the two methods differ fundamentally in the choice of optimization criteria. In Def.~\ref{definition:un}, particularly~\eqref{equation: S_1'S_2T}, our method explicitly maximizes the joint entropy of the newly constructed distribution over $(S'_1, S_2, T)$, which is claimed as follows and proved in Appendix~\ref{app: compareson with broja}.

\begin{lemma}
  \label{le:max entropy}
  Let $\Delta$ be as in~\eqref{equ:broja table}. The unique information in Def.~\ref{definition:un} is {equivalent to}:
  \begin{align}
    \operatorname{Un}(S_1\to T|S_2)=I(S'_1;T' \mid S'_2 ), \text{ s.t. } \displaystyle (S'_1,S'_2,T' ) = \arg\max_{(\hat{S}_1,\hat{S}_2,\hat{T})\in \Delta} H (\hat{S}_1,\hat{S}_2,\hat{T}).
  \end{align}
\end{lemma}
In contrast, $I_{\cap}^{\text{BROJA}}$ is based on minimizing the conditional mutual information $I(S'_1 ; T \mid S_2)$. This distinction in the optimization objective leads to differing behaviors in certain cases.

To further illustrate the differences between our proposed measures $I_{\cap}^{\text{Exp}}$ and $I_{\cap}^{\text{BROJA}}$, we compare using specific examples.
An example {used in~\cite{james2018dit}} to critique $I_{\cap}^{\text{BROJA}}$ is as follows: Consider a distribution of sources~$S_1,S_2$ and a target $T$ as in Table \ref{table:reduce or}.
\begin{table}[htbp]
  \caption{\label{table:reduce or}
  Table of Joint Probability Distribution.}
  \begin{ruledtabular}
    \begin{tabular}{cccccc}

      \textbf{Pr} & $\mathbf{S_1}$ & $\mathbf{S_2}$ & $\mathbf{T}$   \\ \hline
      $1/2$ & 0 & 0 & 0   \\
      $1/4$ & 1 & 0 & 1  \\
      $1/4$ & 0 & 1 & 1   \\
    \end{tabular}
  \end{ruledtabular}
\end{table}

In this distribution, we can intuitively see that when $S_1$ or $S_2$ is equal to~$1$, it follows that $T$~must also be equal to 1, whereas the other source variable cannot completely distinguish the value of~$T$.
This implies that each source provides nonzero unique information about~$T$.
At the same time, when both $S_1=0$ and $S_2=0$ (which occurs with probability $1/2$),
either source reduces the uncertainty of~$T$ in the same way.
Hence, it is also natural to expect a nonzero amount of redundant information
shared by $S_1$ and $S_2$ about~$T$.

Table~\ref{table:result of reduce or} reports the PID atoms computed by different
redundancy measures over the distribution in Table~\ref{table:reduce or} using the \texttt{dit} Python package~\cite{james2018dit} (except $I_{\cap}^{\text{*}}$).
Several measures, including
$I_{\cap}^{\text{WB}}$, $I_{\cap}^{\text{MMI}}$,
$I_{\cap}^{\text{BROJA}}$, and $I_{\cap}^{\text{Harder}}$,
assign zero unique information to both sources.
In these cases, the mutual information between single source and target is effectively attributed
to redundancy, leading to a decomposition in which the two sources are treated
as informationally interchangeable with respect to~$T$.
While this allocation is mathematically consistent with the definitions of these
measures, it contradicts the intuitive asymmetry present in the distribution,
where each source can independently determine~$T$ in distinct events.
\begin{table}[htbp]
  \caption{\label{table:result of reduce or}
  Behavior of $I_{\cap}^{\text{Exp}}$ and other measures on the Distribution in Table \ref{table:reduce or}.}
  \begin{ruledtabular}
    \begin{tabular}{cccccccccc}
      & $I_{\cap}^{\text{WB}}$ & $I_{\cap}^{\text{MMI}}$ & $I_{\cap}^{\wedge}$ &
      % $I_{\cap}^{\text{GH}}$ &
      $I_{\cap}^{\text{Ince}}$ & $I_{\cap}^{\text{FL}}$ & $I_{\cap}^{\text{BROJA}}$ & $I_{\cap}^{\text{Harder}}$ & $I_{\cap}^{\text{dep}}$ & $I_{\cap}^{\text{Exp}}$\\
      \hline
      $\operatorname{Red}(S_1,S_2\to T) $  & 0.311 & 0.311 & 0  & 0  & -0.085 &   0.311 & 0.311  &  0.074  &  0.074\\

      $\operatorname{Un}(S_1\to T|S_2) $  & 0 & 0 & 0.311  &   0.311  & 0.396  &  0  &0 &  0.238 & 0.238\\
      $\operatorname{Un}(S_2\to T|S_1) $ &  0 &   0  &   0.311  & 0.311 & 0.396   &   0  &  0 & 0.238  &  0.238 \\

      $\operatorname{Syn}(S_1,S_2\to T) $ & 0.689  &  0.689  &  0.377  &0.377 & 0.292  &0.689    &0.689    &  0.451 &  0.451 \\
    \end{tabular}
  \end{ruledtabular}
\end{table}

By contrast, the measures
$I_{\cap}^{\wedge}$, $I_{\cap}^{\text{Ince}}$, and $I_{\cap}^{\text{FL}}$
do identify nonzero unique information, in agreement with the above intuition.
However, they do so at the expense of redundancy:
$I_{\cap}^{\wedge}$ and $I_{\cap}^{\text{Ince}}$ assign zero redundancy,
while $I_{\cap}^{\text{FL}}$ yields a negative redundancy value,
which violates basic nonnegativity requirements.
In these approaches, the entire mutual information between each individual source
and the target is effectively classified as unique information.
As a result, these measures suppress the shared contribution that both sources
provide to reducing the uncertainty of~$T$.

Among the considered measures, only $I_{\cap}^{\text{dep}}$ and the proposed
$I_{\cap}^{\text{Exp}}$ allocate the information atoms in a way that reflects
both aspects of the system: a nonzero redundant component capturing the shared
predictive power of the sources, and nonzero unique components capturing their
distinct contributions.
But it should be noticed that although $I_{\cap}^{\text{dep}}$ coincides with $I_{\cap}^{\text{Exp}}$ for this
simple example, it does not satisfy additivity and therefore diverges from
$I_{\cap}^{\text{Exp}}$ in more complex settings.

Overall, this example illustrates that an appropriate PID measure must balance
the allocation between redundancy and unique information, and that
$I_{\cap}^{\text{Exp}}$ achieves this balance while retaining desirable
axiomatic and computational properties, offering a more reliable quantitative tool for PID.

\subsection{Analyzing the Ising Model}
\label{subsec:ising-method}
The two-dimensional Ising model is widely used in statistical physics to study phase transitions and collective phenomena in complex systems.
It consists of a two-dimensional grid with side length $L$, each of which holds a binary variable $\sigma_i \in \{-1, +1\},i\in\{1,\ldots,L^2\}$ representing a ``spin'' pointing down or up. The spins interact only with their immediate neighbors, and the system tends to favor configurations where neighboring spins are aligned.

The energy (Hamiltonian) of the given spin configuration $\boldsymbol{\sigma}=\{\sigma_1,\dots,\sigma_{L^2}\}$ is defined as:
\begin{equation}
  \label{equ:hamilton}
  {\mathcal{E}}(\boldsymbol{\sigma}) = -J \sum_{\langle i,j \rangle} \sigma_i \sigma_j,
\end{equation}
where $\langle i,j \rangle$ denotes unordered nearest-neighbor pairs, and $J>0$ is the coupling constant. The larger its value, the more the two adjacent spins tend to be arranged in the same direction.
Aligned spin configurations ($\sigma_i = \sigma_j$) minimize the system’s energy, and the Boltzmann distribution {(described next)} implies that these low-energy states dominate the system’s behavior at low temperatures.

This energy function plays a central role in determining the system's statistical behavior. Specifically, the probability of each configuration $\boldsymbol{\sigma}$ is defined via the Boltzmann distribution, which assigns higher likelihood to lower-energy states:
\begin{equation}
  P(\boldsymbol{\sigma}) = \frac{1}{Z} \exp\left(-\frac{\mathcal{E}(\boldsymbol{\sigma})}{T}\right),
\end{equation}
where $T$ is the temperature, and $Z = \sum_{\boldsymbol{\sigma}} \exp\left(-\mathcal{E}(\boldsymbol{\sigma})/T\right)$ is a normalization constant to make sure the total probability adds up to 1. At low temperatures, the system prefers low-energy configurations with large domains of aligned spins (ordered phase). At high temperatures, thermal fluctuations dominate and spins become more random (disordered phase).

This model exhibits a sharp phase transition at a critical temperature $T_c \approx 2.269$ (for $J = 1$ on a 2D square lattice), separating ordered and disordered phases. This makes the Ising model a paradigmatic system to study collective behavior, not only in physics but also in applications such as neuroscience, biology, and social systems.

To approximate the Boltzmann distribution through local updates, we use Glauber dynamics \cite{glauber1963time}, which defines a flip probability based on the change in local energy.
This implies that spins tend to flip when doing so lowers the system's energy, though thermal fluctuations can still induce energy-increasing flips, particularly at higher temperatures.
Each complete update step visits every site on the grid exactly once in random order, flipping each spin (from $+1$ to $-1$ or vice versa) with a probability given by:
\begin{align}
  \label{equ:glauber}
  \Pr(\sigma_i \to -\sigma_i)=\frac{1}{1+\exp(\Delta E_i/T)},
\end{align}
where energy change $\Delta E_i$ for flipping spin $\sigma_i$ depends on its interaction with the four nearest neighbors:
\begin{align}
  \Delta E_i = 2J\sigma_{i}\sum_{j \in \text{neighbors}(i)}\sigma_{j}.
\end{align}

This spin system provides a well-controlled setting where the level of statistical dependence between variables varies with temperature. It is thus ideal for testing information decomposition methods from Sections~\ref{sec:two-source-formula}–\ref{sec:multi-measures} that aim to characterize such dependencies.
Specifically, we test whether our PID definitions can (i) detect the emergence of collective order and
(ii) identify temperature regions where PID components exhibit pronounced structure.
We then compare these trends with standard thermodynamic observables (magnetization, susceptibility, and specific heat) to assess whether PID-derived quantities provide complementary signatures of criticality.

The numerical experiments were done in \texttt{Python}.
We use a $128\times128$ grid with periodic (toric) boundaries. Experiments span $50$ temperatures from $2.0$ to $2.8$, spaced by approximately $0.016$.
At each temperature, we first stabilize the system by updating it $20,000$ times.
After stabilization, we record the grid's state over an additional $80,000$ update steps, taking one snapshot after each complete step. Algorithm~\ref{alg:ising-main} outlines this process clearly by calling the single updating algorithm (Glauber sweep) detailed in Algorithm~\ref{alg:glauber-step}, which is the one full update attempt over all sites in random order.
\begin{algorithm}[H]
  \caption{Ising Simulation Procedure at Temperature $T$}
  \label{alg:ising-main}
  \KwIn{Grid size $L=128$, Burn-in steps $B=20,000$, Recorded steps $N=80,000$}
  Randomly initialize each site as $+1$ or $-1$;
  \For{$b=1$ \KwTo $B$}{
    {Perform Glauber sweep (Algorithm~\ref{alg:glauber-step});}
  }
  \For{$n=1$ \KwTo $N$}{
    {Perform Glauber sweep (Algorithm~\ref{alg:glauber-step});}
    Record the current grid configuration;
  }
  \Return $N$ recorded grid configurations;
\end{algorithm}
\begin{algorithm}[H]
  \caption{Single Update Step (Glauber Sweep)}
  \label{alg:glauber-step}
  \KwIn{Current grid configuration, Temperature $T$}
  \For{each~$i\in[L^2]$ (in a random order)}{
    Calculate energy change $\Delta E_i$ using site $i$ and its four neighbors;
    Flip the site's state with probability $1/(1+e^{\Delta E_i/T})$;
  }
\end{algorithm}

\begin{figure}[htbp]
  \centering

  \begin{minipage}[t]{0.32\textwidth}
    \centering
    \includegraphics[width=\linewidth]{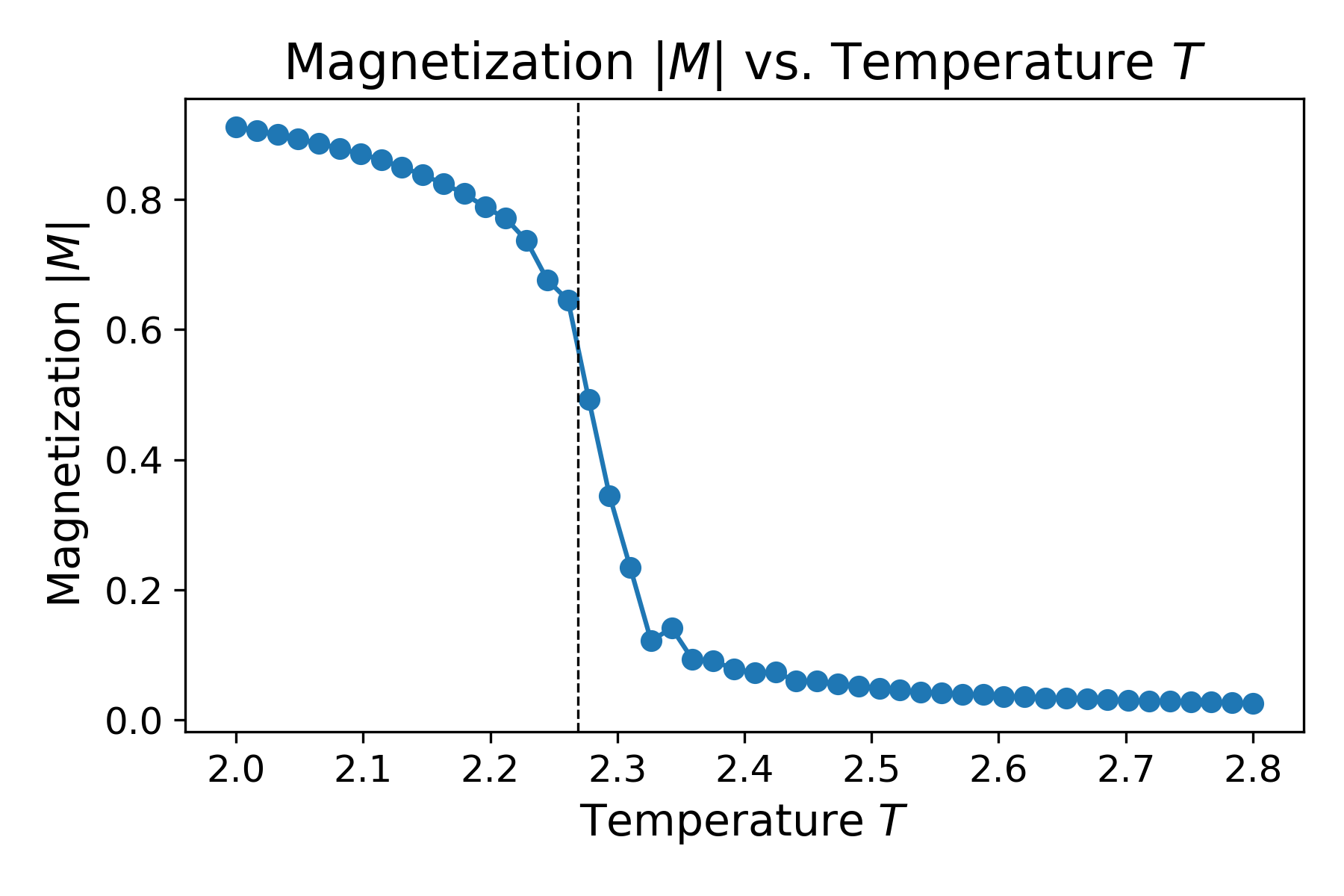}
    \caption*{(A)}
  \end{minipage}
  \hfill
  \begin{minipage}[t]{0.32\textwidth}
    \centering
    \includegraphics[width=\linewidth]{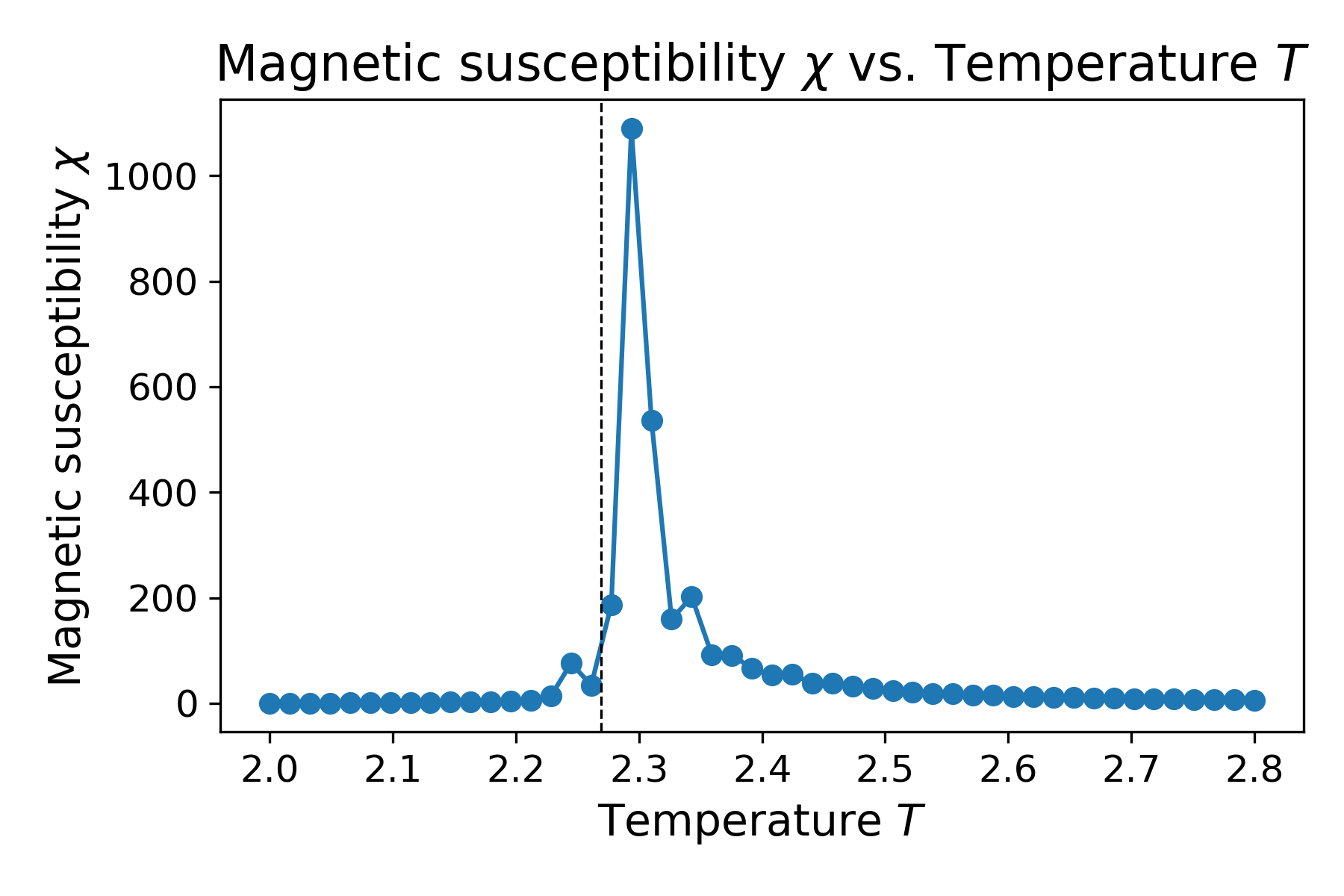}
    \caption*{(B) }
  \end{minipage}
  \hfill
  \begin{minipage}[t]{0.32\textwidth}
    \centering
    \includegraphics[width=\linewidth]{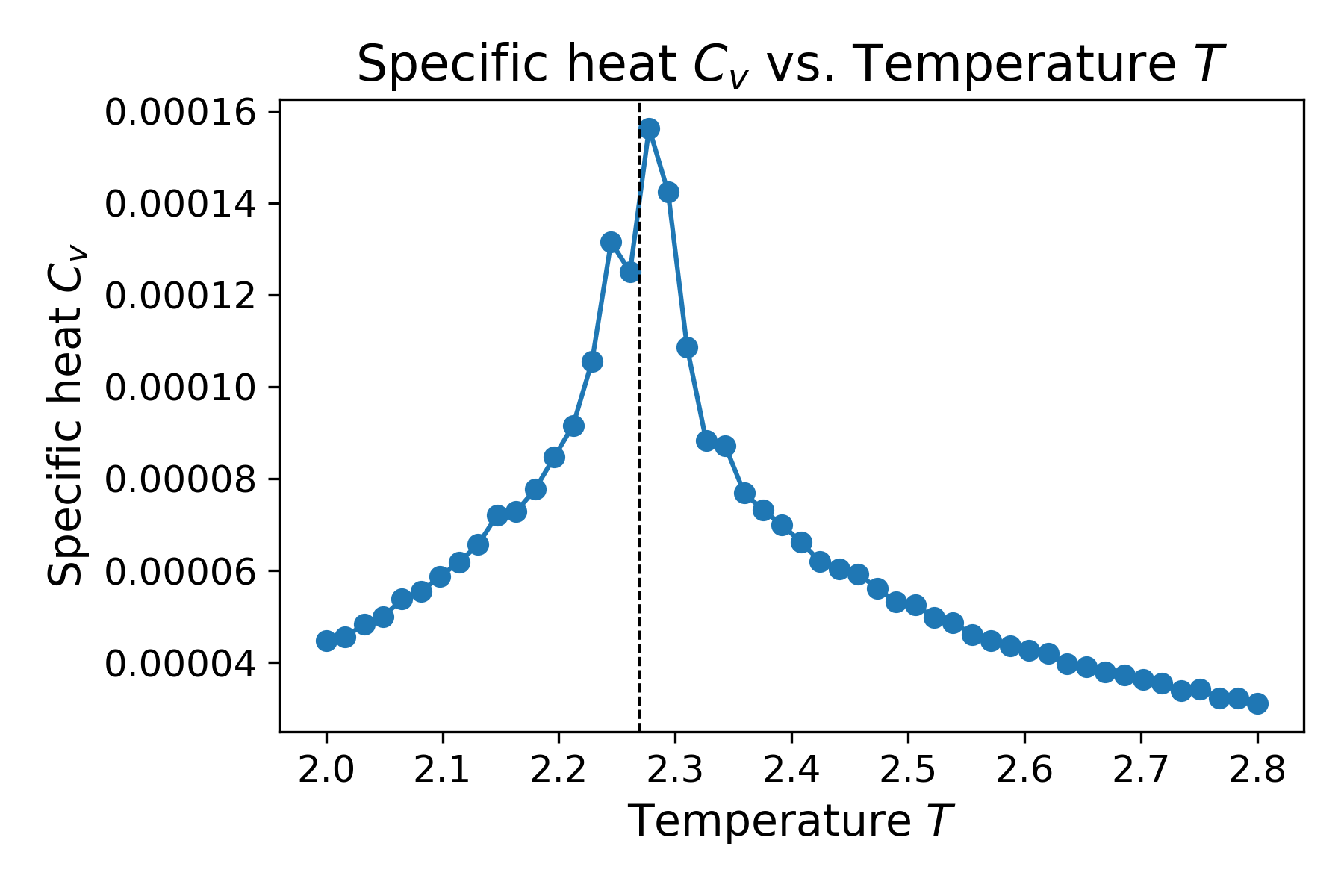}
    \caption*{(C) }
  \end{minipage}

  \caption{Temperature dependence of (A) magnetization $|M| $, (B) susceptibility $\chi$, and (C) specific heat $C_v$. The peaks in $\chi$ and $C_v$ identify the critical region of the phase transition,
    which serves as a reference for interpreting the information-theoretic quantities
  in Fig.~\ref{fig: 3 pid result}.}
  \label{fig:indexes}
\end{figure}

From each recorded grid configuration, we compute three key measurements:

\begin{itemize}\setlength\itemsep{4pt}
  \item \textbf{Magnetization}
    $(M=\tfrac1{L^{2}}\sum_{i}\sigma_{i})$.
    Measures the average value of all spins on the lattice.  A nonzero $M$ indicates a collectively ordered state, while $M=0$ corresponds to a disordered state.

  \item \textbf{Magnetic susceptibility}
    ($\chi = L^{2}\,\mathrm{Var}(M)/T$).
    Quantifies how strongly the magnetization fluctuates under a small external perturbation.  As the system approaches its critical temperature $T_c$, these fluctuations grow without bound.

  \item \textbf{Specific heat}
    ($C_v = \mathrm{Var}(\mathcal{E}(\boldsymbol{\sigma}))/T^{2}$).
    Captures the variance of the energy $\mathcal{E}(\boldsymbol{\sigma})$ with respect to temperature.  At the phase transition, $C_v$ exhibits a pronounced peak reflecting large energy fluctuations.
\end{itemize}
The results for these observables are shown in Fig.~\ref{fig:indexes}.

\paragraph*{Experiment 1 (local PID around a central site).}
To perform a detailed PID analysis, we selected $50$ interior sites uniformly at random (fixed across all temperatures).
For each chosen site $C$ and its four nearest neighbors $(U,D,L,R)$, we estimated the empirical joint distribution of the $2^5$ local spin configurations at each temperature from the recorded samples.
This yields $50$ empirical distributions $P_T^{(m)}(U,D,L,R,C)$ per temperature ($m=1,\dots,50$).

Because full multivariate PID lattices with three or more sources are affected by the incompleteness issues discussed in Section~\ref{sec:limitations}, we focus primarily on two-source decompositions. In the first experiment, we selected the left and right neighbors as the two sources; in the second one, we grouped the top and bottom neighbors as one source and the left and right neighbors as another.
In both experiments, the target variable is~$C$.
These settings allow us to assess the influence of neighbor pairs on the central variable without entangling higher-order decomposition issues.
In particular, the first two settings match the Ising PID setups studied by Sootla \emph{et al.}~\cite{sootla2017analyzing}, who analyzed the same tasks using the $I_{\cap}^{\text{BROJA}}$.
The third experiment was conducted to test the general measures of unique information and total synergistic effect (Def~\ref{definition:general un} and Def.~\ref{def: TSE}) for the four neighbors, while intentionally avoiding the interpretation of the full multi-variable PID lattice and its remaining atoms.
The experimental details of Tasks A-C are as follows:
\begin{itemize}
  \item[(A)] \emph{Left-Right decomposition.}
    Treat $C$ as the target and $L$ and $R$ as two source variables. We computed unique and synergistic information (Def.~\ref{definition:un}, Lemma~\ref{lemma:red}, and~\ref{lemma:syn}) from each source, and the mutual information $I(C;LR)$ between sources and the target.
  \item[(B)] \emph{Vertical-Horizontal decomposition.}
    Combining vertical ($U,D$) and horizontal ($L,R$) neighbors into two composite sources $(UD)$ and $(LR)$, again targeting $C$. The same PID components and mutual information $I(C;(UD),(LR))$ were calculated.
  \item[(C)] \emph{Four source decomposition.}
    Treating each neighbor ($U,D,L,R$) as a separate source and~$C$ as the target. We computed the general unique information (Def.~\ref{definition:general un}) contributed by each neighbor individually and total synergistic effect (Def.~\ref{def: TSE}) by all of them.
\end{itemize}

\begin{figure}[htbp]
  \centering

  \begin{minipage}[t]{0.32\textwidth}
    \centering
    \includegraphics[width=\linewidth]{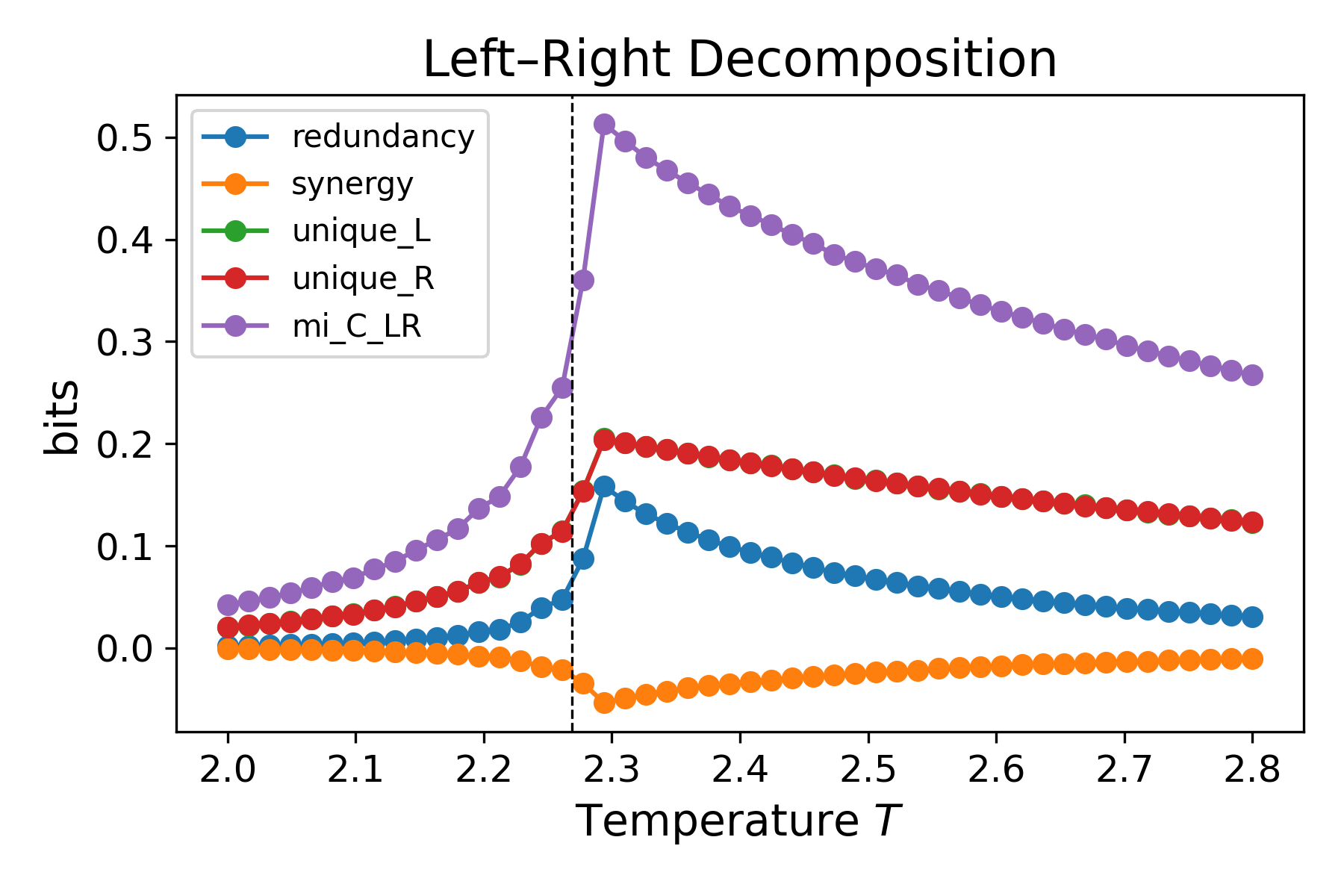}
    \caption*{(A)}
  \end{minipage}
  \hfill
  \begin{minipage}[t]{0.32\textwidth}
    \centering
    \includegraphics[width=\linewidth]{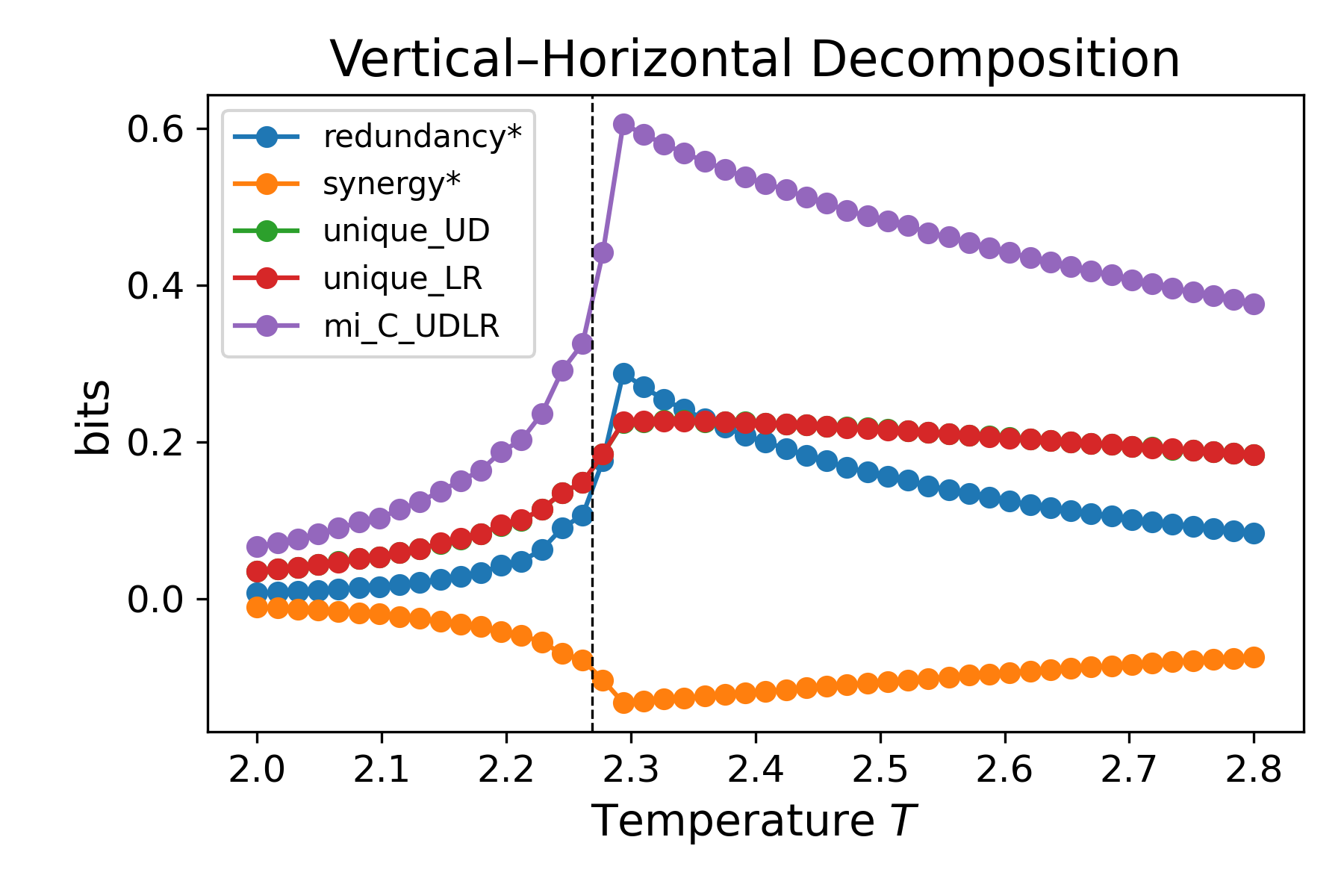}
    \caption*{(B)}
  \end{minipage}
  \hfill
  \begin{minipage}[t]{0.32\textwidth}
    \centering
    \includegraphics[width=\linewidth]{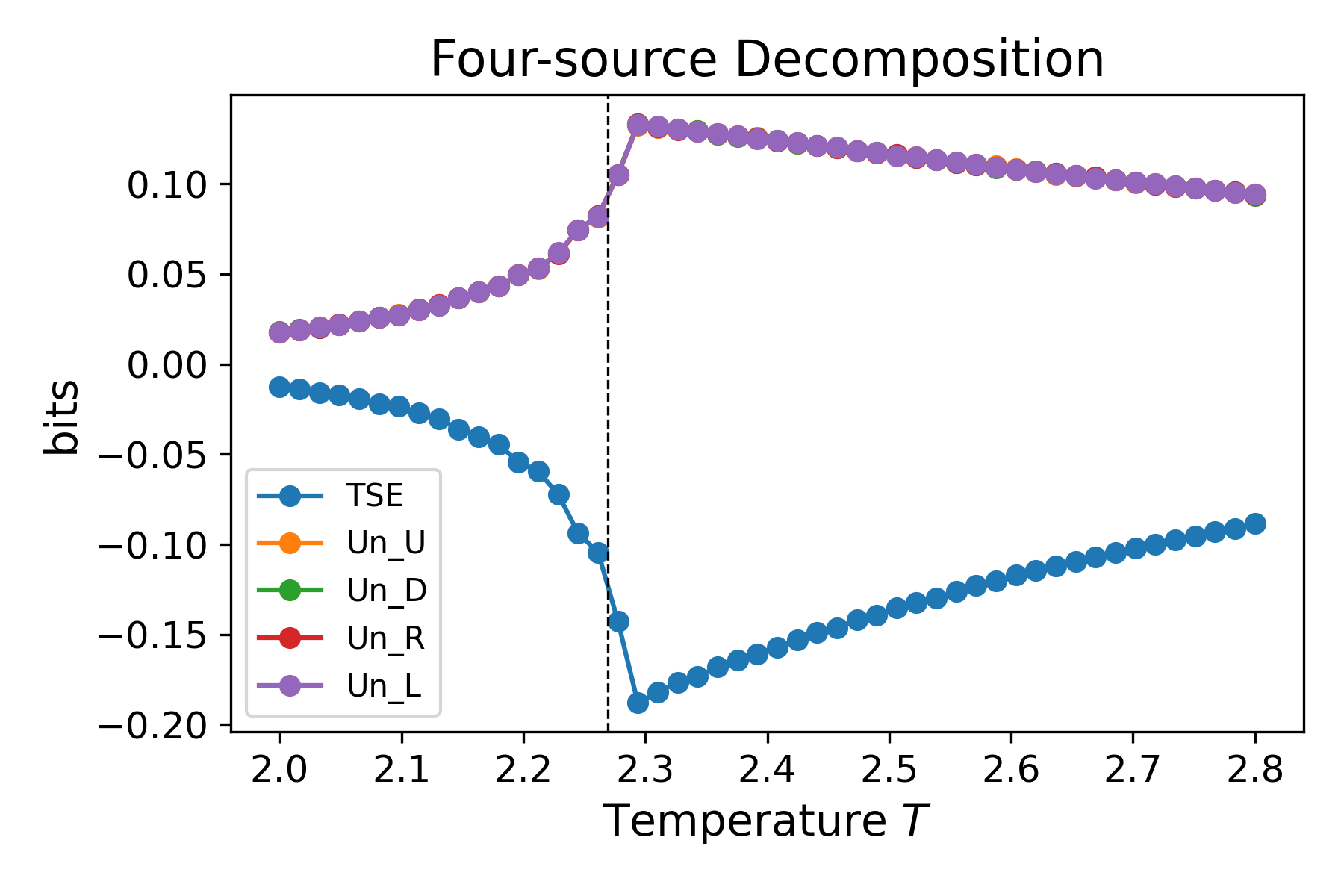}
    \caption*{(C)}
  \end{minipage}

  \caption{ Temperature dependence of PID-related quantities averaged over 50 lattice sites.
    (A) Left--right decomposition,
    (B) vertical--horizontal decomposition,
    and (C) four-source setting.
    The PID measures exhibit pronounced structure near the critical temperature,
    indicating a strong relation between collective physical behavior and
  informational interactions among neighboring spins.}
  \label{fig: 3 pid result}
\end{figure}

PID measures were averaged over the 50 sites at each temperature, yielding five temperature-dependent curves per decomposition type (Figs.~\ref{fig: 3 pid result}).

Comparative analysis of these PID results reveals critical insights. From Figures~\ref{fig: 3 pid result}.(A) and~\ref{fig: 3 pid result}.(B), both mutual information consistently peaks near $T_c\approx 2.269$, aligning with prior studies~\cite{barnett2013information,sootla2017analyzing}.
Specifically, {Task A} measures the mutual information between a site and its two adjacent sites, while Task B measures the four adjacent sites, so as the joint mutual information obtained in Task B is higher.
Past studies have also confirmed this point. Barnett et al.
\cite{barnett2013information} observed that the maximum value of the mutual information between two adjacent sites is less than 0.3, while the maximum value of the mutual information between two neighbors and the center point in \cite{sootla2017analyzing} is about 0.5, which is consistent with Task A, while the maximum value of the mutual information in Task B in this study is 0.8, which is consistent with intuition.

For the PID measures proposed in this paper, redundant information peaks sharply at the critical temperature, underscoring the dominant role of shared information in collective behaviour.  Unlike earlier reports, the \emph{unique} information we measure is clearly \emph{non-zero} and follows almost the same temperature profile as redundancy (its value is identical in each individual figure due to symmetry); in fact, it surpasses all other PID terms near criticality and attains its maximum exactly at the phase-transition point.
Synergistic information, by contrast, exhibits the opposite tendency: it decreases as the system approaches $T_c$ and reaches a pronounced (negative) minimum there.  Away from criticality---at very low or very high temperatures---synergy decays to zero, consistent with the reduced influence of neighbours on the central spin.

Sootla \emph{et al.}~\cite{sootla2017analyzing} studied the same two-source Ising settings using the $I_{\cap}^{\text{BROJA}}$ and report vanishing unique information, with a synergy curve that largely mirrors the behavior of mutual information.
By contrast, our results differ in two aspects: (i) unique information is sizable and temperature-dependent rather than identically zero; and (ii) the synergy curve is inverted, attaining a minimum (rather than a maximum) near $T_c$.

These differences reflect the allocation principles implicit in the respective PID definitions.
In the $I_{\cap}^{\text{BROJA}}$-based PID used in~\cite{sootla2017analyzing}, the mutual information between single source and target can be absorbed into redundancy in a way that yields zero unique information for each neighbor, which matches the behavior also observed in Table~\ref{table:result of reduce or}.
In our framework, we instead quantify each neighbor's non-repetitive contribution to predicting the central spin, which naturally leads to nonzero unique information that varies with temperature.

A possible interpretation of the observed trends is as follows.
Because the Glauber update probability depends on the \emph{sum} of neighboring spins, a change in any single neighbor can alter the central site's update propensity in a way that cannot be fully replicated by the other neighbors, making it plausible that unique information co-varies with redundancy and peaks in the critical region.
At the same time, the local neighborhood near criticality is dominated by increasingly coherent, large-scale patterns, which may suppress XOR-like combinatorial effects and thus drive synergy toward a minimum.
We emphasize that this discussion is meant to provide intuition consistent with the measured curves, while the quantitative comparison is given by the PID profiles and their correlations with standard thermodynamic observables.

When the two neighbors are merged into a single composite source, redundancy increases further near $T_c$ as shown in Fig.~\ref{fig: 3 pid result}(B), reinforcing the view that criticality enhances shared predictive structure in local updates.
The four-source results in Fig.~\ref{fig: 3 pid result}(C) mirror the two-source trends, supporting that the proposed multivariate measures extend the two-source case in a coherent way while avoiding reliance on a full multivariate lattice interpretation.

Finally, Pearson correlations between the averaged PID metrics and the three standard physical observables across all temperatures (Table~\ref{tab:corr-matrix}) quantify the links between information-theoretic and thermodynamic indicators of criticality. Here $\rho$ denotes the Pearson correlation coefficient,
$\rho=\mathrm{Cov}(X,Y)/\sqrt{\mathrm{Var}(X)\mathrm{Var}(Y)}$.

\begin{table}[htbp]
  \centering
  \caption{Pearson correlation matrix of physical observables vs. PID measures}
  \label{tab:corr-matrix}
  \begin{ruledtabular}
    \begin{tabular}{ccccc|cccc|cc}
      & \multicolumn{4}{c}{Task A: Left–Right(L,R)} & \multicolumn{4}{c}{Task B: Vertical–Horizontal(UD,LR)} & \multicolumn{2}{c}{Task C: Four-source}\\
      & Red & Syn & Un$_\mathrm{L/R}$  & $I_{C,LR}$ %
      & Red & Syn & Un$_\mathrm{UD/LR}$  & $I_{C,UDLR}$ %
      & TSE & Un$_\mathrm{U/D/R/L}$ \\
      \hline
      $\lvert M\rvert$ & $-0.60$ & $0.57$ & $-0.86$  & $-0.82$
      & $-0.70$ & $0.84$ & $-0.94$  & $-0.87$
      & $0.79$  & $-0.90$ \\
      $\chi$           & $0.79$  & $-0.80$& $0.54$    & $0.60$
      & $0.59$  & $-0.41$ & $0.28$  & $0.40$
      & $-0.48$ & $0.34$  \\
      $C_v$            & $0.45$  & $-0.53$& $0.15$   & $0.20$
      & $0.37$  & $-0.20$& $-0.09$ & $0.13$
      & $-0.28$ & $0.09$  \\
    \end{tabular}
  \end{ruledtabular}
\end{table}

The correlation table reveals three clear patterns linking information theoretic measures to physical indicators.
For the absolute magnetization $|M|$, it is strongly \emph{positively} correlated with synergy and TSE (with Pearson coefficient $\rho \approx 0.57$--$0.84$) yet strongly \emph{negatively} correlated with both redundancy and all unique--information terms ($|\rho| \gtrsim 0.70$).\ %
This indicates that as the system moves away from criticality and magnetic order sets in, collective synergy and TSE become more prominent, while both shared and source-specific information diminish.
Besides, for the magnetic susceptibility $\chi$, which peaks at the phase transition, exhibits the opposite signature: it is positively correlated with redundant information ($\rho \approx 0.71$--$0.79$ in Task A and B), which is more than mutual information, and also negatively with synergy and TSE ($-0.80$ in Task A).
Thus, near criticality the lattice's magnetic susceptibility is accompanied by an influx of redundant information and a suppression of synergistic effects, consistent with our PID based interpretation of the phase transition.\ %
However, different from the first two indicators, the heat capacity $C_v$ displays only weak to moderate links ($|\rho| \lesssim 0.33$) to any information metric, confirming that energy fluctuations couple less directly to the local information structure than do spin-based observables.
In addition to the comparison with physical indicators,
it is notable that using larger neighborhoods or composite sources (e.g., Task~B) can amplify correlations (especially for $|M|$, where $|\rho|$ rises to $\sim 0.9$), underscoring that collective dynamics become more detectable when broader local contexts are analyzed.
Overall, these results quantitatively substantiate the claim that PID derived quantities offer sensitive and complementary probes of critical phenomena beyond traditional physical indicators and basic mutual information measure.

Besides, in this study, we also tested the PID measures for cross-temporal information flows. For the same three tasks, we keep the neighbor variables at each moment as the source variables but take the center variables at the next step (sweep) as the target variables to test whether there are differences in information flow at different moments. The results show that all the results from the cross-temporal experiment are consistent with the results of the experiment we present.

\paragraph*{Experiment 2 (synergistic detection of emergent spatial order).}

While Experiment~1 focused on how neighboring spins contribute information to the instantaneous state of a central site, this setting primarily probes \emph{local predictability}.
Near criticality, however, the Ising model is characterized by the emergence of pronounced \emph{mesoscopic} spatial structures, namely clusters of aligned spins spanning intermediate length scales between microscopic spin interactions and macroscopic order.
The geometry and statistics of these spin clusters---often described in terms of domain formation and fractal domain walls---change sharply in the critical region and play a central role in the macroscopic behavior of the system
\cite{onsager1944crystal,janke2005fractal}.
Such intermediate-scale clustering is therefore commonly viewed as a geometric signature of emergent behavior in the Ising model.

Motivated by this perspective, we treat the presence of local collective order itself as the target variable.
Rather than predicting the state of a single spin, we ask how pairs of adjacent spins jointly inform whether a surrounding spatial region exhibits coherent alignment.

\begin{figure}[htbp]
  \centering
  \begin{minipage}[t]{0.32\textwidth}
    \centering
    \includegraphics[width=\linewidth]{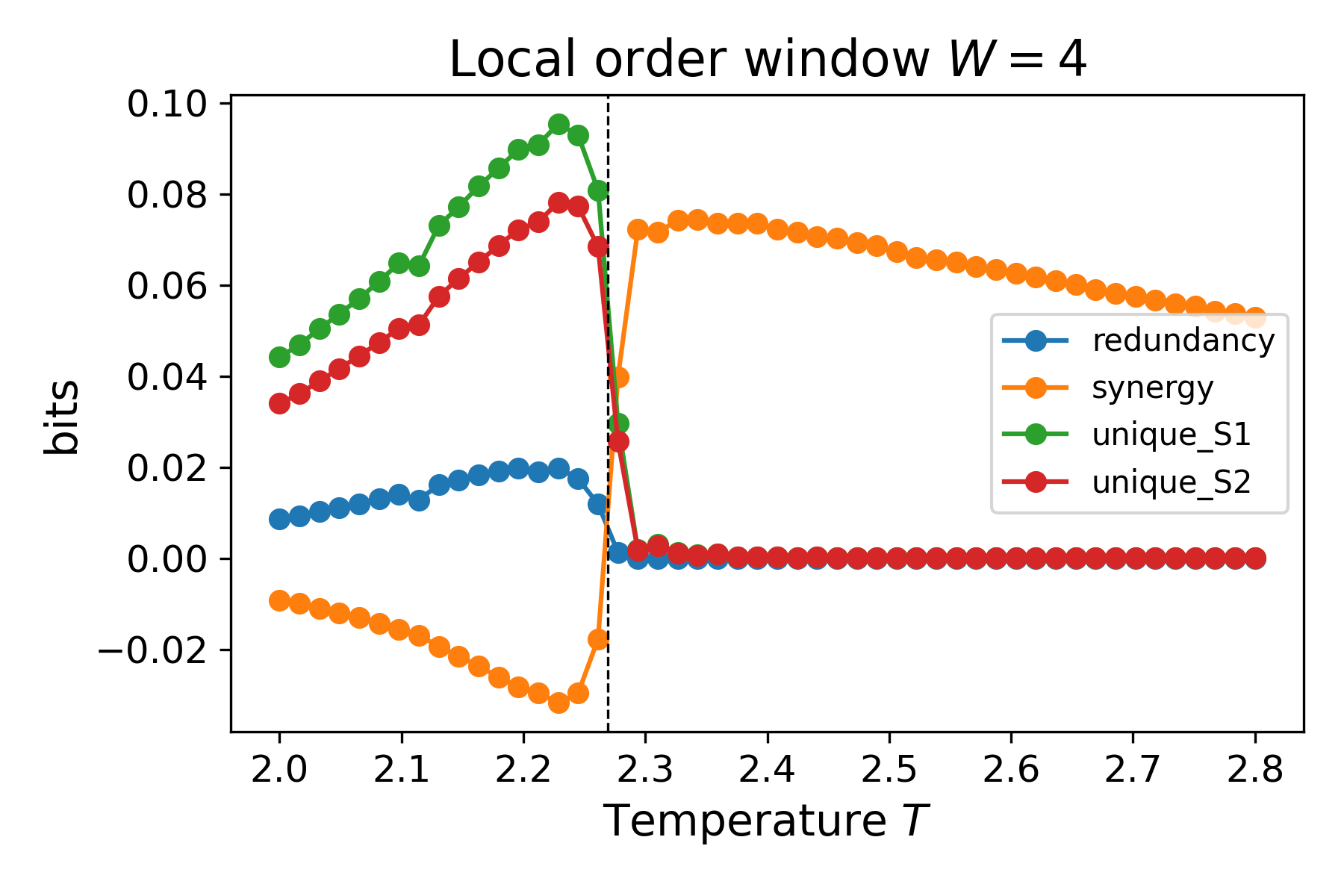}
    \caption*{(A)}
  \end{minipage}
  \hfill
  \begin{minipage}[t]{0.32\textwidth}
    \centering
    \includegraphics[width=\linewidth]{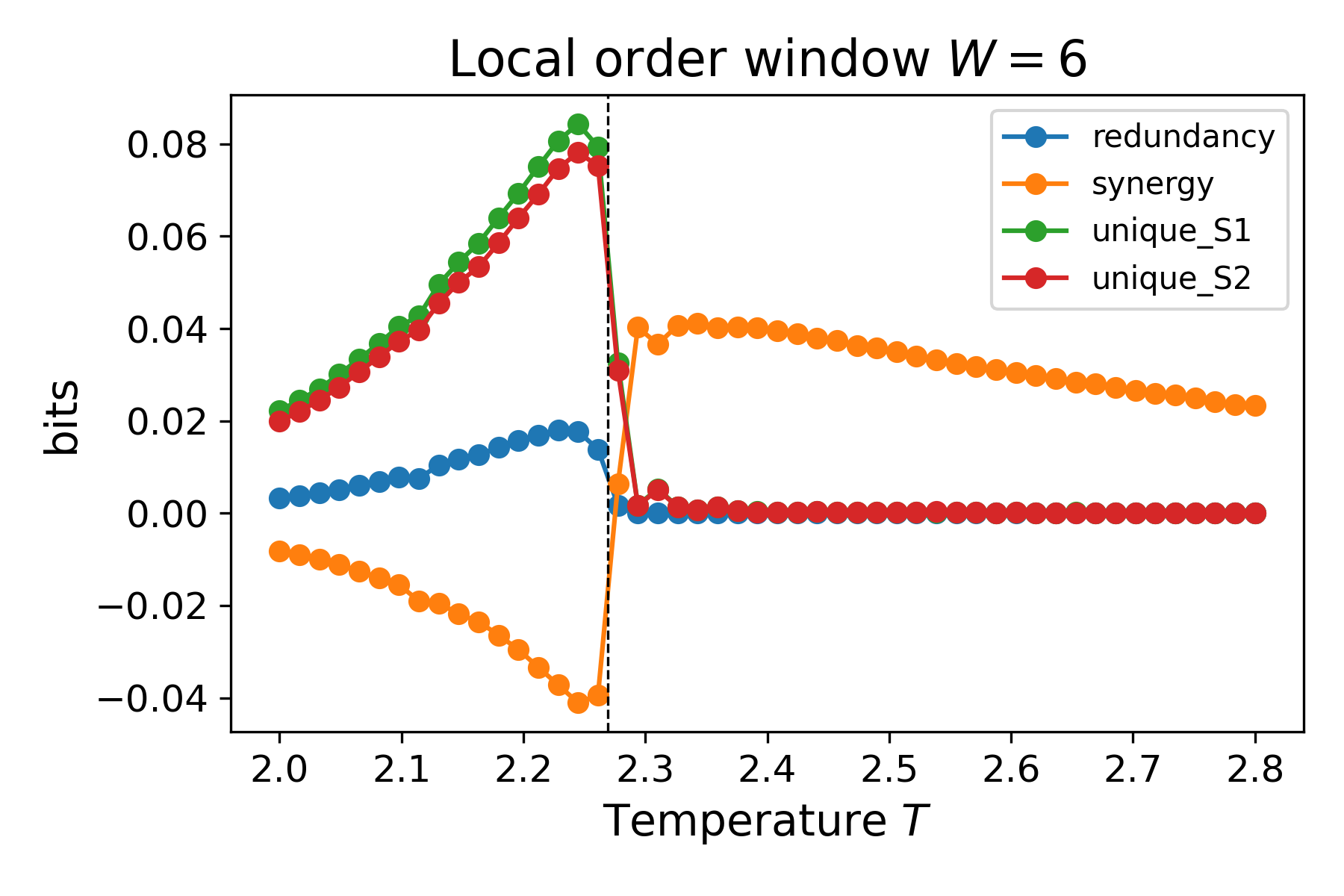}
    \caption*{(B)}
  \end{minipage}
  \hfill
  \begin{minipage}[t]{0.32\textwidth}
    \centering
    \includegraphics[width=\linewidth]{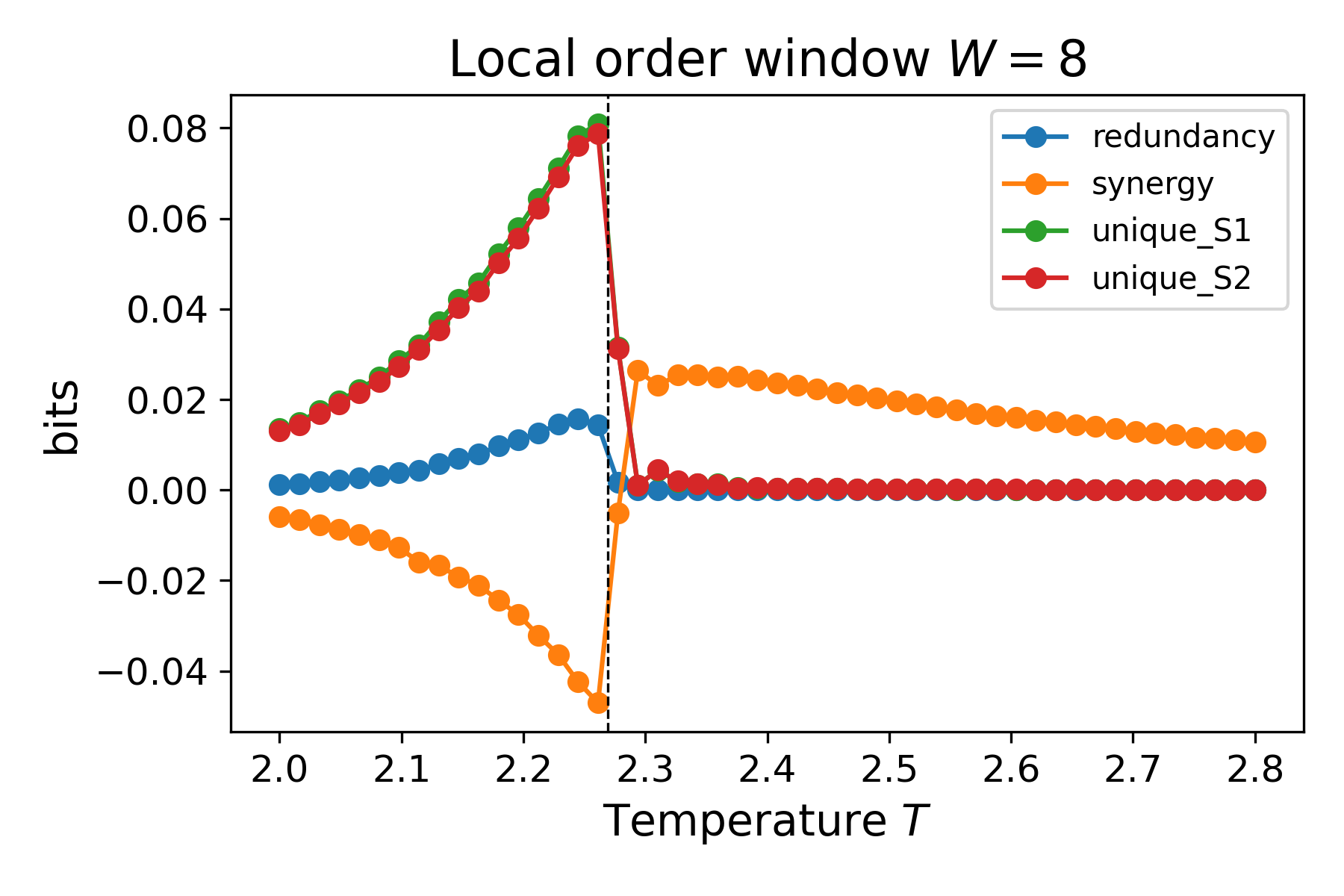}
    \caption*{(C)}
  \end{minipage}
  \caption{Temperature dependence of PID components for detecting emergence.
    Each panel shows redundant, unique, and synergistic information between two adjacent source spins and a coarse-grained target representing the degree of local alignment within the window.
    The vertical dashed line indicates the critical temperature $T_c$.
    Across window sizes, unique and redundant information dominate below $T_c$ and collapse above it, while synergistic information undergoes a sharp transition near criticality and becomes dominant above $T_c$, indicating that the detection of mesoscopic aligned domains increasingly relies on genuinely collective effects.
  }
  \label{fig: exp2 result}
\end{figure}

For each temperature, we randomly select $50$ pairs of adjacent lattice sites.
Each pair $(S_1,S_2)$ consists of two neighboring spins (either horizontal or vertical adjacency), encoded as binary variables. For each selected source pair, we define a square window centered at one of the two spins and compute the absolute mean magnetization within that window,
\[
  \left| \frac{1}{W^2}\sum_{i\in \text{window}} \sigma_i \right|,
\]
where $W$ is the window side length.
This quantity captures the degree of local alignment and serves as a coarse-grained proxy for mesoscopic order associated with the formation of same-spin domains.
To obtain a discrete target variable suitable for PID analysis, the absolute mean magnetization is quantized into three bins:
low ($0$--$0.33$), intermediate ($0.33$--$0.66$), and high local order ($0.66$--$1$).
Figure~\ref{fig: exp2 result} shows the temperature dependence of redundancy, unique information, and synergy for window sizes $W=4,6$ and $8$, respectively.

Across all window sizes, both unique and redundant information display similar qualitative trends in the ordered and pre-critical regimes.
As the temperature approaches the critical point from below, these two components increase monotonically and remain non-negative, before rapidly vanishing above $T_c$.
Unique information grows more steeply and appears as two non-overlapping curves, reflecting geometric asymmetries in the locations of the two source spins within the window, while redundant information increases more slowly and with smaller magnitude.
Taken together, these behaviors indicate that below the phase transition, local alignment is predominantly shaped by direct and partially overlapping contributions from individual spins.
In this regime, neighboring sites influence the development of mesoscopic order in an essentially additive manner, without requiring genuinely collective interactions among sources.

In contrast, synergistic information exhibits a qualitatively distinct signature.
It is suppressed (and slightly negative) throughout the ordered phase, but undergoes a sharp transition near the critical temperature, where it becomes the dominant contribution.
Beyond $T_c$, synergy remains positive and decays gradually with increasing temperature.
This behavior signals a redistribution of informational structure at criticality.
This suggests that predicting whether a mesoscopic aligned domain is present increasingly depends on the joint configuration of multiple spins, rather than on additive or redundant influences from individual sites.
In this sense, the sharp rise of synergy provides a direct informational marker of emergent collective order.

Taken together, these results provide quantitative evidence that the emergence of spatial order in the Ising model is accompanied by a systematic transfer of information from unique and redundant components toward synergistic contributions.
The consistency of this behavior across different window sizes indicates that the observed transition reflects an intrinsic mechanism of emergence rather than an artifact of a particular coarse-graining scale.
Together with Experiment~1, these findings demonstrate that the proposed PID framework captures both predictive and emergent aspects of collective behavior in statistical physical systems.

\section{Discussion}\label{sec:discussion}
\subsection{Summary of Contributions}

In this work, we propose a closed-form solution for PID in the two-source case that satisfies all standard axioms and desirable properties. Our construction eliminates the high-order dependencies that the source variables bring to the target while retaining the marginal distribution between every pair of them, which are the key pairwise relationships. It achieves the maximization of joint entropy while controlling the pair marginal distribution, bringing a different perspective to the information decomposition method.

We then prove that when extending PID to systems with three or more sources, the standard lattice-based framework inherently fails to maintain subsystem consistency. Through both counterexamples and general impossibility results, we proved that no decomposition over antichain lattices can satisfy the full set of axioms. Motivated by this, we abandon the  decomposition over the PID lattice and instead introduce new definitions for multivariate unique and synergistic information that bypass the structural inconsistency.

Our proposed measurement method has advantages over existing PID quantitative methods in terms of the theoretical properties it satisfies. Further, its empirical  behavior is consistent with the known phase transitions and intuitive explanation of the Ising model.
Beyond local predictability, our experiments further demonstrate that the proposed PID measures can detect emergent spatial order by revealing a qualitative redistribution of information toward synergistic contributions near criticality.

\subsection{Structural Inconsistencies of PID}
The core insight from our work is that the difficulty in extending PID beyond two sources is not merely technical, but structural. At the heart of the issue lies the assumption that information atoms can be embedded within a lattice of {antichains}, where redundancy can be recursively distributed via Möbius inversion~\cite{williams2010nonnegative}. While this combinatorial structure formally enables the decomposition of redundancy into all atoms, our results show that such decompositions conflict with the actual behavior of information, leading to overcounting and violations of subsystem consistency.

This contradiction highlights a deeper conceptual flaw: information does not conform to the same topological structure as set inclusion. While the PID lattice is ground on set theory, the structure of multivariate information decomposition exhibits different values at different observation scopes that do not align with simple set unions.
Thus, our counterexamples should not be viewed as adversarial constructions but as demonstrations of this underlying mismatch. They suggest that the PID lattice may be fundamentally mismatched with the structure required for consistent multivariate information accounting.
To resolve this, one must either abandon the goal of defining all atoms simultaneously or discover a fundamentally new topological framework for information decomposition.

Our approach adopts the former strategy: by defining only multivariate unique and synergistic components, we maintain internal consistency and avoid global contradictions. However, we emphasize that this is only a provisional solution. To fully realize the vision of PID, the field must move toward identifying the correct underlying structure---one that generalizes lattices and aligns with the geometry of multivariate information.

\subsection{Limitations and Future Work}
While the proposed definitions of multivariate unique and synergistic information successfully avoid the inconsistencies of lattice-based PID, they only capture two of the many potential types of information atoms. In particular, redundant information in multivariate systems that satisfies all axioms remains an open problem. It cannot be solved by simple extensions of existing ideas.
We plan to extend our framework to address this issue in future work.

Moreover, although our theoretical formulations are well-grounded and validated on synthetic systems, the application of these ideas to real-world data remains largely unexplored. Future work will involve applying our decomposition to empirical domains such as neuroscience, systems biology, and control systems, where high-order informational interactions are common.

\section*{Acknowledgements}
We thank the editor and the reviewers for many helpful comments and suggestions. This work was supported in part by the Air Force Office of Scientific Research (AFOSR) under Grant No. FA9550-23-1-0160.

\section*{Data Availability}

The data supporting the findings of this study are generated by numerical simulations.
The code used to generate all simulation data and reproduce the figures in this article
is publicly available in Ref.~\cite{lyu2026pidcode}.

\newpage
\appendix
\section{Proof of Corollary~\ref{corollary: two result}}
\label{app: proof of corollary}

\begin{proof}
  For the system $(S_1, S_2, T)$ and $(S_1, T)$, according to Lemma~\ref{lemma: subsystem consistency}, let $\boldA=\{S_1,S_2\}$, $\boldB=\{S_1\}$, and $\boldC=\{S_1\}$, then we have
  \begin{align}
    \label{equ:decompose_of_0}
    \Pi^{T}_{1}(\!\bigl\{\!\{1\}\!\bigl\}\!) = \Pi^{T}_{12}(\!\bigl\{\!\{1\}\{2\}\!\bigl\}\!) +\Pi^{T}_{12}(\!\bigl\{\!\{1\}\!\bigl\}\!).
  \end{align}
  Similarly, for the system $(S_1, S_2, T)$ and $(S_2, T)$, we have
  \begin{align}
    \Pi^{T}_{2}(\!\bigl\{\!\{2\}\!\bigl\}\!) = \Pi^{T}_{12}(\!\bigl\{\!\{1\}\{2\}\!\bigl\}\!) +\Pi^{T}_{12}(\!\bigl\{\!\{2\}\!\bigl\}\!).\nonumber
  \end{align}

  Then, following the same approach, we focus on the system $(S_1,S_2,S_3,T)$ and $(S_1, T)$, i.e., we let $\boldA=\{S_1,S_2,S_3\}$, $\boldB=\{S_1\}$, and $\boldC=\{S_1\}$.
  Then, by~Lemma~\ref{lemma: subsystem consistency} we have
  \begin{align}\label{equ:decompose_of_1}
    \Pi^{T}_{1}&(\!\bigl\{\!\{1\}\!\bigl\}\!) =\Pi^{T}_{123}(\!\bigl\{\!\{1\}\{2\}\{3\}\!\bigl\}\!) +\Pi^{T}_{123}(\!\bigl\{\!\{1\}\{2\}\!\bigl\}\!)\nonumber\\
    &+\Pi^{T}_{123}(\!\bigl\{\!\{1\}\{3\}\!\bigl\}\!)+\Pi^{T}_{123}(\!\bigl\{\!\{1\}\{23\}\!\bigl\}\!)+\Pi^{T}_{123}(\!\bigl\{\!\{1\}\!\bigl\}\!).
  \end{align}
  Similarly, for the system $(S_1, S_2,S_3,T)$ and $(S_2, T)$, we have
  \begin{align}
    \Pi^{T}_{2}&(\!\bigl\{\!\{2\}\!\bigl\}\!) =\Pi^{T}_{123}(\!\bigl\{\!\{1\}\{2\}\{3\}\!\bigl\}\!) +\Pi^{T}_{123}(\!\bigl\{\!\{1\}\{2\}\!\bigl\}\!)\nonumber\\
    &+\Pi^{T}_{123}(\!\bigl\{\!\{2\}\{3\}\!\bigl\}\!)+\Pi^{T}_{123}(\!\bigl\{\!\{2\}\{13\}\!\bigl\}\!)+\Pi^{T}_{123}(\!\bigl\{\!\{2\}\!\bigl\}\!),\nonumber
  \end{align}
  where the information atoms contained in both $\Pi^{T}_{1}(\!\bigl\{\!\{1\}\!\bigl\}\!)$ and $\Pi^{T}_{2}(\!\bigl\{\!\{2\}\!\bigl\}\!)$ are $\Pi^{T}_{123}(\!\bigl\{\!\{1\}\{2\}\{3\}\!\bigl\}\!) $ and $\Pi^{T}_{123}(\!\bigl\{\!\{1\}\{2\}\!\bigl\}\!)$.
  Hence, we have
  \begin{align}
    \Pi^{T}_{12}(\!\bigl\{\!\{1\}\{2\}\!\bigl\}\!) &= \Pi^{T}_{123}(\!\bigl\{\!\{1\}\{2\}\{3\}\!\bigl\}\!) +\Pi^{T}_{123}(\!\bigl\{\!\{1\}\{2\}\!\bigl\}\!),\nonumber
  \end{align}
  where $\{\Pi^{T}_{12}(\!\bigl\{\!\{1\}\{2\}\!\bigl\}\!)\}$ and $\{\Pi^{T}_{123}(\!\bigl\{\!\{1\}\{2\}\{3\}\!\bigl\}\!)$, $ \Pi^{T}_{123}(\!\bigl\{\!\{1\}\{2\}\!\bigl\}\!)\}$ are the only atom(s) that appear in both $I(S_1,T)$ (i.e., $\Pi^{T}_{1}(\!\bigl\{\!\{1\}\!\bigl\}\!)$) and $I(S_2,T)$ (i.e., $\Pi^{T}_{2}(\!\bigl\{\!\{2\}\!\bigl\}\!)$) from the decompositions of the systems $(S_1, S_2,T)$ and $(S_1, S_2,S_3,T)$.
  Therefore,~\eqref{equ:cross scale} is proved.

  Then, by~\eqref{equ:cross scale}, \eqref{equ:decompose_of_0}, and \eqref{equ:decompose_of_1}, we have
  \begin{align*}
    \Pi^{T}_{12}(\!\bigl\{\!\{1\}\!\bigl\}\!) \!=\! \Pi^{T}_{123}(\!\bigl\{\!\{1\}\!\{3\}\!\bigl\}\!)\!+\!\Pi^{T}_{123}(\!\bigl\{\!\{1\}\!\{23\}\!\bigl\}\!)\!+\!\Pi^{T}_{123}(\!\bigl\{\!\{1\}\!\bigl\}\!),
  \end{align*}
  which equals~\eqref{equ:cross scale_2}.
\end{proof}

\section{Proof of Lemma~\ref{le:redefinition of redundant and synergistic information}.}
\label{app: proof of def2}

\begin{proof}
  By Lemma~\ref{lemma:red}, we have
  % \begin{align}
  % \label{equ:proof red1}
  $\operatorname{Red}(S_1',S_2\to T)
  = I(S_1;T) - \operatorname{Un}(S_1 \to T | S_2),$
  % \nonumber\\
  which by Def.~\ref{definition:un} equals $H(S_1)-H(S_1|T)-I(S_1';T|S_2).$
  Since by \eqref{equation: S_1'S_2T},
  \begin{align}
    \label{equ:proof red2}
    \Pr(S_1'=s_1,T=t)&= \sum_{s_2} \Pr(S_1'=s_1,S_2=s_2,T=t)\nonumber \\
    &=\sum_{s_2}\Pr(S_1=s_1|T=t)\Pr(S_2=s_2|T=t)\Pr(T=t) \nonumber \\
    &=\Pr(S_1=s_1,T=t),
  \end{align}
  we have:
  \begin{align}
    \operatorname{Red}(S_1',S_2\to T) &=  H(S_1')-H(S_1'|T)-I(S_1';T|S_2) \nonumber\\
    &= H(S_1')-H(S_1'|T)-(H(S_1'|S_2)-H(S_1'|S_2,T)) \nonumber\\
    &= I(S_1';S_2) -H(S_1'|T) +H(S_1'|S_2,T) = I(S_1';S_2), \nonumber
  \end{align}
  where the last step follows since $S_1'-T-S_2$ is a Markov chain from~\eqref{equ:proof red2} and~\eqref{equation: S_1'S_2T}.

  By Lemma~\ref{lemma:syn},
  $\operatorname{Syn}(S_1,S_2\to T) = I(S_1;T|S_2) - \operatorname{Un}(S_1 \to T | S_2) =H(T|S_2)-\operatorname{Un}(S_1 \to T | S_2) - H(T|S_1,S_2),$
  which by Def.~\ref{definition:un}, equals $H(T|S_2)-I(S_1';T|S_2) - H(T|S_1,S_2)=H(T|S_1',S_2)-H(T|S_1,S_2).$
\end{proof}

\section{Closed-form formulation of the information atoms}
\label{app:closed-form formulation}
For two sources $S_1,S_2$ and target $T$, our unique information definition can be written in closed form as a standard conditional mutual information with a constructed variable $S'_1$:
\begin{align*}
  \mathrm{Un}(S_1 \!\to\! T \mid S_2) \;=\; I(S'_1;T \mid S_2),
\end{align*}
where the constructed joint distribution used for the conditional mutual information is given by Eq.~\eqref{equation: S_1'S_2T} as
\begin{align}
  \Pr(S_1'=s_1,S_2=s_2\mid T=t)\triangleq\Pr(S_1=s_1\mid T=t)\Pr(S_2=s_2\mid T=t).\\
  \text{ with }  \Pr(S'_1=s_1\mid T=t) = \Pr(S_1=s_1\mid T=t).
\end{align}
Equivalently, in summation form,
\begin{align*}
  &\mathrm{Un}(S_1 \!\to\! T \mid S_2)\\
  &= \sum_{s_1,s_2,t} \Pr(S_1'=s_1,S_2=s_2,T=t)\,
  \log \frac{\Pr(T=t\mid S_1'=s_1,S_2=s_2)}{\Pr(T=t\mid S_2=s_2)},\\
  &= \sum_{s_1,s_2,t} \Big(\Pr(S_1=s_1\mid T=t)\, \Pr(S_2=s_2\mid T=t)\, \Pr(T=t)\,
  \\&\;\;\;\;\;\;\log \!\frac{\Pr(S_1=s_1\mid T=t)\, \Pr(S_2=s_2\mid T=t)\, \Pr(T=t)}{\Pr(T=t\mid S_2=s_2)\sum_{t'} (\Pr(S_1=s_1\mid T=t') \Pr(S_2=s_2 \mid T=t')\, \Pr(T=t'))}\Big)
\end{align*}
where all conditionals are taken under the constructed distribution $p(s'_1,s_2,t)$ above. The redundant atom then follows from the identity
\begin{align*}
  I(S_1;T) \;=\; \mathrm{Red}(S_1,S_2 \!\to\! T) \;+\; \mathrm{Un}(S_1 \!\to\! T \mid S_2),
\end{align*}
and symmetrically for $S_2$; this implies
\begin{align*}
  \mathrm{Red}(S_1,S_2 \!\to\! T)
  = I(S_1;T) - \mathrm{Un}(S_1 \!\to\! T \mid S_2)
  = I(S_2;T) - \mathrm{Un}(S_2 \!\to\! T \mid S_1).
\end{align*}

\section{Proof of Theorem~\ref{theorem:proof of two}.}
\label{app: proof of satisfaction of two}
\begin{proof}
  The required axioms and properties are proved as follows.

  \emph{\underline{Proof of Axiom~\ref{axiom:mutual constrains}:}}
  Follows immediately from Lemma~\ref{lemma:red} and Lemma~\ref{lemma:syn}.

  \emph{\underline{Proof of Axiom~\ref{axiom: commutativity} (Commutativity):}}

  By switching~${S_1}$ and~${S_2}$ in Lemma~\ref{lemma:red} we have that $\operatorname{Red}(S_2, S_1 \to T) = I(S_2;T) - \operatorname{Un}(S_2 \to T \mid S_1)$.
  based on Lemma~\ref{le:redefinition of redundant and synergistic information}, this equals to $I({S_1};{S'_2})$, where~${S'_2}$ is defined analogously~${S'_1}$ in~\eqref{equation: S_1'S_2T}, i.e.,
  \begin{align}\label{equation:S’_2T}
    \Pr(S_1=s_1,S_2'=s_2|T=t)\triangleq\Pr(S_1=s_1\mid T=t)\Pr(S_2=s_2 \mid T=t).
    %\Pr({S’_1}=x) &= \sum_{y\in\mathcal{{S_2}}}\Pr({S_2}=y)\Pr(A_y=x).
  \end{align}

  {Then, since the right-hand sides of~\eqref{equation: S_1'S_2T} and~\eqref{equation:S’_2T} are identical, we conclude that, for any given value of $t$, the pairs $(S_1', S_2)$ and $(S_1, S_2')$ share the same conditional joint distribution. Consequently, they also share the same (unconditional) joint distribution.}
  % Then, since the joint distributions of $(S_1', S_2)$ and $(S_1, S_2')$, denoted by $\mathcal{D}{(S_1', S_2)}$ and $\mathcal{D}{(S_1, S_2')}$, are generated using the same procedure according to~\eqref{equation: S_1'S_2T} and~\eqref{equation:S’_2T}, which has the same right-hand side.
  Then, we conclude that
  %Take $\mathcal{D}_{({S'_1},{S_2})}$ and $\mathcal{D}_{({S’_2},{S_1})}$ from Def.~\ref{def:WholeTable},
  for~${S'_1}$ and~${S'_2}$ as above, we have that $I({S'_1};{S_2}) = I({S_1};{S'_2})$.
  % \begin{align}
  % \label{equ:I=I}
  %     I({S’_1};{S_2}) = I({S_1};{S'_2})
  % \end{align}
  This, combined with Lemma~\ref{le:redefinition of redundant and synergistic information}, implies the commutativity of~$\operatorname{Red}$.
  % , i.e., $\operatorname{Red}({S_1},{S_2} \to T) = \operatorname{Red}({S_2},{S_1} \to T).$

  \emph{\underline{Proof of Axioms~\ref{axiom: Monotonicity} (Monotonicity) and~\ref{axiom: Self-redundancy} (Self-redundancy):}}

  According to Def.~\ref{definition:un}, $\operatorname{Un}$ is equivalent to a conditional mutual information and hence is nonnegative. Therefore, $\operatorname{Red}({S_1},{S_2}\to T) \le I({S_1};T)$ by Lemma~\ref{lemma:red}. Similarly, by Lemma~\ref{lemma:red} we have~$\operatorname{Red}({S_2},{S_1}\to T)=I({S_2};T)-\operatorname{Un}({S_2}\to T|{S_1})$, and hence $\operatorname{Red}({S_2},{S_1}\to T)\le I({S_2};T)$. Since~$\operatorname{Red}$ is symmetric by Axiom~\ref{axiom: commutativity}, it follows that
  $\operatorname{Red}({S_1},{S_2}\to T)\le \min\{I({S_1};T),I({S_2};T)\}.$
  %Lemma~\ref{lem: Commutativity} we have~$\operatorname{Red}({S_2},{S_1}\to T)=I({S_1};{S’_2})$.

  % combining Axiom~\ref{axiom: commutativity} with Def.~\ref{definition:red} (the latter over

  % implies that $\operatorname{Red}({S_1},{S_2}\to T)=\operatorname{Red}({S_2},{S_1}\to T)$

  % , which implies the following
  % %the expectation of mutual information, and hence which is nonnegativity.
  % %Then, we can easily conclude:
  % \begin{corollary}[Nonnegativity of Unique Information]
  % The unique information in Def.~\ref{definition:un} is nonnegtive, such that:
  %     \begin{align}
  %         \operatorname{Un}({S_1} \to T | {S_2}) \ge 0
  %     \end{align}
  % \end{corollary}
  % Since we have the nonnegativity unique information, we can show
  %Therefore, Axiom~\ref{axiom: Monotonicity and Self-redundancy} follows directly from Def.~\ref{definition:red}.

  \emph{\underline{Proof of Lemma~\ref{Lemma: Nonnegativity} (Nonnegativity):}}
  % , i.e.~\eqref{equ: Nonnegativity}:

  % Proved in
  % \ifthenelse{\boolean{IncludeAppendix}}{Appendix~\ref{proof:un<I}.}
  % {\cite{lyu2024explicit}.}
  Unique information (Def.~\ref{definition:un}) can be written as:
  \begin{align*}
    \operatorname{Un}({S_1} \to T | {S_2})= I({S'_1};T|{S_2})=H({S'_1}|{S_2})-H({S'_1}|T,{S_2})
    =H({S'_1}|{S_2})-H({S_1}|T).
  \end{align*}
  Then, $\operatorname{Un}({S_1} \to T | {S_2})\le I(S_1,T)$ is equivalent to
  $H(S'_1|S_2) - H(S_1|T)\le I(S_1,T)$,
  %That means
  which is equivalent to $H(S'_1|S_2)\le H(S_1)$.
  The latter is true since $\Pr(S_1'=s_1)= \sum_{t}\Pr(S_1'=s_1,T=t)= \sum_{t}\Pr(S_1=s_1,T=t) =\Pr(S_1=s_1)$, for every~$s_1\in\mathcal{S}_1$, by~\eqref{equ:proof red2}.
  % \begin{corollary}[Nonnegativity of Synergistic Information]
  % \label{corollary:Nonnegativity of Synergistic Information}
  % If $H(T|{S_1},{S_2}) = 0$, then Synergistic information (Def.~\ref{definition:syn}) is nonnegative, i.e., $\operatorname{Syn}({S_1},{S_2}\to T) \ge 0$.
  % \end{corollary}

  %\subsubsection{Compatible withShannon’s formula}
  %It is easy to conclude the following corollary just from the definitions:
  %\begin{corollary}Definitions~\ref{definition:un},~\ref{definition:red}, and~\ref{definition:syn} satisfy:
  %\begin{align}\operatorname{Un}({S_1} \to T | {S_2}) + \operatorname{Red}({S_1},{S_2}\to T) = I({S_1},T) \nonumber \\
  %\operatorname{Un}({S_2} \to T | {S_1}) + \operatorname{Red}({S_2},{S_1}\to T) = I({S_2},T) \nonumber \\
  %\operatorname{Un}({S_1} \to T | {S_2}) + \operatorname{Syn}({S_1},{S_2}\to T) = H(T|{S_2}) - H(T|{S_1},{S_2})\nonumber \\
  %\operatorname{Un}({S_2} \to T | {S_1}) + \operatorname{Syn}({S_2},{S_1}\to T) = H(T|{S_1}) - H(T|{S_1},{S_2})
  %\end{align}
  %\end{corollary}

  %means the redundant information in Def.~\ref{definition:red} satisfy:
  % \begin{align}
  % \label{equ:Commutativity of Redundant Information}

  % \end{align}

  \emph{\underline{Proof of Property~\ref{property: Additivity} (Additivity):}}

  % Proved in
  % \ifthenelse{\boolean{IncludeAppendix}}{Appendix~\ref{proof:Additivity of Unique Information}.}
  % {\cite{lyu2024explicit}.}
  By Def.~\ref{definition:un}, we have $
  \operatorname{Un}((S_1,\bar{S_1}) \to (T,\bar{T}) | (S_2,\bar{S_2})) =  I((S_1,\bar{S_1})' ; (T,\bar{T}) | (S_2,\bar{S_2})),$
  where by \eqref{equation: S_1'S_2T},
  \begin{align}
    \label{equ:proof of superposition_2}
    &\Pr((S_1,\bar{S_1})'=(s_1,\bar{s_1}),(S_2,\bar{S_2})=(s_2,\bar{s_2})|(T,\bar{T})=(t,\bar{t}))\nonumber \\
    &=\Pr((S_1,\bar{S_1})=(s_1,\bar{s_1})|(T,\bar{T})=(t,\bar{t}))\cdot \Pr((S_2,\bar{S_2})=(s_2,\bar{s_2})|(T,\bar{T})=(t,\bar{t})).
  \end{align}
  Since $S_1,S_2,T$ are independent of $\bar{S_1},\bar{S_2},\bar{T}$, we have
  \begin{align*}
    \eqref{equ:proof of superposition_2}
    &=\Pr(S_1=s_1|T=t)\Pr(\bar{S_1}=\bar{s_1}|\bar{T}=\bar{t}) \Pr(S_2=s_2|T=t) \Pr(\bar{S_2}=\bar{y}|\bar{T}=\bar{t}) \\
    &\overset{\eqref{equation: S_1'S_2T}}{=}\Pr(S'_1=s_1,S_2=s_2|T=t) \Pr(\bar{S_1}'=\bar{s_1},\bar{S_2}=\bar{s_2}|\bar{T}=\bar{t}),
  \end{align*}
  which implies that the distribution of $S'_1,S_2$ given $T=t$ and $\bar{S_1}',\bar{S_2}$ given $\bar{T}=\bar{t}$ are also independent for every~$t,\bar{t}\in\mathcal{T}\times \bar{\mathcal{T}}$. Therefore, we have
  $
  \operatorname{Un}((S_1,\bar{S_1}) \to (T,\bar{T}) | (S_2,\bar{S_2})) =  I(S'_1,T|S_2) + I(\bar{S_1}',\bar{T}|\bar{S_2})
  = \operatorname{Un}(S'_1\to T|S_2) + \operatorname{Un}(\bar{S_1}'\to \bar{T}|\bar{S_2}).$

  Since mutual information and conditional entropy are additive in the above sense, by Lemma~\ref{lemma:red} and Lemma~\ref{lemma:syn}, $\operatorname{Red}$ and~$\operatorname{Syn}$ are additive as well.

  \emph{\underline{Proof of Property~\ref{property: Continuity} (Continuity):}}

  Recall that Lemma~\ref{le:redefinition of redundant and synergistic information} states that $\operatorname{Red}(S_1,S_2\to T)=I(S'_1;S_2)$. Therefore, since $I(S'_1;S_2)$ is a continuous function of the distribution of $(S'_1,S_2)$ (denoted as $\mathcal{D}_{S'_1,S_2}$), it suffices to prove that
  %the continuity of $\operatorname{Red}(S_1,S_2\to T)$ by showing
  the mapping from the distribution of $(S_1,S_2,T)$ to $(S'_1,S_2)$, i.e.,~$\mathcal{F}(\mathcal{D}_{S_1,S_2,T})=\mathcal{D}_{S'_1,S_2}$ is continuous, which holds by~\eqref{equation: S_1'S_2T}.
  %by Def.~\ref{definition:un} is true.
  Observe that by Lemma~\ref{lemma:red} and Lemma~\ref{lemma:syn},
  $\operatorname{Un}$ and $\operatorname{Syn}$ are continuous as well.
  %(By Def.~\ref{definition:red} and Def.~\ref{definition:syn}, the continuity of~$\operatorname{Un}$ and $\operatorname{Syn}$ can also be derived.)

  %(Apply a perturbation $\sigma$ to the original distribution $\mathcal{D}_{({S_1},{S_2},T)}$ to have the new probability $\Pr({S’_1},{S’_2},T'=x,y,z) = \Pr({S_1},{S_2},T=x,y,z)+\sigma$, which still forms a distribution $\mathcal{D}_{({S’_1},{S’_2},T')}$ but is slightly different from $\mathcal{D}_{({S_1},{S_2},T)}$. )
  % Proved in
  % \ifthenelse{\boolean{IncludeAppendix}}{Appendix~\ref{proof:continuity of do operation}.}
  % {\cite{lyu2024explicit}.}

  \emph{\underline{Proof of Property~\ref{property: Independent Identity} (Independent Identity):}}

  %By Property~\ref{property: Independent Identity}, given
  % Proved in \ifthenelse{\boolean{IncludeAppendix}}{Appendix~\ref{proof:Independent Identity}.}
  % {\cite{lyu2024explicit}.}
  By Def.~\ref{definition:un}, we have~$\operatorname{Un}(S_1 \to T | S_2) = I(S'_1;T|S_2)$, where
  \begin{align}
    \label{equ:proof:Independent Identity_1}
    \Pr(S'_1=s_1,S_2=s_2|T=t) =\Pr(S_1=s_1|T=t)\Pr(S_2=s_2|T=t).
  \end{align}
  Since~$T=(S_1,S_2)$, \eqref{equ:proof:Independent Identity_1} equals zero whenever~$t\ne (s_1,s_2)$, and otherwise
  \begin{align*}
    % \label{equ:proof:Independent Identity_2}
    \eqref{equ:proof:Independent Identity_1} &=\Pr(S_1=s_1|(S_1,S_2)=(s_1,s_2))\Pr(S_2=s_2|(S_1,S_2)=(s_1,s_2))=1.\nonumber
  \end{align*}
  % Also, since $S_1$ and $S_2$ are independent, we have
  % \begin{align*}
  % \end{align*}
  Therefore, we have
  $\Pr(S_1'=s_1,S_2=s_2) = \Pr(T=t) = \Pr(S_1=s_1,S_2=s_2),
  $
  which implies that $S_1'$ and $S_2$ are independent. Then by Lemma \ref{le:redefinition of redundant and synergistic information},
  $
  \operatorname{Red}(S_1 ,S_2 \to T) = I(S_1';S_2)=0.
  $\qedhere
\end{proof}

\section{Proof of Lemma~\ref{lemma: counter example}}
\label{app: proof of counter exp}

\begin{proof}
  %In $(\bar{S}_1,\bar{S}_2,\bar{S}_3,\bar{T})$, let $\bar{S}_1$ and $\bar{S}_2$ be two independent~$\text{Bernoulli}(1/2)$ variables, let $\bar{S}_3 = \bar{S}_1 \oplus \bar{S}_2$, and let~$\bar{T}=(\bar{S}_1,\bar{S}_2,\bar{S}_3)$. Therefore, we have
  Since~$\bar{S}_1$ and~$\bar{S}_2$ are independent, it follows that
  \begin{align}
    \label{equ:I(T;S_1,S_2,S_3)}
    I(\bar{T};\bar{S}_1,\bar{S}_2,\bar{S}_3)=2.
  \end{align}
  Our proof idea is to use Property~\ref{property: Independent Identity} to obtain the values of all PI-atoms in all two-source {sub}systems $(\bar{S}_1,\bar{S}_2,\bar{T}),
  (\bar{S}_1,\bar{S}_3,\bar{T})$ and $(\bar{S}_2,\bar{S}_3,\bar{T})$, and then show that their sum will be greater than the joint mutual information of the system $(\bar{S}_1,\bar{S}_2,\bar{S}_3,\bar{T})$. For simplicity, throughout the following proof, we adopt the convention that all statements are considered for distinct~$i,j,k \in \{1,2,3\}$.

  First, by Property~\ref{property: Independent Identity} (Independent Identity and Remark~\ref{remark:def equal}), and since $\bar{T} =(\bar{S}_1,\bar{S}_2,\bar{S}_3)\overset{\text{det}}{=}(\bar{S}_i,\bar{S}_j)$ we have that
  \begin{align}
    \label{equ:all_three_is_zero}
    \Pi^{\bar{T}}_{ij}(\bigl\{\{i\}\{j\}\bigl\}) = 0.
  \end{align}
  Considering that
  $\Pi^{\bar{T}}_{ij}(\bigl\{\{i\}\{j\}\bigl\})= \Pi^{\bar{T}}_{123}(\bigl\{\{1\}\{2\}\{3\}\bigl\}) + \Pi^{\bar{T}}_{123}(\bigl\{\{i\}\{j\}\bigl\})$,
  which is identical to~\eqref{equ:cross scale}, and by Axiom~\ref{axiom: Monotonicity} (Monotonicity) and Lemma~\ref{Lemma: Nonnegativity} (Nonnegativity), it follows that
  \begin{align}
    \label{equ:pi12=0}
    \Pi^{\bar{T}}_{123}(\bigl\{\{1\}\{2\}\{3\}\bigl\}) = \Pi^{\bar{T}}_{123}(\bigl\{\{i\}\{j\}\bigl\}) =  0.
  \end{align}
  Similarly, \eqref{equ:Information Atoms' relationship_2} implies that $I(\bar{T};\bar{S}_i)= \Pi^{\bar{T}}_{ij}(\bigl\{\{i\}\{j\}\bigl\}) + \Pi^{\bar{T}}_{ij}(\bigl\{\{i\}\bigl\})$, and since $ I(\bar{T};\bar{S}_i) =1$ and due to~\eqref{equ:all_three_is_zero}, it follows that $\Pi^{\bar{T}}_{ij}(\bigl\{\{i\}\bigl\})$ equals $1$,
  which by Corollary~\ref{corollary: two result} (specifically~\eqref{equ:cross scale_2}), {also equals}
  \begin{align}
    \label{equ:pi-red}
    % \eqref{equ:un=1} &=
    \Pi^{\bar{T}}_{123}(\bigl\{\{i\}\bigl\}) + \Pi^{\bar{T}}_{123}(\bigl\{\{i\}\{jk\}\bigl\}) + \Pi^{\bar{T}}_{123}(\bigl\{\{i\}\{k\}\bigl\}).
  \end{align}
  Then, by \eqref{equ:pi12=0} and \eqref{equ:pi-red}, we have
  \begin{align}
    \label{equ:sun=1}
    \Pi^{\bar{T}}_{123}(\bigl\{\{i\}\bigl\})+ \Pi^{\bar{T}}_{123}(\bigl\{\{i\}\{jk\}\bigl\}) = 1,
  \end{align}
  and hence,
  \begin{align*}
    I(\bar{T};\bar{S}_1,\bar{S}_2,\bar{S}_3) &\ge \Pi^{\bar{T}}_{123}(\bigl\{\{1\}\bigl\}) + \Pi^{\bar{T}}_{123}(\bigl\{\{1\}\{23\}\bigl\})  \\&+ \Pi^{\bar{T}}_{123}(\bigl\{\{2\}\bigl\})\nonumber+ \Pi^{\bar{T}}_{123}(\bigl\{\{2\}\{13\}\bigl\}) \\&+ \Pi^{\bar{T}}_{123}(\bigl\{\{3\}\bigl\}) + \Pi^{\bar{T}}_{123}(\bigl\{\{3\}\{12\}\bigl\}) =3,
  \end{align*}
  which contradicts~\eqref{equ:I(T;S_1,S_2,S_3)}.
\end{proof}

\section{Proof of Lemma~\ref{lemma:NoUniversalSubset}}
\label{app: proof of two system}
\begin{proof}
  By the definition of System~1, we have
  \begin{align*}
    I(\hat{T}; \hat{S}_1, \hat{S}_2, \hat{S}_3) &= H(\hat{T}) =H(X_1,X_5,X_9)= 3.
  \end{align*}
  Observe that each one of $x_1, x_5, x_9$ can be predicted by the other two bits in its respective XOR Eq.~\eqref{equation:S1XOR}.
  Therefore, given $\hat{S}_2$ and $\hat{S}_3$ one can determine $x_1$. Thus, the information about $x_1$ is uniquely shared by $\hat{S}_2$ and $\hat{S}_3$ in a synergistic way. This corresponds to the PID atom $\Pi^{\hat{T}}_{123}(\{\{1\}\{23\}\}) = H(x_1) = 1$.
  Similarly, $x_5$ = $x_4 \oplus x_6$ implies that $\hat{S}_1$ and $\hat{S}_3$ together determine $x_5$, yielding $\Pi^{\hat{T}}_{123}(\{\{2\}\{13\}\}) = H(x_5) = 1$, and $x_9$ = $x_7 \oplus x_8$ implies that $\hat{S}_1 $ and $\hat{S}_2$ together determine $x_9$, yielding $\Pi^{\hat{T}}_{123}(\{\{3\}\{12\}\}) = H(x_9) = 1$.
  Then, all remaining atoms are zero in this system.
  A similar argument for System~2 is given shortly.

  To justify the above claims regarding System~1 in formal terms, we
  consider the following three sub-target variables:
  \[
    \hat{T}_1 = x_1, \quad \hat{T}_2 = x_5, \quad \hat{T}_3 = x_9,
  \]
  where \( \hat{T} = (\hat{T}_1, \hat{T}_2, \hat{T}_3) \); the three sub-targets are mutually independent.

  We operate in three steps, where in the first we identify the zero PI-atoms, in the second we identify the nonzero ones, and in the third we combine the conclusions.

  \textbf{Step 1:} Establishing zero PI-atoms.
  By Axiom~\ref{axiom: Self-redundancy} we have~$\operatorname{Red}(\hat{S}_i \to \hat{T}) = I(\hat{S}_i;\hat{T})$, and since
  $I(\hat{S}_2; \hat{T}_1) = I(\hat{S}_3; \hat{T}_1) = 0$,
  it follows that
  \begin{align}
    \label{equ:I(S_i;T_1)=0}
    \Pi^{\hat{T}_1}_{2}(\{\{2\}\}) = \Pi^{\hat{T}_1}_{3}(\{\{3\}\}) = 0.
  \end{align}

  Next, by PID Axiom~\ref{axiom: Monotonicity} (monotonicity), we have $\Pi^{\hat{T}_1}_{ij}(\{\{i\}\{j\}\}) \le \Pi^{\hat{T}_1}_{i}(\{\{i\}\})$, where $\Pi^{\hat{T}_1}_{i}(\{\{i\}\})=0$ for every~$i\ne 1$ by~\eqref{equ:I(S_i;T_1)=0}.
  Therefore, since Lemma~\ref{Lemma: Nonnegativity} (nonnegativity) implies that $\Pi^{\hat{T}_1}_{ij}(\{\{i\}\{j\}\}) \ge 0$, we have
  \begin{align}
    \label{equ:pi12=0_1}
    \Pi^{\hat{T}_1}_{ij}(\{\{i\}\{j\}\}) &= 0, \quad \forall i \neq j \in \{1,2,3\}.
  \end{align}

  Similarly, applying Lemma~\ref{lemma: subsystem consistency} (subsystem consistency), i.e. \eqref{equ:cross scale}, to any system $(\hat{S}_i,\hat{S}_j,\hat{T}_1)$ and to $(\hat{S}_1,\hat{S}_2,\hat{S}_3,\hat{T}_1)$, we have
  \begin{align}
    \Pi^{\hat{T}_1}_{ij}(\{\{i\}\{j\}\}) &= \Pi^{\hat{T}_1}_{123}(\{\{1\}\{2\}\{3\}\}) + \Pi^{\hat{T}_1}_{123}(\{\{i\}\{j\}\}),  %\forall i \neq j \in \{1,2,3\},
    \label{equ:cross_scale_2}
  \end{align}
  and by Axiom~\ref{axiom: Monotonicity} (monotonicity) and Lemma~\ref{Lemma: Nonnegativity} (nonnegativity) we obtain:
  \begin{align}
    0\le \Pi^{\hat{T}_1}_{123}(\{\{1\}\{2\}\{3\}\}) \le \Pi^{\hat{T}_1}_{ij}(\{\{i\}\{j\}\}) \overset{\eqref{equ:pi12=0_1}}{=} 0, \quad \forall i \neq j, \nonumber
  \end{align}
  which implies that
  \begin{align}\label{equation:PIzero123}
    \Pi^{\hat{T}_1}_{123}(\{\{1\}\{2\}\{3\}\})=0.
  \end{align}
  Then, by~\eqref{equ:pi12=0_1} and~\eqref{equ:cross_scale_2}, we have
  \begin{align}
    \label{equ:all_zero}
    \Pi^{\hat{T}_1}_{123}(\{\{i\}\{j\}\}) = 0, \quad \forall i \neq j.
  \end{align}

  Finally, observe that $H(\hat{T}_1,\hat{S}_1|\hat{S}_2, \hat{S}_3)=0$, i.e., the entire system entropy is provided by~$\hat{S}_2,\hat{S}_3$.
  Therefore, all PID atoms that do not include either $\{\hat{S}_2\}$, or
  $\{\hat{S}_3\}$ or $\{\hat{S}_2,\hat{S}_3\}$
  are zero, i.e.,
  \begin{align}
    \label{equ:other_zero}
    \Pi^{\hat{T}_1}_{123}(\{\{1\}\}) &=   \Pi^{\hat{T}_1}_{123}(\{\{12\}\})=\Pi^{\hat{T}_1}_{123}(\{\{13\}\})\nonumber \\&=\Pi^{\hat{T}_1}_{123}(\{\{123\}\})=\Pi^{\hat{T}_1}_{123}(\{\{12\}\{13\}\})=0.
  \end{align}

  \textbf{Step 2:} Determining non-zero PI-atoms.
  First, we have $I(\hat{T}_1; \hat{S}_1) = 1$, and then, we use PID Axiom~\ref{axiom: Self-redundancy} ($I(\hat{T}_1; \hat{S}_1)=\Pi^{\hat{T}_1}_{1}(\{\{1\}\})$) and Lemma~\ref{lemma: subsystem consistency}, i.e.,~\eqref{equ:subsystem} to get
  \begin{align}
    \Pi^{\hat{T}_1}_{1}\!(\{\!\{1\}\!\})&=\Pi^{\hat{T}_1}_{123}(\{\{1\}\{2\}\{3\}\}) \!+\! \Pi^{\hat{T}_1}_{123}(\{\{1\}\{23\}\}) \\ \nonumber
    &\phantom{=\;}+ \Pi^{\hat{T}_1}_{123}(\{\{1\}\{3\}\}) + \Pi^{\hat{T}_1}_{123}(\{\{1\}\{2\}\}) +\Pi^{\hat{T}_1}_{123}(\{\{1\}\}) = 1.
  \end{align}
  Combining this with \eqref{equation:PIzero123}, \eqref{equ:all_zero}, and \eqref{equ:other_zero}, we obtain:
  \[
    \Pi^{\hat{T}_1}_{123}(\{\{1\}\{23\}\}) = 1.
  \]
  The same reasoning applies symmetrically to \( (\hat{S}_1, \hat{S}_2, \hat{S}_3, \hat{T}_2) \) and \( (\hat{S}_1, \hat{S}_2, \hat{S}_3, \hat{T}_3) \), yielding $\Pi^{\hat{T}_2}_{123}(\{\{2\}\{13\}\}) = 1$
  and $\Pi^{\hat{T}_3}_{123}(\{\{3\}\{12\}\}) = 1$.

  \textbf{Step 3:} Final conclusion.
  Since \( \hat{T} = (\hat{T}_1, \hat{T}_2, \hat{T}_3) \) and the three sub-targets are independent, the final conclusion for the system \( (\hat{S}_1, \hat{S}_2, \hat{S}_3, \hat{T}) \) is:
  \begin{align*}
    \Pi^{\hat{T}}_{123}(\{\{i\}\{jk\}\}) = 1, \quad \text{for all distinct}~i,j,k \in [3].
  \end{align*}

  We now turn to discuss System 2 (\(\bar{S}_1,\bar{S}_2,\bar{S}_3,\bar{T}\)).
  Recall that System~2 contains two independent $\operatorname{Bernoulli}(1/2)$ variables~$x_1,x_2$ and their exclusive or~$x_3=x_1\oplus x_2$.
  Then, $\bar{S}_1 = x_1$,
  $\bar{S}_2 = x_2$,
  $\bar{S}_3 = x_3$,
  $\bar{T}   = (x_1, x_2, x_3),$
  and
  \begin{align*}
    I(\bar{T}; \bar{S}_1, \bar{S}_2, \bar{S}_3) &= H(\bar{T}) =H(x_1,x_2,x_3)= 2.
  \end{align*}
  First, by Property~\ref{property: Independent Identity} (Independent Identity and Remark~\ref{remark:def equal}), and since $\bar{T} =(\bar{S}_1,\bar{S}_2,\bar{S}_3)\overset{\text{det}}{=}(\bar{S}_i,\bar{S}_j)$ for every distinct $i,j\in[3]$, we have
  \begin{align}
    \label{equ:all_two_is_zero}
    \Pi^{\bar{T}}_{ij}(\bigl\{\{i\}\{j\}\bigl\}) = 0.
  \end{align}
  Considering that
  $\Pi^{\bar{T}}_{ij}(\bigl\{\{i\}\{j\}\bigl\})= \Pi^{\bar{T}}_{123}(\bigl\{\{1\}\{2\}\{3\}\bigl\}) + \Pi^{\bar{T}}_{123}(\bigl\{\{i\}\{j\}\bigl\})$,
  which is identical to~\eqref{equ:cross scale}, and by Axiom~\ref{axiom: Monotonicity} (Monotonicity) and Lemma~\ref{Lemma: Nonnegativity} (nonnegativity), we have
  \begin{align}
    \label{equ:pi12=0_2}
    \Pi^{\bar{T}}_{123}(\bigl\{\{1\}\{2\}\{3\}\bigl\}) = \Pi^{\bar{T}}_{123}(\bigl\{\{i\}\{j\}\bigl\}) =  0.
  \end{align}
  Similarly, \eqref{equ:Information Atoms' relationship_2} implies that $I(\bar{T};\bar{S}_i)= \Pi^{\bar{T}}_{ij}(\bigl\{\{i\}\{j\}\bigl\}) + \Pi^{\bar{T}}_{ij}(\bigl\{\{i\}\bigl\})$, and moreover since $ I(\bar{T};\bar{S}_i) =1$ and~\eqref{equ:all_two_is_zero}, it follows that $\Pi^{\bar{T}}_{ij}(\bigl\{\{i\}\bigl\})=1$,
  which by Corollary~\ref{corollary: two result}, specifically~\eqref{equ:cross scale_2}, equals
  \begin{align}
    \label{equ:pi-red_2}
    \Pi^{\bar{T}}_{123}(\bigl\{\{i\}\bigl\}) + \Pi^{\bar{T}}_{123}(\bigl\{\{i\}\{jk\}\bigl\}) + \Pi^{\bar{T}}_{123}(\bigl\{\{i\}\{k\}\bigl\}).
  \end{align}
  Then, by \eqref{equ:pi12=0_2} and \eqref{equ:pi-red_2}, we have
  \begin{align}
    \label{equ:all_zero_2}
    \Pi^{\bar{T}}_{123}(\bigl\{\{i\}\bigl\})+ \Pi^{\bar{T}}_{123}(\bigl\{\{i\}\{jk\}\bigl\}) = 1.
  \end{align}
  Similar to System 1, observe that $H(\bar{T},\bar{S}_1|\bar{S}_2, \bar{S}_3)=0$, i.e., the entire system entropy is provided by~$\bar{S}_2,\bar{S}_3$.
  Therefore, all PID atoms that do not include either $\{\bar{S}_2\}$, or
  $\{\bar{S}_3\}$, or $\{\bar{S}_2,\bar{S}_3\}$
  are zero, i.e.,
  \begin{align}\label{equation:Sys1Zeros}
    \Pi^{\bar{T}}_{123}(\{\{1\}\}) &=   \Pi^{\bar{T}}_{123}(\{\{12\}\})=\Pi^{\bar{T}}_{123}(\{\{13\}\})\nonumber \\&=\Pi^{\bar{T}}_{123}(\{\{123\}\})=\Pi^{\bar{T}}_{123}(\{\{12\}\{13\}\})=0.
  \end{align}
  Taking~\eqref{equ:all_zero_2} with $i=1$, we obtain:
  \begin{align*}
    \Pi^{\bar{T}}_{123}(\{\{1\}\{23\}\}) =1.
  \end{align*}
  Next, observing that~$H(\bar{T},\bar{S}_2|\bar{S}_1,\bar{S}_3)=H(\bar{T},\bar{S}_3|\bar{S}_1,\bar{S}_2)=0$, it is readily verified (similar to~\eqref{equation:Sys1Zeros}) that
  \begin{align*}
    \Pi^{\bar{T}}_{123}(\{\{2\}\}) &=   \Pi^{\bar{T}}_{123}(\{\{12\}\})=\Pi^{\bar{T}}_{123}(\{\{23\}\})\nonumber \\&=\Pi^{\bar{T}}_{123}(\{\{123\}\})=\Pi^{\bar{T}}_{123}(\{\{12\}\{23\}\})=0,
  \end{align*}
  and that
  \begin{align*}
    \Pi^{\bar{T}}_{123}(\{\{3\}\}) &=   \Pi^{\bar{T}}_{123}(\{\{23\}\})=\Pi^{\bar{T}}_{123}(\{\{13\}\})\nonumber \\&=\Pi^{\bar{T}}_{123}(\{\{123\}\})=\Pi^{\bar{T}}_{123}(\{\{23\}\{13\}\})=0.
  \end{align*}
  Finally, taking~\eqref{equ:all_zero_2} with $i=2$ and~$i=3$, we have
  $\Pi^{\bar{T}}_{123}(\{\{2\}\{13\}\})=\Pi^{\bar{T}}_{123}(\{\{3\}\{12\}\}) =1$.

  Clearly, System~1 and System~2 above satisfy condition~$(i)$ with respect to the bijection~$\psi$ which maps an antichain in System~1 to an antichain in System~2 having identical indices to all sources (e.g.,~$\psi(\{\{\hat{S}_1\}\}=\{\{\bar{S}_1\}\}$).
    Condition~$(ii)$ is also clearly satisfied.
  \end{proof}

  \section{Proof of the satisfaction for multivariate information measures}
  \label{app: proof of satisfaction}
  \begin{theorem}
    \label{theorem: un bound}
    Def.~\ref{definition:general un} for $\operatorname{Un}(S_i \to T | \boldS_{[N]\setminus i})$ satisfies Axiom~\ref{axiom:bound for un}.
  \end{theorem}

  \begin{proof}
    Commutativity follows immediately from Def.~\ref{definition:general un} and~\eqref{equ:s_1s_N}.
    For monotonicity, by Def.~\ref{definition:general un}, for any $\boldA \subsetneq \boldS_{[N]\setminus i}$,
    \begin{align*}
      \operatorname{Un}(S_i\to T|\boldA) &= I(S_i'; T|\boldA) =H(S_i'|\boldA)-H(S_i'|T,\boldA)\\
      &=H(S_i'|\boldA) - H(S_i'| T)\quad\mbox{[$S_i'- T-\boldA$ is a Markov chain by~\eqref{equ:s_1s_N}.]}\\
      &\ge H(S_i'|\boldS_{[N]\setminus i}) - H(S_i'| T) \quad\mbox{[Since conditioning reduces entropy.]}\\
      &=H(S_i'|\boldS_{[N]\setminus i}) - H(S_i'| T,\boldS_{[N]\setminus i})\;\mbox{[$S_i'- T-\boldA$ is a Markov chain by~\eqref{equ:s_1s_N}.]}\\
      &=I(S'_i;T|\boldS_{[N]\setminus i}),
    \end{align*}
    which by Def.~\ref{definition:general un}, is~$\operatorname{Un}(S_i \to T|\boldS_{[N]\setminus i})$.
    Therefore, we have $$\operatorname{Un}(S_i \to T|\boldS_{[N]\setminus i}) \le \min_{\boldA \subsetneq \boldS_{[N]\setminus i}}\{\operatorname{Un}(S_i\to T|\boldA)\}.$$
    For the bound, by Def.~\ref{definition:general un},
    \begin{align*}
      \operatorname{Un}(S_i \to T|\boldS_{[N]\setminus i}) &= I(S'_i;T|\boldS_{[N]\setminus i})= H(S_i'|\boldS_{[N]\setminus i}) - H(S_i'|\boldS_{[N]\setminus i} ,T)\\
      &=H(S_i'|\boldS_{[N]\setminus i}) - H(S_i'| T) \quad\mbox{[$S_i'- T-\boldS_{[N]\setminus i}$ is a Markov chain by~\eqref{equ:s_1s_N}.]}\\
      &\le H(S_i')- H(S_i'| T) \quad\mbox{[Since conditioning reduces entropy.]}\\
      &= I(S_i';T) = I(S_i;T),
    \end{align*}
    where the last step follows since
    \begin{align}
      \label{eq:same marginal}
      \Pr(S_i'=s_i,T=t) &= \sum_{\bolds_{[N]\setminus i}\in\cS_{[N]\setminus i}}\Pr(S_i'=s_i,\boldS_{[N]\setminus i}=\bolds_{[N]\setminus i},T=t)\\
      &\overset{\eqref{equ:s_1s_N}}{=}\sum_{\bolds_{[N]\setminus i}\in\cS_{[N]\setminus i}}\Pr(S_i=s_i|T=t)\Pr(\boldS_{[N]\setminus i}=\bolds_{[N]\setminus i}|T=t) \Pr(T=t)\\
      &=\Pr(S_i=s_i,T=t).
    \end{align}
    Also, since
    \begin{align*}
      \operatorname {Un}(S_i \to T|\boldS_{[N]\setminus i}) \overset{\eqref{equ:general un}}{=} I(S'_i;T|\boldS_{[N]\setminus i})\le H(T|\boldS_{[N]\setminus i}) \; \mbox{[By definition of mutual information}.],
    \end{align*}
    it follows that $\operatorname {Un}(S_i \to T|\boldS_{[N]\setminus i}) \le \min\{I(S_i;T),H(T|\boldS_{[N]\setminus i})\}.$
  \end{proof}

  \begin{theorem}
    \label{theorem: syn bound}
    Def.~\ref{definition:general syn} ($\operatorname{Syn}(S_1, \dots, S_N\to T)$) satisfies Axiom~\ref{axiom:bound for syn}.
  \end{theorem}

  \begin{proof}
    Commutativity follows immediately from Def.~\ref{definition:general syn} and~\eqref{equation: S[N]'T}.

    For monotonicity, without loss of generality, we prove the claim for the identity permutation $\sigma(i)=i$ for all $i\in[N]$.
    We consider a nontrivial wrapping of the sources $S_i,\dots,S_j$ such that at least two source variables remain after the combination.

    By Definition~5, the monotonicity condition
    \[
      \operatorname{Syn}(S_1,\dots,S_i,\dots,S_j,\dots,S_N\to T)
      \le
      \operatorname{Syn}(S_1,\dots,(S_i,\dots,S_j),\dots,S_N\to T)
    \]
    is equivalent to
    \begin{align}
      \label{eq:ie syn}
      &H(T|\boldS'_{[N]\setminus 1},\dots,\boldS'_{[N]\setminus N}) - H(T|S_1,\dots,S_N) \nonumber\\
      &\le
      H(T|\boldS'_{[N]\setminus 1},\!\dots\!,\boldS'_{[N]\setminus i-1},
        \boldS'_{[N]\setminus\{i,\dots,j\}},
      \boldS'_{[N]\setminus j+1},\!\dots\!,\boldS'_{[N]\setminus N})
      \!-\!
      H(T|S_1,\!\dots\!,(S_i,\dots,S_j),\!\dots\!,S_N),
    \end{align}
    where $\boldS'_{[N]\setminus\{i,\dots,j\}}=(S_1,\dots,S_{i-1},S_{j+1},\dots,S_N)'$.

    The equality
    \[
      H(T|S_1,\dots,S_N)
      =
      H(T|S_1,\dots,(S_i,\dots,S_j),\dots,S_N)
    \]
    holds since wrapping does not change the joint distribution of $(T,S_1,\dots,S_N)$.

    To establish~\eqref{eq:ie syn}, it therefore suffices to show
    \begin{align}
      \label{equ:proof of syn}
      H(T|\boldS'_{[N]\setminus 1},\ldots,\boldS'_{[N]\setminus N})
      \le
      H(T|\boldS'_{[N]\setminus 1},\ldots,\boldS'_{[N]\setminus i-1},
        \boldS'_{[N]\setminus\{i,\dots,j\}},
      \boldS'_{[N]\setminus j+1},\ldots,\boldS'_{[N]\setminus N}).
    \end{align}

    Equation~\eqref{equ:proof of syn} follows because, in the first step, each conditioning variable
    $\boldS'_{[N]\setminus i},\ldots,\boldS'_{[N]\setminus j}$ is replaced by the coarser variable
    $\boldS'_{[N]\setminus\{i,\dots,j\}}$.
    This replacement removes conditioning sub-variables and hence cannot decrease the conditional entropy.
    Equation~\eqref{equ:proof of syn} can be justified more explicitly using the marginal-preservation property.
    By~\eqref{eq:same marginal}, we have
    \begin{align*}
      (\boldS'_{[N]\setminus \{i,\dots,j\}},T)
      &=
      \bigl((S_1,\dots,S_{i-1},S_{j+1},\dots,S_N)',\,T\bigr)  \\
      &=
      (S_1,\dots,S_{i-1},S_{j+1},\dots,S_N,\,T),
    \end{align*}
    and
    \begin{align}
      (\boldS'_{[N]\setminus i},T)
      =
      (\boldS_{[N]\setminus i},T)
      =
      (S_1,\dots,S_{i-1},S_{i+1},\dots,S_j,S_{j+1},\dots,S_N,\,T).
    \end{align}
    Therefore,
    \[
      I\!\left(T;\boldS'_{[N]\setminus \{i,\dots,j\}} \,\middle|\, \boldS'_{[N]\setminus i}\right)=0,
    \]
    which implies that replacing $\boldS'_{[N]\setminus i},\ldots,\boldS'_{[N]\setminus j}$ by the coarser variable
    $\boldS'_{[N]\setminus \{i,\dots,j\}}$ cannot decrease the conditional entropy.
    Consequently,
    \begin{align*}
      H(T|\boldS'_{[N]\setminus 1},\ldots,\boldS'_{[N]\setminus N})
      \le
      H(T|\boldS'_{[N]\setminus 1},\ldots,
        \boldS'_{[N]\setminus i-1},
        \boldS'_{[N]\setminus \{i,\dots,j\}},
        \boldS'_{[N]\setminus j+1},\ldots,
      \boldS'_{[N]\setminus N}).
    \end{align*}
    Repeated occurrences of the same conditioning variable are then merged without affecting the conditional entropy.

    This monotonicity behavior is consistent with the intended scope of Axiom~6: when $N\ge 3$ and the wrapping operation still leaves at least two source variables, wrapping can aggregate synergistic effects of different orders.
    In particular, synergistic contributions that were previously attributable to smaller subsets of sources may be absorbed into the synergy term of the whole source set in the wrapped system; as a result, the total synergistic information can increase under wrapping.

    Finally, to show the bound, by Def.~\ref{definition:general syn} we have
    \begin{align*}
      \operatorname{Syn}(S_1, \dots, S_N\to T)&= (T|\boldS'_{[N]\setminus 1},\dots,\boldS'_{[N]\setminus N}) - H(T|\boldS_{[N]}) \\
      &\le \min_{n\in[N]}\{H(T|\boldS'_{[N]\setminus n})- H(T|\boldS_{[N]}) \}\\
      &\le \min_{\boldA \subsetneq \boldS_{[N]}}\{H(T|\boldA)-H(T|\boldS_{[N]})\} \;\mbox{[Since conditioning reduces entropy]},
    \end{align*}
    where the inequality is due to the fact that the conditional entropy decreases after reducing the given variable.
  \end{proof}

  \section{Proof of lemma~\ref{le:max entropy}}
  \label{app: compareson with broja}

  \begin{proof}
    We begin with the definition of $S_1'$ given in~\eqref{equation: S_1'S_2T}, which defines random variables $(S_1', S_2, T)$ as:
    \[
      \Pr(S_1' = s_1, S_2 = s_2, T = t) = \Pr(S_1 = s_1 \mid T = t) \cdot \Pr(S_2 = s_2 \mid T = t) \cdot \Pr(T = t).
    \]
    We now verify that $(S_1', S_2, T) \in \Delta$:

    First, compute the marginals:
    \begin{align*}
      \Pr(T = t, S'_1 = s_1)  &=\sum_{s_2} \Pr(S'_1=s_1, S'_2=s_2, T=t) \\
      &= \Pr(S_1 = s_1 \mid T=t) \Pr(T=t) \sum_{s_2} \Pr(S_2=s_2 \mid T=t) \\
      &= \Pr(S_1=s_1 \mid T=t) \Pr(T=t)\\
      &=\Pr(T = t, S_1 = s_1)
    \end{align*}
    which shows $\Pr(T = t, S_1' = s_1) = \Pr(T = t, S_1 = s_1)$.
    Thus, $(S_1', S_2, T)$ satisfies both marginal constraints and hence $(S_1', S_2, T) \in \Delta$.

    Next, we argue that this $(S_1', S_2, T)$ is the maximizer of $H(\hat{S}_1, \hat{S}_2, \hat{T})$ over $\Delta$. This follows from the classical principle of maximum entropy under marginal constraints: given fixed marginal distributions, the maximum entropy joint distribution assumes conditional independence~\cite{jaynes1957information}. Specifically, among all $(\hat{S}_1, \hat{S}_2, \hat{T})$ with fixed $(\hat{S}_1, \hat{T})$ and $(\hat{S}_2, \hat{T})$ marginals, the entropy $H(\hat{S}_1, \hat{S}_2, \hat{T})$ is maximized when $\hat{S}_1$ and $\hat{S}_2$ are conditionally independent given $\hat{T}$, i.e.,
    \begin{align*}
      \Pr(\hat{S}_1=s_1, \hat{S}_2=s_2, \hat{T}=t) &= \frac{\Pr(\hat{S}_1=s_1,  \hat{T}=t) \Pr( \hat{S}_2=s_2, \hat{T}=t)}{\Pr(\hat{T}=t)} \\
      &= \Pr(\hat{S}_1=s_1|  \hat{T}=t) \Pr( \hat{S}_2=s_2| \hat{T}=t) \Pr(\hat{T}=t).
    \end{align*}
    Hence, our construction of $(S_1', S_2, T)$ via~\eqref{equation: S_1'S_2T} achieves the maximum entropy among all $(\hat{S}_1, \hat{S}_2, \hat{T}) \in \Delta$.

    Finally, since {Def.~\ref{definition:un}} defines the unique information as
    \[
      \mathrm{Un}(S_1 \rightarrow T \mid S_2) = I(S_1'; T \mid S_2),
    \]
    and our constructed $(S_1', S_2, T)$ coincides with the one used in this definition, we conclude
    \[
      \mathrm{Un}(S_1 \rightarrow T \mid S_2) = I(S_1'; T' \mid S'_2),
    \]
    with $(S_1', S'_2, T') = \arg\max_{(\hat{S}_1, \hat{S}_2, \hat{T}) \in \Delta} H(\hat{S}_1, \hat{S}_2, \hat{T})$.

  \end{proof}

  \bibliography{ref}% Produces the bibliography via BibTeX.

  \end{document}